\documentclass[a4paper,12pt]{article}
\usepackage{amssymb, amsmath, graphicx, stmaryrd}
\usepackage[color,curve]{xypic}
\usepackage{enumerate}
\usepackage{color}
\usepackage{tikz}
\usepackage{tikz-cd}  

\DeclareRobustCommand{\gobblefour}[4]{}
\newcommand*{\SkipTocEntry}{\addtocontents{toc}{\gobblefour}}

\providecommand{\abs}[1]{\lvert#1\rvert}

\providecommand{\norm}[1]{\lVert#1\rVert}

\providecommand{\Spin}{\textnormal{Spin}}

\providecommand{\CCl}{\mathbb{C}\textnormal{l}}

\providecommand{\Tr}{\textnormal{Tr}}

\providecommand{\id}{\textnormal{id}}

\providecommand{\Ker}{\textnormal{Ker}}
\providecommand{\IIm}{\textnormal{Im}}

\providecommand{\Hom}{\textnormal{Hom}}

\providecommand{\Aut}{\textnormal{Aut}}
\providecommand{\Ker}{\textnormal{Ker}}

\providecommand{\DiffVect}{\textnormal{DiffVect}}

\providecommand{\rk}{\textnormal{rk}}

\providecommand{\ch}{\textnormal{ch}}
\providecommand{\PD}{\textnormal{PD}}
\providecommand{\cpt}{\textnormal{cpt}}
\providecommand{\ev}{\textnormal{ev}}
\providecommand{\odd}{\textnormal{odd}}
\providecommand{\GL}{\textnormal{GL}}

\providecommand{\Tor}{\textnormal{Tor}}

\providecommand{\Top}{\textnormal{Top}}

\providecommand{\SO}{\textnormal{SO}}

\providecommand{\PPP}{\mathbb{P}}

\providecommand{\fl}{\textnormal{fl}}

\providecommand{\Fred}{\textnormal{Fred}}
\providecommand{\N}{\mathbb{N}}
\providecommand{\NN}{\mathcal{N}}
\providecommand{\Z}{\mathbb{Z}}
\providecommand{\Q}{\mathbb{Q}}
\providecommand{\R}{\mathbb{R}}
\providecommand{\C}{\mathbb{C}}
\providecommand{\CC}{\mathcal{C}}
\providecommand{\Cc}{\mathfrak{C}}

\providecommand{\Ss}{\mathcal{S}}
\providecommand{\HH}{\mathcal{H}}

\providecommand{\cl}{\textnormal{cl}}
\providecommand{\CS}{\textnormal{CS}}

\providecommand{\dR}{\textnormal{dR}}
\providecommand{\Td}{\textnormal{Td}}
\providecommand{\colim}{\textnormal{colim}}
\providecommand{\A}{\mathcal{A}}
\providecommand{\B}{\mathcal{B}}
\providecommand{\G}{\mathcal{G}}

\providecommand{\ppar}{\textnormal{par}}
\providecommand{\K}{\mathcal{K}}
\providecommand{\U}{\mathfrak{U}}
\providecommand{\uu}{\mathfrak{u}}
\providecommand{\V}{\mathfrak{V}}
\providecommand{\UU}{\textnormal{U}}
\providecommand{\PU}{\textnormal{PU}}
\providecommand{\Ll}{\mathcal{L}}
\providecommand{\LL}{\mathfrak{L}}
\providecommand{\VB}{\textnormal{VB}}
\providecommand{\VBN}{\textnormal{VB}\nabla}
\providecommand{\NIVB}{\textnormal{NIVB}}
\providecommand{\NIVBN}{\textnormal{NIVB}\nabla}
\providecommand{\Gr}{\textnormal{\textbf{G}r}}
\providecommand{\simpc}{\textnormal{sc}}
\providecommand{\Tau}{\mathcal{T}}
\providecommand{\Kint}{K\textnormal{-int}}
\newcommand{\dblsetminus}{\mathbin{{\setminus}\mspace{-5mu}{\setminus}}}

\tikzset{
	curvarr/.style={
		to path={ -- ([xshift=2ex]\tikztostart.east)
			|- (#1) [near end]\tikztonodes
			-| ([xshift=-2ex]\tikztotarget.west)
			-- (\tikztotarget)}
	}
}

\makeatletter
\newcommand\tint{\mathop{\mathpalette\tb@int{t}}\!\int}
\newcommand\bint{\mathop{\mathpalette\tb@int{b}}\!\int}
\newcommand\tb@int[2]{%
  \sbox\z@{$\m@th#1\int$}%
  \if#2t%
    \rlap{\hbox to\wd\z@{%
      \hfil
      \vrule width .35em height \dimexpr\ht\z@+1.4pt\relax depth -\dimexpr\ht\z@+1pt\relax
      \kern.05em % a small correction on the top
    }}
  \else
    \rlap{\hbox to\wd\z@{%
      \vrule width .35em height -\dimexpr\dp\z@+1pt\relax depth \dimexpr\dp\z@+1.4pt\relax
      \hfil
    }}
  \fi
}
\makeatother

\allowdisplaybreaks

\begin{document}

\begin{titlepage}
\titlepage
$ $\vskip 0.5cm
\centerline{ \bf \LARGE Twisted differential K-characters }
\vskip 0.7cm
\centerline{ \bf \LARGE and D-branes }
\vskip 1.7truecm

\begin{center}
{\bf \large Fabio Ferrari Ruffino -- Juan Carlos Rocha Barriga}
\vskip 1.5cm
\em 
Departamento de Matem\'atica -- Universidade Federal de S\~ao Carlos \\
Rod.\ Washington Lu\'is, Km 235 -- C.P.\ 676 - 13565-905 \\
S\~ao Carlos (SP), Brasil

\vskip 2.5cm

\large \bf Abstract
\end{center}

\normalsize We analyse in detail the language of partially non-abelian Deligne cohomology and of twisted differential $K$-theory, in order to describe the geometry of type II superstring backgrounds with D-branes. This description will also provide the opportunity to show some mathematical results of independent interest. In particular, we begin classifying the possible gauge theories on a D-brane or on a stack of D-branes using the intrinsic tool of long exact sequences. Afterwards, we recall how to construct two relevant models of differential twisted $K$-theory, paying particular attention to the dependence on the twisting cocycle within its cohomology class. In this way we will be able to define twisted $K$-homology and twisted Cheeger-Simons $K$-characters in the category of simply-connected manifolds, eliminating any unnatural dependence on the cocycle. The ambiguity left for non simply-connected manifolds will naturally correspond to the ambiguity in the gauge theory, following the previous classification. This picture will allow for a complete characterization of D-brane world-volumes, the Wess-Zumino action and topological D-brane charges within the $K$-theoretical framework, that can be compared step by step to the old cohomological classification. This has already been done for backgrounds with vanishing $B$-field; here we remove this hypothesis.

\vskip1.5cm

\vskip1.5\baselineskip

\vfill
 \hrule width 5.cm
\vskip 2.mm
{\small 
\noindent }
\begin{flushleft}
%\vspace{.5cm}
ferrariruffino@gmail.com, juanrocha@dm.ufscar.br
\end{flushleft}
\end{titlepage}

\newtheorem{Theorem}{Theorem}[section]
\newtheorem{Lemma}[Theorem]{Lemma}
\newtheorem{Corollary}[Theorem]{Corollary}
\newtheorem{ThmDef}[Theorem]{Theorem - Definition}

\newtheorem{Rmk}[Theorem]{Remark}
\newtheorem{Rmks}[Theorem]{Remarks}
\newtheorem{Def}[Theorem]{Definition}
\newtheorem{Not}[Theorem]{Notation}

%%%%%%%%%%%%%%%%%%%%%%%%%%%%%%%%%%%%%%%%%%%%%%%%%%%%%%%%%%%%%%%%%%%%%%%%%%%%%%%%%%%%%%%%%%%%%%%

\tableofcontents

%%%%%%%%%%%%%%%%%%%%%%%%%%%%%%%%%%%%%%%%%%%%%%%%%%%%%%%%%%%%%%%%%%%%%%%%%%%%%%%%%%%%%%%%%%%%%%%

\section{Introduction}

In \cite{BFS} and \cite{FR2} we showed how to classify the possible gauge theories on a D-brane or on a stack of D-branes in type II superstring theory. In particular, the Freed-Witten anomaly cancellation imposes a suitable interaction between the $A$-field and the $B$-field on the world-volume, so that the $A$-field does not always represent an ordinary gauge theory. For this reason, a classification of the possible natures of such a field was necessary, and we realized it using the language of partially non-abelian Deligne cohomology. Nevertheless, we achieved this aim ``by hand'', through the local representation of the fields in a suitable (good) cover, since we were not able to use a more intrinsic machinery when certain twistings were present. In the first part of this paper we reproduce that classification in a completely intrinsic way, only using the tool of long exact sequences in Deligne cohomology.

Afterwards, we link the previous description of the gauge theories to the following more general discussion, concerning the rigorous geometrical and topological foundations of type II superstring theory. As we already summarized in \cite{FR4}, there are two fundamental pictures that describe and classify D-brane charges and the Ramond-Ramond fields in this framework. The first one relies on classical cohomology. In particular, a D-brane world-volume is a submanifold, which becomes a singular cycle via a suitable triangulation, and the Poincar\'e dual of the underlying homology class is the topological charge. The Ramond-Ramond fields are classified by ordinary differential cohomology, for which the Deligne cohomology provides a concrete model \cite{Brylinski}, and the Wess-Zumino action turns out to be the holonomy of a differential cohomology class along the world-volume. The other fundamental classification scheme relies on $K$-theory (see \cite{Evslin, MM, MW, OS, FS, Valentino} among many others; see also \cite{DFM} for a more refined proposal, that we do not consider here, and the recent preprint \cite{GS} for a detailed analysis of the Ramond-Ramond fields). In particular, a D-brane world volume defines a $K$-homology class in the space-time through the Chan-Patton bundle, and its Poincar\'e dual is by definition the topological charge. The Ramond-Ramond fields are classified by a differential $K$-theory class, that is supposed to couple with the world-volume, so that the Wess-Zumino action is well-defined. In \cite{FR4} we tried to clarify how to correctly define the world-volume through the notion of differential $K$-character \cite{BM, FR}, in order to get such a coupling. In this way we managed to draw a complete parallel between the two classification schemes. Nevertheless, since we considered ordinary $K$-theory, we supposed that the $B$-field was vanishing. Here we remove this hypothesis, therefore we develop an analogous construction for twisted $K$-theory and its differential extension.

In this setting, we pay particular attention to the unnatural dependence on the twisting cocycle, developing a model of twisted K-homology and of Cheeger-Simons characters that only depends on the twisting cohomology class as far as it is possible. Coherently, we show that the ambiguity left in such definitions, when certain hypotheses are not satisfied, in spite of being a limit of this model from a purely mathematical point of view, has an interesting physical meaning, since it naturally fits in the classification of the gauge theories we started from. Moreover, we show that the Freed-Witten anomaly cancellation, that has been introduced originally in order for the world-sheet action to be well-defined, is also the condition for the Wess-Zumino action to be well-defined, since it is essentially the hypothesis imposed on the twisting classes in the definition of Cheeger-Simons $K$-character.

The paper is organized as follows. In chapter \ref{TwistedVBSec} we deal with partially non-abelian Deligne cohomology and we use it to classify the gauge theories on a D-brane. In chapter \ref{DiffTwistedKTh} we review and analyse two relevant models of differential twisted K-theory, also completing some constructions that we did not find explicitly in literature. In chapter \ref{TwKChar} we provide the definition of twisted Cheeger-Simons characters and of twisted K-homology from the viewpoint sketched above. We conclude with chapter \ref{DBraneCWZ}, in which we apply the previous material to complete the rigorous gemetrical description of type II supestring background with D-branes.

%%%%%%%%%%%%%%%%%%%%%%%%%%%%%%%%%%%%%%%%%%%%%%%%%%%%%%%%%%%%%%%%%%%%%%%%%%%%%%%%%%%%%%%%%%%%%%%

\section{Partially non-abelian Deligne cohomology}\label{TwistedVBSec}

We are going to recall the notion of twisted vector bundle with connection and to insert it within the framework of partially non-abelian Deligne cohomology. In this way we will be able to classify the possible gauge theories on a D-brane or on a stack of D-branes, using the intrinsic language of long exact sequences.

\subsection{Brief review on the Deligne cohomology}\label{ReviewCoh}

Given a smooth manifold $X$, we consider the complex of sheaves:
\begin{equation}\label{ComplexSp}
	S^{p}_{X} := \underline{\UU}(1) \overset{\!\tilde{d}}\longrightarrow \Omega^{1}_{\R} \overset{d}\longrightarrow \cdots \overset{d}\longrightarrow \Omega^{p}_{\R},
\end{equation}
where $\underline{\UU}(1)$ is the sheaf of smooth $\UU(1)$-valued functions, $\Omega^{k}_{\R}$ is the sheaf of real $k$-forms, $d$ is the exterior differential and $\tilde{d}f := \frac{1}{2\pi i} f^{-1}df$. The Deligne cohomology group of degree $p$ on $X$ is by definition the sheaf hypercohomology group of the complex \eqref{ComplexSp}, i.e., $\check{H}^{p}(X, S^{p}_{X})$. It can be concretely described through a good cover $\mathfrak{U} = \{U_{i}\}_{i \in I}$ of $X$ as follows: we consider the double complex whose rows are the $\rm\check{C}$ech complexes of the sheaves involved in \eqref{ComplexSp}, and we consider the cohomology of the associated total complex. This means that a $p$-cocycle consists in a sequence $(\{g_{i_{0} \cdots i_{p}}\}, \{(C_{1})_{i_{0} \cdots i_{p-1}}\}, \ldots, \{(C_{p-1})_{i_{0}i_{1}}\}, \{(C_{p})_{i_{0}}\})$, where $C_{k}$ is a $k$-form, satisfying the conditions:
\begin{equation}\label{PGerbesCocycle}
\begin{array}{l}
	(C_{p})_{i_{1}} - (C_{p})_{i_{0}} = d(C_{p-1})_{i_{0}i_{1}}\\
	(C_{p-1})_{i_{1}i_{2}} - (C_{p-1})_{i_{0}i_{2}} + (C_{p-1})_{i_{0}i_{1}} = -d(C_{p-2})_{i_{0}i_{1}i_{2}}\\
	\ldots\\
	\check{\delta}^{p-1}(C_{1})_{i_{0}\ldots i_{p-1}} = \frac{(-1)^{p+1}}{2\pi i} g_{i_{0}\ldots i_{p}}^{-1} dg_{i_{0}\ldots i_{p}} \\
	\check{\delta}^{p}g_{i_{0}\ldots i_{p}} = 1.
\end{array}
\end{equation}
We call $\G := [g, C_{1}, \ldots, C_{p}]$ the corresponding cohomology class. The local forms $C_{p}$ correspond to the Ramond-Ramond potentials in string theory (if we consider this model) and their differentials $dC_{p}$ glue to a global gauge-invariant closed form $G_{p+1}$ (the Ramond-Ramond field strength), which is called \emph{curvature}. Moreover, from the underlying class $[\{g_{i_{0} \cdots i_{p}}\}] \in \check{H}^{p}(\U, \underline{\UU}(1))$, applying the isomorphism $\check{H}^{p}(\U, \underline{\UU}(1)) \simeq H^{p+1}(X; \Z)$, we get the \emph{first Chern class} $c_{1}(\G) \in H^{p+1}(X; \Z)$. The de-Rham cohomology class, represented by the curvature, is the real image of the first Chern class (Dirac quantization condition), therefore the curvature has integral periods. Of course we are free to add any coboundary to the cocycle $(g, C_{1}, \ldots, C_{p})$, the meaningful datum being the corresponding cohomology class, since it is determined by the two real physical observables: the field strength $G_{p+1}$ (corresponding to the field $F$ in electromagnetism) and the holonomy or Wess-Zumino action (providing the additional piece of information that, in electromagnetism, is measured by the phase difference in the Aranhov-Bohm effect). The latter can be computed on a singular $p$-cycle of $X$, integrating the local $k$-forms $C_{k}$ on the $k$-simplicies and summing the results in a suitable way \cite{GT}. It is related to the curvature by a Stokes-type formula, hence, if the class is flat (i.e.\ the curvature vanishes), then its holonomy on a cycle only depends on the underlying homology class.

\paragraph{\textbf{Differential cohomology diagram.}} We set $\hat{H}^{p}(X) := \check{H}^{p-1}(S^{p-1}_{X})$ and we get the following commutative diagram \cite{HS}:
\begin{equation}\label{DiagramDC}
\xymatrix{
	\hat{H}^{\bullet}(X) \ar@{->>}[rr]^{c_{1}} \ar@{->>}[d]_{curv} & & H^{\bullet}(X; \Z) \ar[d]^{\otimes_{\Z} \R} \\
	\Omega_{int}^{\bullet}(X) \ar[rr]^{dR} & & H^{\bullet}_{dR}(X).
}
\end{equation}
Here $c_{1}$ is the first Chern class, $curv$ is the curvature, $dR$ is the de-Rham cohomology class and $\Omega_{int}^{\bullet}(X)$ is the group of closed real forms with integral periods. The surjective map $c_{1}$ in diagram \eqref{DiagramDC} shows that $\hat{H}^{\bullet}(X)$ is a differential refinement of $H^{\bullet}(X; \Z)$.

\paragraph{\textbf{Relative Deligne cohomology.}} Given a smooth closed embedding $\rho \colon Y \hookrightarrow X$, we consider the complexes of sheaves $S^{p}_{X}$ and $\rho_{*}S^{p-1}_{Y}$ on $X$. We recall that, for any open subset $U \subset X$, by definition $(\rho_{*}S^{p-1}_{Y})(U) := S^{p-1}_{Y}(\rho^{-1}U)$. We have the natural morphism
\begin{equation}\label{RhoExclamDiff}
	\rho^{!} \colon S^{p}_{X} \to \rho_{*}S^{p-1}_{Y},
\end{equation}
defined as follows:
\begin{equation}\label{RelativeDeligneCplx}
\xymatrix{
	\underline{\UU}(1)_{X} \ar[r]^{\tilde{d}} \ar[d]^{\rho^{!,0}} & \Omega^{1}_{X, \R} \ar[r]^{d} \ar[d]^{\rho^{!,1}} & \cdots \ar[r]^(.45){d} & \Omega^{p-1}_{X, \R} \ar[r]^{d} \ar[d]^{\rho^{!,p-1}} & \Omega^{p}_{X, \R} \ar[d] \\
	\rho_{*}\underline{\UU}(1)_{Y} \ar[r]^{\tilde{d}} & \rho_{*}\Omega^{1}_{Y, \R} \ar[r]^(.55){d} & \cdots \ar[r]^(.4){d} & \rho_{*}\Omega^{p-1}_{Y, \R} \ar[r] & 0,
}\end{equation}
where $\rho^{!, q}(\omega) = \rho^{*}\omega$ for any $q \geq 1$ and $\rho^{!, 0}(f) = f \circ \rho$. The corresponding cone complex is the following one:
\begin{equation}\label{ConeCplx}
	\check{\pmb{C}}^{\bullet}(\rho) := \check{C}^{\bullet}(S^{p}_{X}) \oplus \check{C}^{\bullet-1}(\rho_{*}S^{p-1}_{Y}) \qquad\qquad \check{\pmb{D}}^{\bullet}(\alpha, \beta) := \bigl( \check{D}^{\bullet}(\alpha), \rho^{!}(\alpha) - \check{D}^{\bullet-1}(\beta) \bigr).
\end{equation}
The cohomology groups of $(\check{\pmb{C}}^{\bullet}(\rho), \check{\pmb{D}}^{\bullet})$ are by definition the relative Deligne cohomology groups of $\rho$.

\begin{Rmk}\label{RelativeLB} \emph{The case $p = 1$ is quite clear geometrically. In fact, a relative Deligne class is represented by a cocyle of the form $(\alpha, \beta)$, where $\alpha = (\{g_{ij}\}, \{A_{i}\}) \in \check{Z}^{1}(S^{1}_{X})$ represents a line bundle with connection $(L, \nabla)$ and $\beta = \{h_{i}\} \in \check{C}^{0}(\underline{\UU}(1)_{Y})$ represents a trivialization (i.e.\ a global non-vanishing section) of $\rho^{*}L$. Let us see why such a construction is natural. Fixing a piece-wise smooth curve $\gamma \colon I \to X$, such that $\gamma(\partial I) \subset Y$, the parallel transport of $\nabla$ along $\gamma$ is not well-defined as a complex number, since $\gamma$ is not a closed curve. Nevertheless, since $\gamma(\partial I) \subset Y$, the fixed trivialization of $L\vert_{Y}$ provides a canonical way to identify the fibres $L_{\gamma(0)}$ and $L_{\gamma(1)}$ with $\C$, hence the parallel transport, as a unitary liner map from $L_{\gamma(0)}$ to $L_{\gamma(1)}$, becomes a well-defined number belonging to $\UU(1)$. Therefore, the 1-Deligne cohomology group of $\rho$ naturally leads to the notion of \emph{relative} holonomy.}
\end{Rmk}

In general, if $(\alpha, \beta)$ represents a relative cocycle of degree $p$, then $\alpha$ represents a Deligne $p$-cohomology class on $X$ such that $\rho^{*}\alpha$ has trivial first Chern class. This means that $\rho^{*}\alpha$ is topologically trivial, but not necessarily trivial as a Deligne class. It follows that $\rho^{*}\alpha$ is cohomologous to a cochain of the form $(1, 0, \ldots, 0, G)$, where $G$ is a global potential. The cochain $\beta$ provides a suitable reparametrization, i.e.\ $\rho^{*}\alpha - \check{D}\beta = (1, 0, \ldots, 0, G)$. By definition, the \emph{curvature} of $[(\alpha, \beta)]$ is the relative form $(F, G) \in \Omega^{p}(\rho)$, where $F$ is the curvature of $\alpha$ in $X$ and $G$ is the global potential on $Y$. Now we can understand why, in \eqref{RelativeDeligneCplx}, the complex on $Y$ has been truncated at degree $p-1$, not $p$. That's because, if we reach $p$ even in the lower row, then the cocycle condition also imposes $G = 0$, hence $\rho^{*}\alpha$ must be trivial (not only topologically). Therefore, the morphism $\rho^{!} \colon S^{p}_{X} \to \rho_{*}S^{p}_{Y}$ (with $p$ on both sides) leads to a proper subgroup of relative $p$-Deligne cohomology, whose elements are called \emph{parallel classes}:
\begin{equation}\label{RelativeDeligneCplxPar}
\xymatrix{
	\underline{\UU}(1)_{X} \ar[r]^{\tilde{d}} \ar[d]^{\rho^{!,0}} & \Omega^{1}_{X, \R} \ar[r]^{d} \ar[d]^{\rho^{!,1}} & \cdots \ar[r]^(.45){d} & \Omega^{p-1}_{X, \R} \ar[r]^{d} \ar[d]^{\rho^{!,p-1}} & \Omega^{p}_{X, \R} \ar[d]^{\rho^{!,p}} \\
	\rho_{*}\underline{\UU}(1)_{Y} \ar[r]^{\tilde{d}} & \rho_{*}\Omega^{1}_{Y, \R} \ar[r]^(.55){d} & \cdots \ar[r]^(.4){d} & \rho_{*}\Omega^{p-1}_{Y, \R} \ar[r]^(.48){d} & \rho_{*}\Omega^{p}_{Y, \R}.
}\end{equation}
We set:
\begin{equation}\label{RelDelignePar}
	\hat{H}^{p+1}(\rho) := \check{H}^{p}(\rho^{!} \colon S^{p}_{X} \to \rho_{*}S^{p-1}_{Y}) \qquad\qquad \hat{H}^{p+1}_{\ppar}(\rho) := \check{H}^{p}(\rho^{!} \colon S^{p}_{X} \to \rho_{*}S^{p}_{Y}).
\end{equation}
We get the following diagram, generalizing \eqref{DiagramDC} to the relative framework:
\begin{equation}\label{DiagramDCRel}
\xymatrix{
	\hat{H}^{\bullet}(\rho) \ar@{->>}[rr]^(.43){c_{1}} \ar@{->>}[d]_{curv} & & H^{\bullet}(\rho; \Z) \ar[d]^{\otimes_{\Z} \R} \\
	\Omega_{int}^{\bullet}(\rho) \ar[rr]^(.45){dR} & & H^{\bullet}_{dR}(\rho).
}
\end{equation}
Moreover, $\hat{H}^{p+1}_{\ppar}(\rho)$ is the subgroup of $\hat{H}^{p+1}(\rho)$ formed by classes with curvature of the form $(F, 0)$. It is possible to define the holonomy of a class $(\alpha, \beta) \in \hat{H}^{p+1}(\rho)$ on relative $p$-cycles, such cycles being defined through the homological version of the cone complex.

\begin{Rmk} \emph{In the case $p=1$, using the notations of remark \ref{RelativeLB}, a class is parallel when the fixed trivialization of $L\vert_{Y}$ is a global \emph{parallel} section with respect to $\nabla$. It follows that $\nabla\vert_{Y} = 0$.}
\end{Rmk}

\subsection{Twisted vector bundles}\label{SecTwVB}

We fix a paracompact topological space $X$ and a good cover $\U = \{U_{i}\}_{i \in I}$, whose existence we assume by hypothesis. Every smooth manifold admits a good cover \cite{BT}. We denote by $\underline{\UU}(r)$ the sheaf of $\UU(r)$-valued continuous functions on $X$ and, when $r = 1$, we denote by $\check{C}^{\bullet}(\U, \underline{\UU}(1))$, $\check{Z}^{\bullet}(\U, \underline{\UU}(1))$ and $\check{H}^{\bullet}(\U, \underline{\UU}(1))$ the corresponding Cech cochains, cocyles and cohomology classes, with respect to the fixed good cover $\U$.

\begin{Def}\label{DefTwistedBdl} Given a cochain $\zeta := \{\zeta_{ijk}\} \in \check{C}^{2}(\U, \underline{\UU}(1))$, a \emph{$\zeta$-twisted vector bundle} of rank $r$ on $X$ is a collection of trivial Hermitian vector bundles $\pi_{i} \colon E_{i} \to U_{i}$ of rank $r$ and of unitary vector bundle isomorphisms $\varphi_{ij} \colon E_{i}\vert_{U_{ij}} \to E_{j}\vert_{U_{ij}}$, such that $\varphi_{ki} \varphi_{jk} \varphi_{ij} = \zeta_{ijk} \cdot \id$.
\end{Def}

Of course, when $\zeta_{ijk} = 1$, we get an ordinary vector bundle by identifying $v$ with $\varphi_{ij}(v)$ for every $v \in E_{i}\vert_{U_{ij}}$. It is easy to prove by direct computation that, if there exists a $\zeta$-twisted vector bundle, then $\zeta$ is necessarily a cocyle, hence the cohomology class $[\zeta] \in \check{H}^{2}(\U, \underline{\UU}(1)) \simeq H^{3}(X; \Z)$ is well-defined. We will show in remark \ref{TorsionRmk} that it is necessarily a torsion class.

\begin{Def}\label{MorphTwistedVB} Given two $\zeta$-twisted vector bundles $E := (\{E_{i}\}, \{\varphi_{ij}\})$ and $F := (\{F_{i}\}, \{\psi_{ij}\})$, a \emph{morphism} from $E$ to $F$ is a collection of vector bundle morphisms $f_{i} \colon E_{i} \to F_{i}$ such that $f_{j} \circ \varphi_{ij} = \psi_{ij} \circ f_{i}$ for every $i, j \in I$. The morphism is called \emph{unitary} if each $f_{i}$ is.
\end{Def}

Of course an \emph{isomorphism} is an invertible morphism, and this is equivalent to requiring that each $f_{i}$ is a vector bundle isomorphism. The following definition easily generalizes to the non-abelian setting the basic tools of Cech cohomology in low degree.

\begin{Def}\label{DefCoch01} A Cech cochain of degree $1$ of the sheaf $\underline{\UU}(r)$ is a collection of continuous functions $\{g_{ij} \colon U_{ij} \to \UU(r)\}$. Similarly, a Cech cochain of degree $0$ is a collection of continuous functions $\{g_{i} \colon U_{i} \to \UU(r)\}$. We denote the set of $p$-cochains, for $p \in \{0, 1\}$, by $\check{C}^{p}(\U, \underline{\UU}(r))$. Given $\zeta := \{\zeta_{ijk}\} \in \check{C}^{2}(\U, \underline{\UU}(1))$, a cochain $\{g_{ij}\} \in \check{C}^{1}(\U, \underline{\UU}(r))$ is called \emph{$\zeta$-cocyle} if $g_{ki} g_{jk} g_{ij} = \zeta_{ijk} \cdot I_{r}$. We denote by $\check{Z}^{1}_{\zeta}(\U, \underline{\UU}(r))$ the set of $\zeta$-cocycles.
\end{Def}

There is a natural action of $0$-cochains on $1$-cochains, defined by $\{h_{i}\} \cdot \{g_{ij}\} := \{h_{i}g_{ij}h_{j}^{-1}\}$. It is easy to prove that such an action determines an equivalence relation in $\check{C}^{1}(\U, \underline{\UU}(r))$, that restricts to an equivalence relation in $\check{Z}^{1}_{\zeta}(\U, \underline{\UU}(r))$.

\begin{Def}\label{DefTwistedTorsion} The \emph{$\zeta$-twisted cohomology set} of degree $1$ and rank $r$, that we denote by $\check{H}^{1}_{\zeta}(\U, \underline{\UU}(r))$, is the quotient of $\check{Z}^{1}_{\zeta}(\U, \underline{\UU}(r))$ by the action of $0$-cochains.
\end{Def}

Clearly, when $\zeta = 1$, we get ordinary non-abelian cohomology of degree $1$, that classifies the isomorphism classes of rank-$r$ vector bundles on $X$. We can easily show that the same happens for any $\zeta$. In fact, given a twisted vector bundle $E := (\{E_{i}\}, \{\varphi_{ij}\})$ of rank $r$, for each $i \in I$ we can fix a set of $r$ pointwise-independent local sections $s_{1, i}, \ldots, s_{r, i} \colon U_{i} \to E_{i}$ of unit norm, determining vector bundle isomorphisms $\xi_{i} \colon E_{i} \to U_{i} \times \C^{r}$, $\lambda^{k}s_{i,k}(x) \mapsto (x, (\lambda^{1}, \ldots, \lambda^{r}))$. The isomorphisms $\varphi_{ij}$ determine local transition functions $g_{ij} \colon U_{ij} \to \UU(r)$ such that $\varphi_{ij}(\xi_{i}^{-1}(x, \lambda)) = \xi_{j}^{-1}(x, g_{ij}(x) \cdot \lambda)$. Equivalently, $g_{ij}(x)$ is the change of basis in $(E_{j})_{x}$ from $\{s_{j,1}(x), \ldots, s_{j,r}(x)\}$ to $\{\varphi_{ij}(s_{i,1}(x)), \ldots, \varphi_{ij}(s_{i,r}(x))\}$. The condition $\varphi_{ki} \varphi_{jk} \varphi_{ij} = \zeta_{ijk} \cdot \id$ is equivalent to $g_{ki} g_{jk} g_{ij} = \zeta_{ijk} \cdot I_{n}$, hence $\{g_{ij}\} \in \check{Z}^{1}_{\zeta}(\U, \underline{\UU}(r))$. Finally, it is straightforward to verify, as for ordinary vector bundles, that the cohomology class $[\{g_{ij}\}] \in \check{H}^{1}_{\zeta}(\U, \underline{\UU}(r))$ only depends on the isomorphism class of $E := (\{E_{i}\}, \{\varphi_{ij}\})$, in such a way that we get the natural bijection $[E] \mapsto [\{g_{ij}\}]$.

\begin{Rmk}\label{TorsionRmk} \emph{The class $[\zeta]$ is necessarily torsion. In fact, computing the determinants, we get $\det(g_{ki})\det(g_{jk})\det(g_{ij}) = \zeta_{ijk}^{r}$; since $\det(g_{ij})$ is a $\UU(1)$-valued function, this shows that $\{\zeta_{ijk}^{r}\}$ is a trivial cocyle, hence $[\zeta]^{r} = 1$ (or $r[\zeta] = 0$, thinking of $H^{3}(X; \Z)$). In particular, the order of $[\zeta]$ divides $r$. One can prove that, for any cocycle representing a torsion class, there exist some corresponding twisted bundles \cite{AS}.}
\end{Rmk}

\paragraph{\textbf{Dependence on the cocycle.}} Let us suppose that $\zeta = \{\zeta_{ijk}\}$ and $\xi = \{\xi_{ijk}\}$ are cohomologous and let us fix $\eta = \{\eta_{ij}\}$ such that $\xi = \zeta \cdot \check{\delta}^{1}\eta$. We get the following isomorphism:
\begin{equation}\label{IsoPhiEta}
\begin{split}
	\Phi_{\eta} \colon & \check{H}^{1}_{\zeta}(\U, \underline{\UU}(r)) \overset{\!\simeq}\longrightarrow \check{H}^{1}_{\xi}(\U, \underline{\UU}(r)) \\
	& [\{g_{ij}\}] \mapsto [\{g_{ij} \cdot \eta_{ij}\}].
\end{split}
\end{equation}
Equivalently, calling $\VB_{\zeta}(X)$ the set of isomorphism classes of $\zeta$-twisted vector bundles:
\begin{equation}\label{IsoPhiEtaBdl}
\begin{split}
	\Phi_{\eta} \colon & \VB_{\zeta}(X) \overset{\!\simeq}\longrightarrow \VB_{\xi}(X) \\
	& [\{E_{i}\}, \{\varphi_{ij}\}] \mapsto [\{E_{i}\}, \{\varphi_{ij} \cdot \eta_{ij}\}].
\end{split}
\end{equation}
This shows that the set $\VB_{\zeta}(X)$ only depends on $[\zeta]$, but in a non-canonical way, since $\Phi_{\eta}$ depends on $\eta$ by construction. More precisely, the choice of $\eta$ is unique up to a cocycle and the equality $\Phi_{\eta} = \Phi_{\chi}$ holds if and only if $\eta^{-1}\chi$ is a coboundary. Therefore, the set of isomorphisms of the form $\Phi_{\eta}$ is a torsor over $\check{H}^{1}(\U, \underline{\UU}(1)) \simeq H^{2}(X; \Z)$, the latter corresponding to the group of (non-twisted) line bundles on $X$. In particular, if $H^{2}(X; \Z) = 0$, then we can define $\VB_{[\zeta]}(X)$ canonically. In general, $\VB_{\zeta}(X)$ depends on the cocycle $\zeta$ up to the tensor product by a line bundle. Hence, only the quotient up to this action, that we denote by $\VB_{[\zeta]}(X)$ anyway, depends on $[\zeta]$ in a canonical way.

If $\V = \{V_{\alpha}\}_{\alpha \in A}$ is a good cover that refines $\U = \{U_{i}\}_{i \in I}$, by definition there exists a function $\phi \colon A \to I$ such that $V_{\alpha} \subset U_{\phi(\alpha)}$ for every $\alpha \in A$. The cocycle $\zeta$, based on $\U$, induces the cocycle $\phi^{*}\zeta = \{\tilde{\zeta}_{\alpha\beta\gamma}\}$, based on $\V$, defined by $\tilde{\zeta}_{\alpha\beta\gamma} := \zeta_{\phi(\alpha)\phi(\beta)\phi(\gamma)}\vert_{V_{\alpha\beta\gamma}}$. We get the isomorphism $\Phi_{\phi} \colon \VB_{\zeta}(X) \overset{\!\simeq}\longrightarrow \VB_{\phi^{*}\zeta}(X)$, $[\{E_{i}\}, \{\varphi_{ij}\}] \mapsto [\{\tilde{E}_{\alpha}\}, \{\tilde{\varphi}_{\alpha\beta}\}]$, where $\tilde{E}_{\alpha} := E_{\phi(\alpha)}\vert_{V_{\alpha}}$ and $\tilde{\varphi}_{\alpha\beta} = \varphi_{\phi(\alpha)\phi(\beta)}\vert_{V_{\alpha\beta}}$. Similarly we get $\Phi_{\phi} \colon \check{H}^{1}_{\zeta}(\U, \underline{\UU}(r)) \overset{\!\simeq}\longrightarrow \check{H}^{1}_{\phi^{*}\zeta}(\V, \underline{\UU}(r))$. Such isomorphisms depend on the function $\phi$, hence they are non-canonical in general. Nevertheless, since the cohomology class represented by $\phi^{*}\zeta$ does not depend on $\phi$, the isomorphism $\Phi_{\phi}$ is canonical when $H^{2}(X; \Z) = 0$. In this case, the sets $\check{H}^{1}_{[\zeta]}(X, \underline{\UU}(r))$ are well-defined and $\VB_{[\zeta]}(X) \simeq \bigsqcup_{r \in \N} \check{H}^{1}_{[\zeta]}(X, \underline{\UU}(r))$ canonically.

\paragraph{\textbf{Non-integral vector bundles.}} The following notion will naturally appear in the classification of the gauge theories on a D-brane, and it is the natural generalization of the ``line bundles with non-integral first Chern class'' defined in \cite{BFS}.
\begin{Def}\label{DefNonIntVB} We call \emph{non-integral vector bundle} a $\zeta$-twisted vector bundle, where $\zeta$ is a \emph{constant} cocycle, i.e.\ $\zeta \in \check{Z}^{2}(\U, \UU(1))$.
\end{Def}
In this case the image of $[\zeta]$ in the cohomology of $\underline{\UU}(1)$ is always a torsion class, but the corresponding $\UU(1)$-cohomology class is not necessarily torsion, unless the transition functions can be chosen constant too. We denote by $\NIVB_{\zeta}(X)$ the set of non-integral vector bundles with twisting class $\zeta$. In this case we have isomorphisms analogous to \eqref{IsoPhiEta} and \eqref{IsoPhiEtaBdl}, but with respect to a $\UU(1)$-cochain $\eta$. It follows that the set of isomorphisms of the form $\Phi_{\eta}$ is a torsor over the image of the natural map $H^{1}(\U, \UU(1)) \to H^{1}(\U, \underline{\UU}(1))$, where $H^{1}(\U, \UU(1)) \simeq H^{1}(X; \R/\Z) \simeq \Hom(H_{1}(X; \Z); \R/\Z)$ and of course $H^{1}(\U, \underline{\UU}(1)) \simeq H^{2}(X; \Z)$. The image is canonically isomorphic to $\Tor\,H^{2}(X; \Z)$. Therefore, if $\Tor\,H^{2}(X; \Z) = 0$ (in particular, if $X$ is simply connected), then $\NIVB_{[\zeta]}(X)$ is canonically defined, with $[\zeta] \in H^{2}(X; \R/\Z)$.

For any $k \geq 1$, we call $\Gamma_{k}$ the subgroup of $\UU(1)$ formed by $k$-th roots of unity and we set $\Gamma := \bigcup_{k \in \N} \Gamma_{k}$. It follows that $\Gamma_{k}$ is the image of the group embedding $\Z_{k} \hookrightarrow \UU(1)$, $a \mapsto e^{2\pi i \frac{a}{k}}$, and $\Gamma$ is the image of $\Q \hookrightarrow \UU(1)$, $q \mapsto e^{2\pi i q}$. It is possible to define non-integral vector bundle with twisting cocycle in $\Gamma$ or $\Gamma_{k}$. This general picture can be useful is some contexts (e.g.\ dealing with fractionary branes), but for our purposes it will be enough to consider $\Z_{2}$ (or $\Gamma_{2}$).
\begin{Def}\label{DefNonIntVBZ2} We call \emph{$\Z_{2}$-non-integral vector bundle} a $\zeta$-twisted vector bundle, where $\zeta \in \check{Z}^{2}(\U, \Gamma_{2})$.
\end{Def}
We denote by $\NIVB_{[2] \, \zeta}(X)$ the set of $\Z_{2}$-non-integral vector bundles with twisting class $\zeta$. The set of isomorphisms of the form $\Phi_{\eta}$ is a torsor over the image of the natural map $H^{1}(X; \Z_{2}) \to H^{1}(X; \underline{\UU}(1))$, where $H^{1}(X; \Z_{2}) \simeq \Hom(H_{1}(X; \Z); \Z_{2})$ and of course $H^{1}(\U, \underline{\UU}(1)) \simeq H^{2}(X; \Z)$. The image is canonically isomorphic to subgroup of $\Tor\,H^{2}(X; \Z)$ formed by classes of order $2$, that we denote by $\Tor_{(2)}H^{2}(X; \Z)$. Therefore, if $\Tor_{(2)}H^{2}(X; \Z) = 0$ (in particular, if $X$ is simply connected), then $\NIVB_{[2] \, [\zeta]}(X)$ is canonically defined, with $[\zeta] \in H^{2}(X; \Z_{2})$. The most meaningful case for us will be $[\zeta] = w_{2}(X)$.

\subsection{Twisted connections}

In section \ref{ReviewCoh} we considered the sheaf complex $S^{p}_{X}$, defined in \eqref{ComplexSp}, whose Cech $p$-hypercohomo\-logy group is by definition the Deligne $p$-cohomology group of the smooth manifold $X$. Now we are interested in $S^{2}_{X}$ as a source of twistings. In fact, as a bundle can be twisted with respect to a 2-cocycle of the sheaf $\underline{\UU}(1)$, similarly a bundle with connection can be twisted with respect to a 2-cocycle of the complex $S^{2}_{X}$ \cite{FR2}.

\begin{Def}\label{DefTwConn} Given a Cech 2-cocycle $(\zeta, \Lambda, B)$ of the complex $S^{2}_{X}$ and a $\zeta$-twisted vector bundle $E := (\{E_{i}\}, \{\varphi_{ij}\})$, a $(\zeta, \Lambda, B)$-twisted \emph{connection} on $E$ is defined by a connection $\nabla_{i}$ on $E_{i}$ for each $i \in I$, compatible with the Hermitian metric and with curvature $\nabla_{i}^{2}$, in such a way that:
\begin{equation}\label{TwistNabla}
	\nabla_{i} - \varphi_{ij}^{*}\nabla_{j} = 2\pi i \Lambda_{ij} \cdot I \qquad\qquad \Tr(\nabla_{i}^{2}) + 2\pi i r B_{i} = 0,
\end{equation}
where $r = \rk(E)$.
\end{Def}
With this definition, the main features of topological twisting hold for connections too. In particular, if we suppose that $(\zeta, \Lambda, B)$ is only a cochain, the existence of a corresponding twisted connection implies that it is a cocycle, as the reader can prove by direct computation. It follows that the \emph{twisting class} $[(\zeta, \Lambda, B)] \in \check{H}^{2}(S^{2}_{X})$ is well-defined, and it must be a \emph{torsion} class, as we are going to show in remark \ref{TorTwistNabla}.

If, for each $i \in I$, we fix a set of $r$ pointwise-independent local sections $s_{1, i}, \ldots, s_{r, i} \colon U_{i} \to E_{i}$ of unit norm, we get the transition functions $g_{ij} \colon U_{ij} \to \UU(r)$ and the local potentials $A_{i} \colon TU_{i} \to i\uu(r)$, where $\uu(r)$ is the Lie algebra of $\UU(r)$, defined by $(\nabla_{i})_{V}s_{\alpha, i} = -2\pi i A_{i}(V)_{\alpha}^{\beta}s_{\beta, i}$. We set $F_{i} := dA_{i} + 2\pi i A_{i} \wedge A_{i}$. The twisting condition \eqref{TwistNabla} is equivalent to:
\begin{equation}\label{NonAbCocycle}
	\left\{ \begin{array}{l} g_{ki} g_{jk} g_{ij} = \zeta_{ijk} \cdot I_{n} \\ A_{j} - g_{ij}^{-1}A_{i}g_{ij} - \frac{1}{2\pi i} g_{ij}^{-1}dg_{ij} = \Lambda_{ij} \cdot I_{n} \\ \frac{1}{r}\Tr(F_{i}) = B_{i}. \end{array} \right.
\end{equation}
Two $(\zeta, \Lambda, B)$-twisted bundles $E := (\{E_{i}\}, \{\varphi_{ij}\}, \{\nabla_{i}\})$ and $E' := (\{E'_{i}\}, \{\varphi'_{ij}\}, \{\nabla'_{i}\})$ are \emph{isomorphic} if there exists a topological isomorphism $\{f_{i} \colon E_{i} \to E'_{i}\}$, as in definition \ref{MorphTwistedVB}, such that $(f_{i})_{*}\nabla_{i} = \nabla'_{i}$ for every $i$. In this case, if we choose local sections $s_{1, i}, \ldots, s_{r, i} \colon U_{i} \to E_{i}$ and $s'_{1, i}, \ldots, s'_{r, i} \colon U_{i} \to E'_{i}$, we get the functions $h_{i} \colon U_{i} \to \UU(r)$, where $h_{i}(x)$ is the change of basis in $(E'_{i})_{x}$ from $\{s'_{i,1}(x), \ldots, s'_{i,r}(x)\}$ to $\{f_{i}(s_{i,1}(x)), \ldots, f_{i}(s_{i,r}(x))\}$, that satisfy the following relations (as for ordinary vector bundles):
\begin{equation}\label{NonAbCobondary}
	\left\{ \begin{array}{l} g'_{ij} = h_{i}^{-1} g_{ij} h_{j} \\ A'_{i} = h_{i}^{-1} A_{i} h_{i} - \frac{1}{2\pi i} h_{i}^{-1}dh_{i}. \end{array} \right.
\end{equation}

\begin{Rmk}\label{TorTwistNabla} \emph{The class $[(\zeta, \Lambda, B)]$ is necessarily torsion (in particular, flat). In fact, computing determinants and traces in \eqref{NonAbCocycle}, we get $\check{D}^{1}(\det g, \Tr A) = (\zeta^{r}, r\Lambda, rB) = r(\zeta, \Lambda, B)$, where $\check{D}^{1}$ is the coboundary of the total complex associated to the $\rm\check{C}$ech double complex induced by $S^{2}_{X}$. It follows that $r[(\zeta, \Lambda, B)] = 0$. In particular, the order of $[(\zeta, \Lambda, B)]$ divides $r$. One can prove that, for any cocycle representing a torsion class, there exist some corresponding twisted bundles with connection \cite{FR2}. This confirms the naturality of definition \ref{DefTwConn}, since it extends to the differential setting all of the fundamental features of topological twisting.}
\end{Rmk}

It follows from the previous remark that, as topological twisting classes are classified by $\Tor \, H^{3}(X; \Z)$, similarly differential twisting classes are classified by $\Tor \, H^{2}(X; \R/\Z)$. If $H_{2}(X; \Z)$ is finitely generated (in particular, if $X$ is compact), then $\Tor \, H^{2}(X; \R/\Z) \simeq H^{2}(X; \Q/\Z)$. Moreover, if $E$ is an ordinary vector bundle, then an ordinary connection $\nabla$ on $E$ is a $(1, 0, \frac{1}{r}\Tr F_{i})$-twisted connection, $\Tr F_{i}$ being a global integral form. Coherently, the class $[(1, 0, \frac{1}{r}\Tr F_{i})]$ is torsion and its order divides $r$.

\begin{Rmk}\label{CohDeligneNotP} \emph{If we consider the $q$-cohomology of \eqref{ComplexSp}, for $q \neq p$, we get the following picture. If $q < p$, then $\check{H}^{q}(S^{p}_{X}) \simeq H^{q}(X; \R/\Z)$. This is due to the fact that the $q$-cochains and the $q$-coboundaries of $S^{p}_{X}$ coincides with the ones of $S^{q}_{X}$, but the cocycle condition is stronger, since it also imposes flatness. If $q > p$, then $\check{H}^{q}(S^{p}_{X}) \simeq H^{q+1}(X; \Z)$, since, the sheaves $\Omega^{\bullet}_{\R}$ being acyclic, the reader can verify that the cohomology class $[(\zeta, \Lambda^{1}, \ldots, \Lambda^{p})] \in \check{H}^{q}(S^{p}_{X})$ is completely determined by $[\zeta] \in \check{H}^{q}(\underline{\UU}(1)) \simeq H^{q+1}(X; \Z)$.}
\end{Rmk}

We now weaken the twisting condition as follows \cite{Park}. We observe that, if $(\zeta, \Lambda, B)$ is a 2-cocycle of $S^{2}_{X}$, then $(\zeta, \Lambda)$ is a 2-cocycle of $S^{1}_{X}$. Conversely, any 2-cocycle $(\zeta, \Lambda)$ of $S^{1}_{X}$ can be completed to a 2-cocycle $(\zeta, \Lambda, B)$ of $S^{2}_{X}$, the $B$-component being unique up to a global form $\tilde{B} \in \Omega^{2}(X)$. It follows from remark \ref{CohDeligneNotP} that $[(\zeta, \Lambda)] \simeq [\zeta] \in \check{H}^{2}(\U, \underline{\UU}(1)) \simeq H^{3}(X, \Z)$.

\begin{Def}\label{DefTwConn2} Given a Cech 2-cocycle $(\zeta, \Lambda)$ of the complex $S^{1}_{X}$ and a $\zeta$-twisted vector bundle $E := (\{E_{i}\}, \{\varphi_{ij}\})$, a $(\zeta, \Lambda)$-twisted \emph{connection} on $E$ is defined by a connection $\nabla_{i}$ on $E_{i}$ for each $i \in I$, compatible with the Hermitian metric, in such a way that:
\begin{equation}\label{TwistNabla2}
	\nabla_{i} - \varphi_{ij}^{*}\nabla_{j} = 2\pi i \Lambda_{ij} \cdot I.
\end{equation}
\end{Def}
Again, if we suppose that $(\zeta, \Lambda)$ is only a cochain, the existence of a corresponding connection implies that it is a cocycle. It follows that the twisting (torsion) class $[(\zeta, \Lambda)] \in \check{H}^{2}(S^{1}_{X}) \simeq H^{3}(X; \Z)$ is well-defined. For any cocycle representing a torsion class, there exists a twisted bundle with connection. With the same notations of equations \eqref{NonAbCocycle}, fixing a set of pointwise-independent local sections of each $E_{i}$ we get:
\begin{equation}\label{NonAbCocycle2}
	\left\{ \begin{array}{l} g_{ki} g_{jk} g_{ij} = \zeta_{ijk} \cdot I_{n} \\ A_{j} - g_{ij}^{-1}A_{i}g_{ij} - \frac{1}{2\pi i} g_{ij}^{-1}dg_{ij} = \Lambda_{ij} \cdot I_{n}, \end{array} \right.
\end{equation}
while condition \eqref{NonAbCobondary} remains unchanged. If $E$ is an ordinary vector bundle, then an ordinary connection $\nabla$ on $E$ is a $(1, 0)$-twisted connection.

\begin{Not} \emph{In section \ref{SecTwVB} we introduced the notation $\VB_{\zeta}(X)$, referring to the set of isomorphism classes of $\zeta$-twisted vector bundles on $X$. Similarly, we call $\VBN_{(\zeta, \Lambda)}(X)$ and $\VBN_{(\zeta, \Lambda, B)}(X)$ the sets of isomorphism classes of $(\zeta, \Lambda)$-twisted and $(\zeta, \Lambda, B)$-twisted vector bundles with connection.}
\end{Not}

For any $(\zeta, \Lambda, B)$ fixed, we have the following commutative diagram:
\begin{equation}\label{DiagTwistVB}
	\xymatrix{
	\VBN_{(\zeta, \Lambda, B)}(X) \ar@{^(->}[rr] \ar@{->>}[dr] & & \VBN_{(\zeta, \Lambda)}(X) \ar@{->>}[dl] \\
	& \VB_{\zeta}(X).
}\end{equation}
Moreover:
\begin{equation}\label{DisjUnionTwistVB}
	\VBN_{(\zeta, \Lambda)}(X) = \bigsqcup_{B: \; [(\zeta, \Lambda, B)] \in \Tor H^{2}(X; \R/\Z)} \VBN_{(\zeta, \Lambda, B)}(X).
\end{equation}
Any two families of local potentials $B$ and $B'$ in the disjoint union differ by a \emph{global} closed form $\tilde{B} \in \Omega^{2}_{\cl}(X)$ such that a finite multiple of $\tilde{B}$ has integral periods. If $H_{2}(X; \Z)$ is finitely generated (in particular, if $X$ is compact), it means that $\tilde{B}$ has rational periods.

\paragraph{\textbf{Non-integral vector bundles with connection.}} The following definition is the differential extension of \ref{DefNonIntVB}.
\begin{Def}\label{DefNonIntVBNabla} We call \emph{non-integral vector bundle with connection} a $(\zeta, 0)$-twisted vector bundle with connection, for any fixed $\zeta$.
\end{Def}
Since $(\zeta, 0)$ must be a cocycle, it follows that $\zeta$ is constant, therefore we are endowing a non-integral (topological) vector bundle, as defined in \ref{DefNonIntVB}, with a connection, the latter satisfying by definition $\nabla_{i} = \varphi_{ij}^{*}\nabla_{j}$. It follows from \eqref{NonAbCocycle2} that the corresponding isomorphism class can be represented in the form $[\{g_{ij}, A_{i}\}]$, with
\begin{equation}\label{NonIntCocycle}
	\left\{ \begin{array}{l} g_{ki} g_{jk} g_{ij} = \zeta_{ijk} \cdot I_{n} \\ A_{j} - g_{ij}^{-1}A_{i}g_{ij} = \frac{1}{2\pi i} g_{ij}^{-1}dg_{ij}, \end{array} \right.
\end{equation}
the latter condition being identical to the one holding for ordinary connections. We denote by $\NIVBN_{\zeta}(X)$ the set of non-integral vector bundles with connection twisted by $(\zeta, 0)$. Similarly we define $\NIVBN_{[2]\,\zeta}(X)$, considering definition \ref{DefNonIntVBZ2}.

For this kind of bundles we can define the Chern classes in the usual way. In fact, the gauge transformations of the curvature are given by $F_{j} = g_{ij}^{-1}F_{i}g_{ij}$, as for ordinary vector bundles (cfr.\ formula \eqref{CurvatureCC} with $B$ global). Therefore, one can define the Chern classes via the elementary symmetric polynomials $P_{k}$, which are invariant by conjugation:
\begin{equation}\label{ChernClasses}
	c_{k}[\{g_{ij}, A_{i}\}] := [P_{k}(F)].
\end{equation}
The cohomology class of $P_{k}(F)$ only depends on the (topological) non-integral bundle, not on the connection. The main difference with respect to ordinary vector bundles is that, when the twisting class is not trivial, a Chern class is not the real image of an integral class in general (i.e.\ it is represented by a form whose periods are not necessarily integral).\footnote{Even if the twisting cocycle is not constant, one can give a good definition of Chern classes \cite{Karoubi}, but we do not need this definition here.} As a particular case, the Chern classes are defined for $\Z_{2}$-non-integral bundles. In this case they are half-integral. A similar consideration holds more generally for $\Z_{n}$-bundles or $\Q$-bundles.

\subsection{Partially non-abelian Deligne cohomology}\label{SecPartNonAbDeligne}

Let us show how relative Deligne cohomology can be suitably generalized in order to describe twisted vector bundles with connection in a quite intrinsic way. We start from the topological setting, then we consider the differential extension, analysing the absolute and the relative versions in each case. Moreover, we show the corresponding long exact sequences.

\paragraph{\textbf{Topological framework -- Absolute version.}} We define the partially non-abelian Cech complex of the sheaf $\underline{\UU}(r)$, for a fixed rank $r$, as follows (we leave the cover $\U$ implicit):
\begin{equation}\label{ComplexUr}
\xymatrix{
	\check{C}^{0}(\underline{\UU}(r)) \ar@{-->}[r]^{\check{\delta}^{0}} & \check{C}^{1}(\underline{\UU}(r)) \ar[r]^{\check{\delta}^{1}} & \check{C}^{2}(\underline{\UU}(r)) \ar[r]^{\check{\delta}^{2}} & \check{C}^{3}(\underline{\UU}(1)) \ar[r]^(.6){\check{\delta}^{3}} & \cdots,
}\end{equation}
where $\check{\delta}^{0}$ is not a morphism but a group action, defined by $\{h_{i}\} \cdot \{g_{ij}\} := \{h_{i}g_{ij}h_{j}^{-1}\}$. As usual we set $\check{\delta}^{1}\{g_{ij}\} := \{g_{ki}g_{jk}g_{ij}\}$. Moreover, $\check{\delta}^{2}\{\zeta_{ijk}\} := \check{\delta}^{2}\{\det\zeta_{ijk}\}$, where $\check{\delta}^{2}$ on the r.h.s.\ is the usual one of the (abelian) complex $\check{C}^{\bullet}(\underline{\UU}(1))$. All of the other groups and boundaries are the ones of $\check{C}^{\bullet}(\underline{\UU}(1))$, but they will be irrelevant for the present discussion. The $1$-degree cohomology, defined as the quotient of the kernel of $\check{\delta}^{1}$ up to the action $\check{\delta}^{0}$, is canonically isomorphic to the pointed set of rank-$r$ vector bundles on $X$ up to isomorphism. The (irrelevant) higher-degree cohomology sets and groups are defined as usual, while the $0$-degree cohomology of \eqref{ComplexUr} is the kernel of $\check{\delta}^{0}$, i.e.\ the set of $0$-cochains such that $\{h_{i}\} \cdot 1 = 1$. We get the global sections of $\underline{\UU}(r)$, as it has to be.  

Moreover, as in section \ref{SecTwVB}, given a cocycle $\zeta \in \check{Z}^{2}(\underline{\UU}(1))$, we call $\check{H}^{1}_{\zeta}(\underline{\UU}(r))$ the set of 1-cochains of \eqref{ComplexUr} whose coboundary is $\zeta \cdot I_{r}$, up to the action of $\check{\delta}^{0}$. We get the canonical bijection $\VB_{\zeta}(X) \simeq \bigsqcup_{r \in \N} \check{H}^{1}_{\zeta}(\underline{\UU}(r))$. Replacing $\zeta$ by a cohomologous cocycle, the non-canonical isomorphism \eqref{IsoPhiEta} holds.

\paragraph{\textbf{Topological framework -- Relative version.}} Given a continuous closed embedding $\rho \colon Y \hookrightarrow X$, we consider the complexes of sheaves $\underline{\UU}(1)_{X}$ and $\rho_{*}\underline{\UU}(r)_{Y}$ on $X$. We have the natural morphism
\begin{equation}\label{RhoExclam}
	\rho^{!} \colon \underline{\UU}(1)_{X} \to \rho_{*}\underline{\UU}(r)_{Y},
\end{equation}
defined by $\rho^{!}(f) := (f \circ \rho) \cdot I_{r}$, inducing the cochain morphism $\rho^{!} \colon \check{C}^{\bullet}(\underline{\UU}(1)_{X}) \to \check{C}^{\bullet}(\rho_{*}\underline{\UU}(r)_{Y})$. In order to define the relative Cech cohomology sets of $\rho^{!}$, we generalize the cone complex to this partially non-abelian framework as follows:
\begin{equation}\label{ComplexConeUrTop}
\begin{split}
\xymatrix{
	\check{C}^{0}(\underline{\UU}(1)) \ar[r]^(.36){\check{\pmb{\delta}}^{0}} & \check{C}^{1}(\underline{\UU}(1)) \times \check{C}^{0}(\underline{\UU}(r)) \ar@{-->}[r]^{\check{\pmb{\delta}}^{1}} & \check{C}^{2}(\underline{\UU}(1)) \times \check{C}^{1}(\underline{\UU}(r)) \ar[r]^(.8){\check{\pmb{\delta}}^{2}} &
} \\ \xymatrix{ \ar[r]^(.2){\check{\pmb{\delta}}^{2}} & \check{C}^{3}(\underline{\UU}(1)) \times \check{C}^{2}(\underline{\UU}(r)) \ar[r]^{\check{\pmb{\delta}}^{3}} & \check{C}^{4}(\underline{\UU}(1)) \times \check{C}^{3}(\underline{\UU}(1)) \ar[r]^(.72){\check{\pmb{\delta}}^{4}} & \cdots,
}\end{split}
\end{equation}
where $\check{\pmb{\delta}}^{1}$ is the action $(\{\eta_{ij}\}, \{h_{i}\}) \cdot (\{\zeta_{ijk}\}, \{g_{ij}\}) := (\{\zeta_{ijk}\} \cdot \check{\delta}^{1}\{\eta_{ij}\}, \{\rho^{!}\eta_{ij}I_{r} \cdot h_{i}^{-1}g_{ij}h_{j}\})$ and $\check{\pmb{\delta}}^{2}(\{\zeta_{ijk}\}, \{g_{ij}\}) = (\check{\delta}^{2}\{\zeta_{ijk}\}, \{\rho^{!}\zeta_{ijk}I_{r} \cdot (g_{ki}g_{jk}g_{ij})^{-1}\})$. Let us analyse the corresponding 2-degree cohomology. The kernel of $\check{\pmb{\delta}}^{2}$ is formed by classes $(\{\zeta_{ijk}\}, \{g_{ij}\})$ such that $\{\zeta_{ijk}\}$ is a cocycle and $g_{ki}g_{jk}g_{ij} = \rho^{!}\zeta_{ijk}$, therefore we get transition functions of $\rho^{!}\zeta_{ijk}$-twisted vector bundles. The action of $(0, \{h_{i}^{-1}\})$ through $\check{\pmb{\delta}}^{1}$ replaces $g_{ij}$ by $h_{i}g_{ij}h_{j}^{-1}$, therefore it changes the representative within the same isomorphism class, and the action of $(\{\eta_{ij}\}, 0)$ corresponds to the isomorphism \eqref{IsoPhiEta}. This means that the group $\check{H}^{2}(\rho^{!})$ naturally encodes twisted vector bundles on $Y$ with respect to any class induced by pull-back from $X$, up to isomorphism and change of representative in the twisting class. We observe that, choosing $Y = X$ and $\rho = \id_{X}$, we get twisted vector bundles on $X$.

\begin{Rmk} \emph{It is important to observe that $\check{H}^{2}(\rho^{!})$ is not twisted, hence it encodes twisted bundles in ordinary cohomology. The same remark holds about the differential extension described below.}
\end{Rmk}

\paragraph{\textbf{Differential extension -- Absolute version I.}} We have seen that, in the topological framework, partially non-abelian Cech cohomology describes in a unitary way twisted vector bundles. We are going to do the same in the differential framework, i.e.\ we introduce partially non-abelian Deligne cohomology in order to describe twisted vector bundles with connection. First of all, we start from \eqref{ComplexUr}, that is the Cech complex of the sheaf $\underline{\UU}(r)$ on $X$, and we extend it to the Cech double-complex of the sheaf complex
\begin{equation}\label{ComplexS1r}
	\xymatrix{
	S^{1,r}_{X} := \underline{\UU}(r) \ar@{-->}[r]^(.58){\tilde{d}} & \Omega^{1}_{i\uu(r)},
}\end{equation}
where $\tilde{d}$ is the action $f \cdot \omega := f\omega f^{-1} + \frac{1}{2\pi i} f^{-1}df$. We define such a double complex as:
\begin{equation}\label{DoubleComplexUr}
\xymatrix{
	\check{C}^{0}(\Omega^{1}_{i\uu(r)}) \ar@{-->}@[blue][r]^{\color{blue} \check{\delta}^{0}} & \check{C}^{1}(\Omega^{1}_{i\uu(r)}) \ar@[green][r]^(.55){\color{green!70!black} \check{\delta}^{1}} & \check{C}^{2}(\Omega^{1}_{\R}) \ar[r]^{\check{\delta}^{2}} & \check{C}^{3}(\Omega^{1}_{\R}) \ar[r]^(.6){\check{\delta}^{3}} & \cdots \\
	\check{C}^{0}(\underline{\UU}(r)) \ar@{-->}@[red][r]^{\color{red} \check{\delta}^{0}} \ar@{-->}@[red][u]_{\color{red} \tilde{d}} & \check{C}^{1}(\underline{\UU}(r)) \ar@[blue][r]^{\color{blue} \check{\delta}^{1}} \ar@{-->}@[blue][u]_{\color{blue} \tilde{d}} & \check{C}^{2}(\underline{\UU}(r)) \ar@[green][r]^{\color{green!70!black} \check{\delta}^{2}} \ar@[green][u]_{\color{green!70!black} \tilde{d}} & \check{C}^{3}(\underline{\UU}(1)) \ar[r]^(.6){\check{\delta}^{3}} \ar[u]_{\tilde{d}} & \cdots,
}\end{equation}
where the first coboundary $\color{red} \check{D}^{0}$ is the action $\{h_{i}\} \cdot (\{g_{ij}\}, \{A_{i}\}) := (\{h_{i}g_{ij}h_{j}^{-1}\}, \{h_{i}A_{i}h_{i}^{-1} + \frac{1}{2\pi i} h_{i}^{-1}dh_{i}\})$ and the second one is defined as ${\color{blue} \check{D}^{1}}(\{g_{ij}\}, \{A_{i}\}) := (\{g_{ij}g_{jk}g_{ki}\}, \{A_{j} - g_{ij}^{-1}A_{i}g_{ij} - \frac{1}{2\pi i}g_{ij}^{-1}dg_{ij}\})$. The third (irrelevant) one is ${\color{green!70!black} \check{D}^{2}}(\{\zeta_{ijk}\}, \{\Lambda_{ij}\}) := (\check{\delta}^{2}\{\det \zeta_{ijk}\}, \check{\delta}^{1}\{\Lambda_{ij}\} + \{\tilde{d}\det \zeta_{ijk}\})$ and all the others are the ones of the complex $\underline{\UU}(1) \to \Omega^{1}_{\R}$. The $1$-degree cohomology of \eqref{DoubleComplexUr}, defined as the quotient of $\check{D}^{1}$ up to the action $\check{D}^{0}$, is canonically isomorphic to the set of rank $r$ vector bundles with connection on $X$. The (irrelevant) higher-degree cohomology sets are defined as usual, while the $0$-degree cohomology is the kernel of $\check{D}^{0}$, i.e.\ the set of $0$-cochains such that $\{h_{i}\} \cdot 1 = 1$. We get the global \emph{constant} sections of $\underline{\UU}(r)$, as in the abelian case. Moreover, we call
\begin{equation}\label{H1ZetaLambda}
	\check{H}^{1}_{(\zeta, \Lambda)}(\U, S^{1,r}_{X})
\end{equation}
the set of 1-cochains of \eqref{DoubleComplexUr} whose 1-coboundary is $(\zeta, \Lambda)$, up to the action of $\check{D}^{0}$. We get the canonical bijection $\VBN_{(\zeta, \Lambda)}(X) \simeq \bigsqcup_{r \in \N} \check{H}^{1}_{(\zeta, \Lambda)}(\U, S^{1,r}_{X})$, obtained fixing local point-wise independent sections and computing the corresponding transition functions and potentials. Moreover, if $(\xi, \Theta) = (\zeta, \Lambda) \cdot \check{\delta}^{1}(\eta, \lambda)$ in $S^{1}_{X}$, we get the following isomorphism, extending \eqref{IsoPhiEta} to the differential setting:
\begin{equation}\label{IsoPhiEtaDiff}
\begin{split}
	\Phi_{(\eta, \lambda)} \colon & \check{H}^{1}_{(\zeta, \Lambda)}(\U, S^{1,r}_{X}) \overset{\!\simeq}\longrightarrow \check{H}^{1}_{(\xi, \Theta)}(\U, S^{1,r}_{X}) \\
	& [\{g_{ij}\}, \{A_{i}\}] \mapsto [\{g_{ij} \cdot \eta_{ij}\}, \{A_{i} + \lambda_{i} \cdot I_{r}\}].
\end{split}
\end{equation}
Equivalently, using the language of twisted bundles with connection (and without fixing the rank $r$):
\begin{equation}\label{IsoPhiEtaDiffVB}
\begin{split}
	\Phi_{(\eta, \lambda)} \colon & \VBN_{(\zeta, \Lambda)}(X) \overset{\!\simeq}\longrightarrow \VBN_{(\xi, \Theta)}(X) \\
	& [\{E_{i}\}, \{\varphi_{ij}\}, \{\nabla_{i}\}] \mapsto [\{E_{i}\}, \{\varphi_{ij} \cdot \eta_{ij}\}, \{\nabla_{i} + \lambda_{i}I\}].
\end{split}
\end{equation}
Fixing $(\zeta, \Lambda)$ and $(\xi, \Theta)$, two choices $(\eta, \lambda)$ and $(\eta', \lambda')$ differ by a cocycle. Moreover, $\Phi_{(\eta, \lambda)} = \Phi_{(\eta', \lambda')}$ if and only if $(\eta, \lambda) - (\eta', \lambda')$ is a coboundary, hence the set of isomorphisms of the form \eqref{IsoPhiEtaDiff} or \eqref{IsoPhiEtaDiffVB} is a torsor over $\check{H}^{1}(S^{1}_{X})$.

\paragraph{\textbf{Differential extension -- Absolute version II.}} If, instead of \eqref{ComplexS1r}, we consider the degree-2 complex
\begin{equation}\label{ComplexS2r}
	\xymatrix{
	S^{2,r}_{X} := \underline{\UU}(r) \ar@{-->}[r]^(.58){\tilde{d}} & \Omega^{1}_{i\uu(r)} \ar[r]^{d \circ \Tr} & \Omega^{2}_{\R},
}\end{equation}
then the double complex \eqref{DoubleComplexUr} is replaced by the following one:
\begin{equation}\label{DoubleComplexUrDegree2}
\xymatrix{
	\check{C}^{0}(\Omega^{2}_{\R}) \ar[r]^{\check{\delta}^{0}} & \check{C}^{1}(\Omega^{2}_{\R}) \ar[r]^{\check{\delta}^{1}} & \check{C}^{2}(\Omega^{2}_{\R}) \ar[r]^{\check{\delta}^{2}} & \check{C}^{3}(\Omega^{1}_{\R}) \ar[r]^(.6){\check{\delta}^{3}} & \cdots \\
	\check{C}^{0}(\Omega^{1}_{i\uu(r)}) \ar@{-->}@[blue][r]^{\color{blue} \check{\delta}^{0}} \ar@[blue][u]_{\color{blue} d \circ \Tr} & \check{C}^{1}(\Omega^{1}_{i\uu(r)}) \ar@[green][r]^(.55){\color{green!70!black} \check{\delta}^{1}} \ar@[green][u]_{\color{green} d \circ \Tr} & \check{C}^{2}(\Omega^{1}_{\R}) \ar[r]^{\check{\delta}^{2}} \ar[u]_{d} & \check{C}^{3}(\Omega^{1}_{\R}) \ar[r]^(.6){\check{\delta}^{3}} \ar[u]_{d} & \cdots \\
	\check{C}^{0}(\underline{\UU}(r)) \ar@{-->}@[red][r]^{\color{red} \check{\delta}^{0}} \ar@{-->}@[red][u]_{\color{red} \tilde{d}} & \check{C}^{1}(\underline{\UU}(r)) \ar@[blue][r]^{\color{blue} \check{\delta}^{1}} \ar@{-->}@[blue][u]_{\color{blue} \tilde{d}} & \check{C}^{2}(\underline{\UU}(r)) \ar@[green][r]^{\color{green!70!black} \check{\delta}^{2}} \ar@[green][u]_{\color{green!70!black} \tilde{d}} & \check{C}^{3}(\underline{\UU}(1)) \ar[r]^(.6){\check{\delta}^{3}} \ar[u]_{\tilde{d}} & \cdots.
}\end{equation}
We call
\begin{equation}\label{H1ZetaLambdaB}
	\check{H}^{1}_{(\zeta, \Lambda, B)}(\U, S^{2,r}_{X})
\end{equation}
the set of 1-cochains of \eqref{DoubleComplexUrDegree2} whose 1-coboundary is $(\zeta, \Lambda, B)$, up to the action of $\check{D}^{0}$. We get the canonical bijection $\VBN_{(\zeta, \Lambda, B)}(X) \simeq \bigsqcup_{r \in \N} \check{H}^{1}_{(\zeta, \Lambda, B)}(\U, S^{2,r}_{X})$. Moreover, if $(\xi, \Theta, C) = (\zeta, \Lambda, B) \cdot \check{\delta}^{1}(\eta, \lambda)$ in $S^{2}_{X}$, we get the following isomorphism:
\begin{equation}\label{IsoPhiEtaDiffB}
\begin{split}
	\Phi_{(\eta, \lambda)} \colon & \check{H}^{1}_{(\zeta, \Lambda, B)}(\U, S^{2,r}_{X}) \overset{\!\simeq}\longrightarrow \check{H}^{1}_{(\xi, \Theta, C)}(\U, S^{2,r}_{X}) \\
	& [\{g_{ij}\}, \{A_{i}\}] \mapsto [\{g_{ij} \cdot \eta_{ij}\}, \{A_{i} + \lambda_{i} \cdot I_{r}\}].
\end{split}
\end{equation}
Equivalently, using the language of twisted bundles with connection, we get \eqref{IsoPhiEtaDiffVB} adding $B$ and $C$ to the twisting cocycles. The set of isomorphisms of the form \eqref{IsoPhiEtaDiffB} is a torsor over the flat part of $\check{H}^{1}(S^{1}_{X})$, i.e.\ over $H^{1}(X; \R/\Z) \simeq \Hom(H_{1}(X; \Z); \R/\Z)$. It follows that, if $H_{1}(X; \Z) = 0$, then $\check{H}^{1}_{[(\zeta, \Lambda, B)]}(S^{2,r}_{X})$ is canonically defined.

\paragraph{\textbf{Differential extension -- Relative version I.}} Given a smooth closed embedding $\rho \colon Y \hookrightarrow X$, we compute the relative Deligne cohomology of degree 2, but in the partially non-abelian setting. We have the natural morphism $\rho^{!} \colon S^{2}_{X} \to \rho_{*}S^{1,r}_{Y}$ defined as follows:
\begin{equation}\label{RelativeDeligneCplxRankR}
\xymatrix{
	\underline{\UU}(1)_{X} \ar[r]^{\tilde{d}} \ar[d]^{\rho^{!,0}} & \Omega^{1}_{X, \R} \ar[r]^(.47){d} \ar[d]^{\rho^{!,1}} & \Omega^{2}_{X, \R} \ar[d] \\
	\rho_{*}\underline{\UU}(r)_{Y} \ar@{-->}[r]^(.45){\tilde{d}} & \rho_{*}\Omega^{1}_{Y, i\uu(r)} \ar[r] & 0,
}\end{equation}
where $\rho^{!, 0}(f) := (f \circ \rho) \cdot I_{r}$ and $\rho^{!, 1}(\omega) := \rho^{*}\omega \cdot I_{r}$. The relative Cech hypercohomology groups of $\rho^{!}$ are by definition the cohomology groups of the corresponding cone complex, that is the total complex associated to the following double complex:

\begin{tiny}
\begin{equation}\label{ComplexConeUr}
\begin{split}
\xymatrix{
	\color{red} \check{C}^{0}(\Omega^{2}_{X, \R}) \ar[r]^(.38){\check{\pmb{\delta}}^{0}} & \color{green!70!black} \check{C}^{1}(\Omega^{2}_{X, \R}) \ar[r]^{\check{\pmb{\delta}}^{1}} & \check{C}^{2}(\Omega^{2}_{X, \R}) \ar[r]^{\check{\pmb{\delta}}^{2}} & \check{C}^{3}(\Omega^{2}_{X, \R}) \ar[r]^(.7){\check{\pmb{\delta}}^{3}} & \cdots \\
	{ \color{yellow!70!black} \check{C}^{0}(\Omega^{1}_{X, \R}) } \ar[r]^(.38){\check{\pmb{\delta}}^{0}} \ar[u]_{d} & { \color{red} \check{C}^{1}(\Omega^{1}_{X, \R}) } \times { \color{blue} \check{C}^{0}(\Omega^{1}_{Y, i\uu(r)}) } \ar@{-->}[r]^{\check{\pmb{\delta}}^{1}} \ar[u]_{d} & { \color{green!70!black}  \check{C}^{2}(\Omega^{1}_{X, \R}) } \times { \color{magenta} \check{C}^{1}(\Omega^{1}_{Y, i\uu(r)}) } \ar[r]^{\check{\pmb{\delta}}^{2}} \ar[u]_{d} & \check{C}^{3}(\Omega^{1}_{X, \R}) \times \check{C}^{2}(\Omega^{1}_{Y, \R}) \ar[r]^(.7){\check{\pmb{\delta}}^{3}} \ar[u]_{d} & \cdots \\
	\check{C}^{0}(\underline{\UU}(1)_{X}) \ar[r]^(.38){\check{\pmb{\delta}}^{0}} \ar[u]_{\tilde{d}} & { \color{yellow!70!black} \check{C}^{1}(\underline{\UU}(1)_{X}) } \times { \color{cyan!70!black} \check{C}^{0}(\underline{\UU}(r)_{Y}) } \ar@{-->}[r]^{\check{\pmb{\delta}}^{1}} \ar@{-->}[u]_{(\tilde{d}, \tilde{d})} & { \color{red} \check{C}^{2}(\underline{\UU}(1)_{X}) } \times { \color{blue} \check{C}^{1}(\underline{\UU}(r)_{Y}) } \ar[r]^{\check{\pmb{\delta}}^{2}} \ar@{-->}[u]_{(\tilde{d}, \tilde{d})} & { \color{green!70!black} \check{C}^{3}(\underline{\UU}(1)_{X}) } \times { \color{magenta} \check{C}^{2}(\underline{\UU}(r)_{Y}) } \ar[r]^(.7){\check{\pmb{\delta}}^{3}} \ar[u]_{(\tilde{d}, \tilde{d})} & \cdots.
}\end{split}
\end{equation}
\end{tiny}

\noindent A 2-cochain is of the form $({\color{red} \{\zeta_{ijk}\}, \{\Lambda_{ij}\}, \{B_{i}\} }, { \color{blue} \{g_{ij}\}, \{A_{i}\} })$, where $(\{\zeta_{ijk}\}, \{\Lambda_{ij}\}, \{B_{i}\})$ is a 2-cochain of $S^{2}_{X}$ and $(\{g_{ij}\}, \{A_{i}\})$ is a 1-cochain of $S^{1,r}_{Y}$. We denote it as $(\zeta, \Lambda, B, g, A)$ for simplicity. The 2-coboundary is by definition $\check{\pmb{D}}^{2}({\color{red} \zeta, \Lambda, B }, {\color{blue} g, A}) = (\check{D}^{2}(\zeta, \Lambda, B), \rho^{!}(\zeta, \Lambda, B) - \check{D}^{1}(g, A)) = ({ \color{green!70!black} \check{\delta}^{2}\zeta, \check{\delta}^{1}\Lambda + \tilde{d}\zeta, \check{\delta}^{0}B - d\Lambda }, { \color{magenta} \{\rho^{*}\zeta \cdot (g_{ij}g_{jk}g_{ki})^{-1}\}, \{\Lambda_{ij}I_{r} - (A_{j} - g_{ij}^{-1}A_{i}g_{ij} - \frac{1}{2\pi i}}$ \linebreak ${ \color{magenta} g_{ij}^{-1}dg_{ij}) \} } )$, therefore a 2-cocycle is formed by a Deligne 2-cocycle $(\zeta, \Lambda, B)$ on $X$ and a representative of a $\rho^{*}(\zeta, \Lambda)$-twisted line bundle on $Y$.

The 1-coboundary $\check{\pmb{D}}^{1}$ is by definition the action $({ \color{yellow!70!black} \eta, \lambda }, { \color{cyan!70!black} h }) \cdot ({\color{red} \zeta, \Lambda, B }, { \color{blue} g, A }) := ({\color{red} \zeta \cdot \check{\delta}^{1}\eta, \Lambda}$ \linebreak ${\color{red} + \check{\delta}^{0}\lambda - \tilde{d}\eta, B + d\lambda }, { \color{blue} (\rho^{*}\eta^{-1}I_{r} \cdot hgh^{-1})^{-1}, \rho^{*}\lambda I_{r} - h A h^{-1} + \tilde{d}h})$. This means that the action of $(0, 0, h)$ is a gauge transformation of a $\rho^{*}(\zeta, \Lambda)$-twisted line bundle, as in \eqref{NonAbCobondary}. Moreover, the action of $(\eta, \lambda, 0)$ corresponds to the isomorphism \eqref{IsoPhiEtaDiff}. It follows that the group $\check{H}^{2}(\rho^{!})$ naturally encodes twisted vector bundles with connection on $Y$, with respect to any class induced by pull-back from $X$, up to isomorphism and change of representative in the twisting class. As in the topological framework, choosing $Y = X$ and $\rho = \id_{X}$, we get twisted vector bundles with connection on $X$.

\paragraph{\textbf{Differential extension -- Relative version II.}} We consider the natural morphism $\rho^{!} \colon S^{2}_{X} \to \rho_{*}S^{2,r}_{Y}$ defined as follows:
\begin{equation}\label{RelativeDeligneCplxRankRPar}
\xymatrix{
	\underline{\UU}(1)_{X} \ar[r]^{\tilde{d}} \ar[d]^{\rho^{!,0}} & \Omega^{1}_{X, \R} \ar[r]^(.47){d} \ar[d]^{\rho^{!,1}} & \Omega^{2}_{X, \R} \ar[d]^{\rho^{!,2}} \\
	\rho_{*}\underline{\UU}(r)_{Y} \ar@{-->}[r]^(.45){\tilde{d}} & \rho_{*}\Omega^{1}_{Y, i\uu(r)} \ar[r]^{d \circ \Tr} & \rho_{*}\Omega^{2}_{Y, \R}.
}\end{equation}
The relative Cech hypercohomology groups of $\rho^{!}$ are by definition the cohomology groups of the corresponding cone complex, that is the total complex associated to the following double complex:

\begin{tiny}
\begin{equation}\label{ComplexConeUrPar}
\begin{split}
\xymatrix{
	\check{C}^{0}(\Omega^{2}_{X, \R}) \ar[r]^(.38){\check{\pmb{\delta}}^{0}} & \check{C}^{1}(\Omega^{2}_{X, \R}) \times \check{C}^{0}(\Omega^{2}_{Y, \R}) \ar[r]^{\check{\pmb{\delta}}^{1}} & \check{C}^{2}(\Omega^{2}_{X, \R}) \times \check{C}^{1}(\Omega^{2}_{Y, \R}) \ar[r]^{\check{\pmb{\delta}}^{2}} & \check{C}^{3}(\Omega^{2}_{X, \R}) \times \check{C}^{2}(\Omega^{2}_{Y, \R}) \ar[r]^(.7){\check{\pmb{\delta}}^{3}} & \cdots \\
	\check{C}^{0}(\Omega^{1}_{X, \R}) \ar[r]^(.38){\check{\pmb{\delta}}^{0}} \ar[u]_{d} & \check{C}^{1}(\Omega^{1}_{X, \R}) \times \check{C}^{0}(\Omega^{1}_{Y, i\uu(r)}) \ar@{-->}[r]^{\check{\pmb{\delta}}^{1}} \ar[u]_{(d, d)} & \check{C}^{2}(\Omega^{1}_{X, \R}) \times \check{C}^{1}(\Omega^{1}_{Y, i\uu(r)}) \ar[r]^{\check{\pmb{\delta}}^{2}} \ar[u]_{(d, d)} & \check{C}^{3}(\Omega^{1}_{X, \R}) \times \check{C}^{2}(\Omega^{1}_{Y, \R}) \ar[r]^(.7){\check{\pmb{\delta}}^{3}} \ar[u]_{(d, d)} & \cdots \\
	\check{C}^{0}(\underline{\UU}(1)_{X}) \ar[r]^(.38){\check{\pmb{\delta}}^{0}} \ar[u]_{\tilde{d}} & \check{C}^{1}(\underline{\UU}(1)_{X}) \times \check{C}^{0}(\underline{\UU}(r)_{Y}) \ar@{-->}[r]^{\check{\pmb{\delta}}^{1}} \ar@{-->}[u]_{(\tilde{d}, \tilde{d})} & \check{C}^{2}(\underline{\UU}(1)_{X}) \times \check{C}^{1}(\underline{\UU}(r)_{Y}) \ar[r]^{\check{\pmb{\delta}}^{2}} \ar@{-->}[u]_{(\tilde{d}, \tilde{d})} & \check{C}^{3}(\underline{\UU}(1)_{X}) \times \check{C}^{2}(\underline{\UU}(r)_{Y}) \ar[r]^(.7){\check{\pmb{\delta}}^{3}} \ar[u]_{(\tilde{d}, \tilde{d})} & \cdots.
}\end{split}
\end{equation}
\end{tiny}

\noindent The only difference with respect to \eqref{ComplexConeUr} is that, in the cocycle condition, we also have the constraint $rB_{i} - \Tr(F_{i}) = 0$, i.e.\ the second component of the curvature must vanish. It follows that a 2-cocycle is formed by a Deligne 2-cocycle $(\zeta, \Lambda, B)$ on $X$ and a representative of a $\rho^{*}(\zeta, \Lambda, B)$-twisted vector bundle on $Y$. The action of $(0, 0, h)$ is a gauge transformation of a $\rho^{*}(\zeta, \Lambda, B)$-twisted vector bundle, as in \eqref{NonAbCobondary}, and the action of $(\eta, \lambda, 0)$ corresponds to the isomorphism \eqref{IsoPhiEtaDiffB}.

\paragraph{\textbf{Summary.}} Generalizing \eqref{RelDelignePar} to this partially non-abelian framework, we set:
\begin{equation}\label{RelDeligneParNAb}
	\hat{H}^{3, r}(\rho) := \check{H}^{2}(\rho^{!} \colon S^{2}_{X} \to \rho_{*}S^{1,r}_{Y}) \qquad\qquad \hat{H}^{3, r}_{\ppar}(\rho) := \check{H}^{2}(\rho^{!} \colon S^{2}_{X} \to \rho_{*}S^{2,r}_{Y}).
\end{equation}
It follows that $\hat{H}^{3, r}(\rho)$ naturally encodes $(\zeta, \Lambda)$-twisted bundles and $\hat{H}^{3, r}_{\ppar}(\rho)$ naturally encodes $(\zeta, \Lambda, B)$-twisted bundles. We observe that $B$ is fixed even in a non-parallel cocycle, since it allows to define the gauge-invariant field-strength $rB - \Tr\, F$, but the twisted bundle itself is not linked to $B$. In particular, the relative curvature of the class $[(\zeta, \Lambda, B, g, A)] \in \hat{H}^{3, r}(\rho)$ is the relative form $(H, rB - \Tr\, F)$, where $H = dB$. Moreover, in the absolute setting, we can extend diagram \eqref{DiagTwistVB} to the following one:

\begin{small}
\begin{equation}\label{DiagTwistVBCohom}
	\xymatrix{
	\VBN_{(\zeta, \Lambda, B)}(X) \ar@{^(->}[rr] \ar@{->>}[dr] \ar[dd]^{\simeq} & & \VBN_{(\zeta, \Lambda)}(X) \ar@{->>}[dl] \ar[dd]^{\simeq} \\
	& \VB_{\zeta}(X) \ar[dd]^(.6){\simeq} \\
	\bigsqcup_{r \in \N} \check{H}^{1}_{(\zeta, \Lambda, B)}(\U, S^{2,r}_{X}) \ar@{^(->}[rr]|\hole \ar@{->>}[dr] & & \bigsqcup_{r \in \N} \check{H}^{1}_{(\zeta, \Lambda)}(\U, S^{1,r}_{X}) \ar@{->>}[dl] \\
	& \bigsqcup_{r \in \N} \check{H}^{1}_{\zeta}(\U, \underline{\UU}(r)).
}\end{equation}
\end{small}

\noindent Therefore, formula \eqref{DisjUnionTwistVB} is equivalent to the following one:
\begin{equation}\label{DisjUnionTwistVBCohom}
	\check{H}^{1}_{(\zeta, \Lambda)}(\U, S^{1,r}_{X}) = \bigsqcup_{B: \; [(\zeta, \lambda, B)] \in \Tor H^{2}(X; \R/\Z)} \check{H}^{1}_{(\zeta, \Lambda, B)}(\U, S^{2,r}_{X}).
\end{equation}

\paragraph{\textbf{Non-integral vector bundles.}} We denote by $\check{H}^{1}_{\zeta}(\U, S^{1,r}_{X})$ the group \eqref{H1ZetaLambda} when $\Lambda = 0$, canonically isomorphic to $\NIVBN_{\zeta}(X)$ (see definition \ref{DefNonIntVBNabla}). If $\xi = \zeta \cdot \check{\delta}^{1}\eta$ as $\UU(1)$-cochains, we get the following isomorphism, analogous to \eqref{IsoPhiEtaDiff}:
\begin{equation}\label{IsoPhiEtaNonInt}
\begin{split}
	\Phi_{\eta} \colon & \check{H}^{1}_{\zeta}(\U, S^{1,r}_{X}) \overset{\!\simeq}\longrightarrow \check{H}^{1}_{\xi}(\U, S^{1,r}_{X}) \\
	& [\{g_{ij}\}, \{A_{i}\}] \mapsto [\{g_{ij} \cdot \eta_{ij}\}, \{A_{i}\}].
\end{split}
\end{equation}
Equivalently, we get the isomorphism analogous to \eqref{IsoPhiEtaDiffVB}. It follows that the set of isomorphisms of the form \eqref{IsoPhiEtaNonInt} is a torsor over $\check{H}^{1}(\U, \UU(1)) \simeq H^{1}(X; \R/\Z) \simeq \Hom(H_{1}(X; \Z); \R/\Z)$. Therefore, if $H_{1}(X; \Z) = 0$ (in particular, if $X$ is simply connected), then the semi-group $(\NIVB\nabla_{[\zeta]}(X), \oplus)$ is canonically defined for any $[\zeta] \in H^{2}(X; \R/\Z)$. We observe that the Chern classes \eqref{ChernClasses} are invariant under the tensor product with a flat twisted line bundle, since the latter has vanishing real first Chern class. Since this is exactly the ambiguity in the definition of $\NIVBN_{[\zeta]}(X)$ when $H_{1}(X; \Z)$ is not vanishing, it follows that the Chern classes only depend on $[\zeta]$ in any case.

The cohomological description of non-integral vector bundles can be realized as follows. Topologically (i.e.\ without connection) we replace \eqref{RhoExclam} by
\begin{equation}\label{RhoExclamFlat}
	\rho^{!}_{\fl} \colon \UU(1)_{X} \to \rho_{*}\underline{\UU}(r)_{Y},
\end{equation}
defined in the same way, but considering only constant cochains on $X$. We get the cone complex analogous to \eqref{ComplexConeUrTop}. It follows that $\check{H}^{1}(\rho^{!}_{\fl})$ naturally encodes non-integral vector bundles on $Y$, twisted with respect to any constant class induced by pull-back from $X$. If we introduce a $(\zeta, 0)$-connection, then we replace \eqref{RelativeDeligneCplxRankR} by $\rho^{!}_{\fl} \colon \UU(1)_{X} \to \rho_{*}S^{1,r}_{Y}$, defined by:
\begin{equation}\label{RelativeDeligneCplxRankRFlat}
\xymatrix{
	\UU(1)_{X} \ar[r] \ar[d]^{\rho^{!,0}} & 0 \ar[d] \\
	\rho_{*}\underline{\UU}(r)_{Y} \ar@{-->}[r]^(.45){\tilde{d}} & \rho_{*}\Omega^{1}_{Y, i\uu(r)}.
}\end{equation}
The set $\check{H}^{2}(\rho^{!}_{\fl})$ naturally encodes non-integral vector bundles with connection on $Y$, with respect to any class induced by pull-back from $X$. We can also consider parallel classes, i.e.\ we introduce a $(\zeta, 0, B)$-connection, where $B$ is necessarily global. In this case we replace \eqref{RelativeDeligneCplxRankRPar} by $\rho^{!}_{\fl} \colon (S^{2}_{X})_{\fl} \to \rho_{*}S^{2,r}_{Y}$, defined by:
\begin{equation}\label{RelativeDeligneCplxRankRParFlat}
\xymatrix{
	\UU(1)_{X} \ar[r] \ar[d]^{\rho^{!,0}} & 0 \ar[r] \ar[d] & \Omega^{2}_{X, \R} \ar[d] \\
	\rho_{*}\underline{\UU}(r)_{Y} \ar@{-->}[r]^{\tilde{d}} & \rho_{*}\Omega^{1}_{Y, i\uu(r)} \ar[r]^{d \circ \Tr} & \rho_{*}\Omega^{2}_{Y, \R}.
}\end{equation}
Again $\check{H}^{2}(\rho^{!}_{\fl})$ is the required set. We remark that the first line of \eqref{RelativeDeligneCplxRankRFlat} and the first line of \eqref{RelativeDeligneCplxRankRParFlat} are not equivalent. In fact, the 2-cohomology of the former is $H^{2}(X; \R/\Z)$, which is the flat part of $\hat{H}^{2}(X)$, while the 2-cohomology of the latter is $H^{2}(X; \R/\Z) \oplus \Omega^{2}_{X, \R}$, that is not a subgroup of $\hat{H}^{2}(X)$.\footnote{The component $\Omega^{2}_{X, \R}$ is not quotiented out up to integral forms, even in cohomology.} This is due to the fact that a $\UU(1)$-class already provides the complete information about the corresponding Deligne class, therefore the choice of $B$ is a piece of information more.

Similar considerations hold replacing $\UU(1)_{X}$ by $\Z_{2}$ (more generally, by $\Z_{n}$ or $\Q$). In particular, when $\Lambda = 0$ and $\zeta$ and $\xi$ are cohomologous as $\Gamma_{2}$-cochains, the set of isomorphisms of the form \eqref{IsoPhiEtaNonInt} is a torsor over $\check{H}^{1}(\U, \Z_{2}) \simeq H^{1}(X; \Z_{2})$. Therefore, if $H^{1}(X; \Z_{2}) = 0$ (in particular, if $X$ is simply connected), then the semi-group $(\NIVB\nabla_{[2] \, [\zeta]}(X), \oplus)$ is canonically defined for any $[\zeta] \in H^{2}(X; \Z_{2})$.

\paragraph{\textbf{Remarks on canonicity.}} We have seen that the set $\VBN_{(\zeta, \Lambda)}(X)$ is independent of the twisting cocycle (within a fixed cohomology class) up to the action of $\hat{H}^{2}(X)$, while the sets $\VBN_{(\zeta, \Lambda, B)}(X)$ and $\NIVBN_{\zeta}(X)$ are independent of the cocycle up to the action of $H^{1}(X; \R/\Z)$, i.e.\ the flat part of $\hat{H}^{2}(X)$. Topologically, the sets $\VB_{\zeta}(X)$ and $\NIVB_{\zeta}(X)$ are independent of the cocycle up to the action of $H^{2}(X; \Z)$ and $\Tor \, H^{2}(X; \Z)$ respectively. As we already pointed out, an interesting consequence is that, if $X$ is simply connected, then the sets $\VBN_{[(\zeta, \Lambda, B)]}(X)$, $\NIVBN_{[\zeta]}(X)$ and $\NIVB_{[\zeta]}(X)$ are canonically defined. On the contrary, $\VB\nabla_{[(\zeta, \Lambda)]}(X)$ is almost never well defined.

Let us analyse more in detail the case $H_{1}(X; \Z) = 0$. Considering formula \eqref{DisjUnionTwistVB} or \eqref{DisjUnionTwistVBCohom}, we have that each term of the disjoint union only depends on the cohomology class $[(\zeta, \Lambda, B)]$, but the disjoint union is not canonically determined by the class $[(\zeta, \Lambda)]$. In this case $\hat{H}^{2}(X) \simeq \Omega^{2}_{int}(X)$, since only the curvature is meaningful, and the action of $F \in \Omega^{2}_{int}(X)$ on $\VB\nabla_{(\zeta, \Lambda)}(X)$ sends $\VB\nabla_{(\zeta, \Lambda, B)}(X)$ to $\VB\nabla_{(\zeta, \Lambda, B + F)}(X)$ by tensor product with a line bundle with curvature $F$ (unique up to isomorphism). This is the canonical isomorphism between two fixed representatives of $\VB\nabla_{[(\zeta, \Lambda, B)]}(X)$, since $(\zeta, \Lambda, B)$ and $(\zeta, \Lambda, B + F)$ are cohomologous, but in the whole disjoint union it is a non-trivial automorphism. The situation changes when we consider $(\zeta, \Lambda)$ necessarily of the form $(\zeta, 0)$, i.e.\ when we consider $\NIVBN_{\zeta}(X)$. In this case only the flat part of $\hat{H}^{2}(X)$ acts, and it vanishes when $H_{1}(X; \Z) = 0$, therefore the disjoint union depends canonically on $[\zeta]$ as well. Moreover, we have the canonical projection $\NIVB\nabla_{[\zeta]}(X) \twoheadrightarrow \NIVB_{[\zeta]}(X)$. If $H^{2}(X; \Z) = 0$ too, then we also get the canonical projection $\VB\nabla_{[(\zeta, \Lambda, B)]}(X) \twoheadrightarrow \VB_{[\zeta]}(X)$.

\subsection{Twisted bundles vs non-twisted cohomology}

The sets $\VB_{[\zeta]}(X)$, $\VBN_{[(\zeta, \Lambda)]}(X)$, $\VBN_{[(\zeta, \Lambda, B)]}(X)$ and $\NIVBN_{[\zeta]}(X)$, even when they are defined only up to the action of a non-trivial (abelian) cohomology group, can be described through non-twisted cohomology as well, and this is the information that will lead to a natural description of the possible gauge theories on a D-brane or on a stack of D-branes (more natural than in \cite{FR2}).

\paragraph{\textbf{Abelian framework -- Absolute topological version.}} In order to make the exposition clearer, we start from the abelian setting, since cohomology is more natural in this context. In particular, we consider non-integral line bundles, i.e.\ line bundles twisted by a constant cocycle $\zeta$, and we show that they are classified up to the action of $H^{1}(X; \R/\Z)$ by the 1-cohomology of the sheaf $\underline{\UU}(1)/\UU(1)$. In fact, given a class $[\{g_{ij}\}] \in \check{H}^{1}(\underline{\UU}(1)/\UU(1))$, the cocycle condition imposes that $\zeta_{ijk} := g_{ki}g_{jk}g_{ij}$ is constant, since it must vanish up to the quotient by $\UU(1)$. Moreover, each $g_{ij}$ is defined up to a constant transition function, therefore the cohomology class $[\{g_{ij}\}]$ represents a non-integral line bundle up to a flat (non-twisted) line bundle,\footnote{We should write more precisely $[\{[g_{ij}]\}]$, where the inner square brakects correspond to the quotient by $\UU(1)$.} i.e.\ up to the action of $H^{1}(X; \R/\Z)$. Since, multiplying $g_{ij}$ by any constant transition function $\eta_{ij}$, we get the twisting cocycle $\zeta \cdot \check{\delta}^{1}\eta$, it follows that the only meaningful information on the twisting is the cohomology class $[\zeta_{ijk}] \in \check{H}^{2}(\UU(1))$, therefore $\check{H}^{1}(\underline{\UU}(1)/\UU(1))$ is canonically isomorphic to the rank-1 subset of $\NIVB_{[\zeta]}(X)$. Actually, the same argument can be applied to classes of any degree, i.e.\ to $[\{g_{i_{0} \ldots i_{p}}\}] \in \check{H}^{p}(\underline{\UU}(1)/\UU(1))$, with twisting class $[\{\zeta_{i_{0} \ldots i_{p+1}}\}] \in \check{H}^{p+1}(\UU(1))$, where $\{\zeta_{i_{0} \ldots i_{p+1}}\} = \check{\delta}^{p}\{g_{i_{0} \ldots i_{p}}\}$. In order to describe this picture with the language of homological algebra, we consider the exact sequence of sheaves $0 \to \UU(1) \to \underline{\UU}(1) \to \underline{\UU}(1)/\UU(1) \to 0$, inducing the corresponding long exact sequence
\begin{equation}\label{DiagramT}
	\xymatrix{
	\cdots \ar[r] & \check{H}^{p}(\underline{\UU}(1)) \ar[r] & \check{H}^{p}(\underline{\UU}(1)/\UU(1)) \ar[r]^(.53){t} & \check{H}^{p+1}(\UU(1)) \ar[r] & \cdots.
}\end{equation}
The Bockstein map $t$ associates to $[\{g_{i_{0} \ldots i_{p}}\}]$ its twisting class $[\{\zeta_{i_{0} \ldots i_{p+1}}\}]$, the latter coinciding with the image in $\UU(1)$ of the first Chern class, as the reader can verify both by direct computation and through homological algebra. More precisely, from the long exact sequence induced by $0 \to \R \to \underline{\R} \to \underline{\UU}(1)/\UU(1) \to 0$, we can see that $\check{H}^{p}(\underline{\UU}(1)/\UU(1)) \simeq H^{p+1}(X; \R)$, the latter isomorphism being precisely the first Chern class, whose image in $H^{p+1}(X; \UU(1))$ is the twisting class. We observe that the morphism $\check{H}^{p}(\underline{\UU}(1)) \to \check{H}^{p}(\underline{\UU}(1)/\UU(1))$ in the sequence \eqref{DiagramT} can be generalized to $\check{H}^{p}_{\zeta}(\underline{\UU}(1)) \to \check{H}^{p}(\underline{\UU}(1)/\UU(1))$ for any constant twisting cocycle $\zeta$. In this case, the image is contained in $t^{-1}[\zeta]$, that, by exactness, is a coset of the image of $\check{H}^{p}(\underline{\UU}(1))$. This is the reason why $\check{H}^{p}(\underline{\UU}(1)/\UU(1))$ encodes any twisted class, up to its natural ambiguity.

\paragraph{\textbf{Abelian framework -- Absolute differential version.}} If we introduce a connection on a non-integral line bundle, we have to consider the complex
\begin{equation}\label{ComplexS1Twidle}
	\tilde{S}^{p}_{X} := \underline{\UU}(1)/\UU(1) \overset{\!\tilde{d}}\longrightarrow \Omega^{1}_{\R} \overset{d}\longrightarrow \cdots \overset{d}\longrightarrow \Omega^{p}_{\R}
\end{equation}
and the corresponding Deligne cohomology group $\check{H}^{p}(\tilde{S}^{p}_{X})$. An element of this group is a class $[g, \Lambda^{1}, \ldots, \Lambda^{p}]$, satisfying the usual cocycle condition in Deligne cohomology, except for the fact that  $\check{\delta}^{p}g = \zeta$, with $\zeta$ constant. We get the curvature $F$, locally described by $F = d\Lambda^{p}$, which is not necessarily an integral form. Its de-Rham cohomology class corresponds to the real first Chern class of $[g]$. Composing the projection $[g, \Lambda^{1}, \ldots, \Lambda^{p}] \mapsto [g]$ with the map $t$ in sequence \eqref{DiagramT}, we get the map $\tilde{t} \colon \check{H}^{p}(\tilde{S}^{p}_{X}) \to H^{p+1}(X; \UU(1))$. Since we have the natural projection $S^{p}_{X} \to \tilde{S}^{p}_{X}$, that quotients out $g$ up to constant functions, we get the following exact sequence, whose blue segment refines \eqref{DiagramT} to the differential setting:

\vspace{-5pt}
\begin{scriptsize}
\begin{equation}\label{ExSeqTildeS1}
	\xymatrix{
	0 \ar[r] & H^{p}(X; \UU(1)) \ar[r] & {\color{blue} \check{H}^{p}(S^{p}_{X})} \ar@[blue][r] & {\color{blue} \check{H}^{p}(\tilde{S}^{p}_{X})} \ar@[blue][r]^(.4){\color{blue} \tilde{t}} & {\color{blue} H^{p+1}(X; \UU(1))} \ar[r] & \Tor \, H^{p+2}(X; \Z) \ar[r] & 0.
}
\end{equation}
\end{scriptsize}
\vspace{-10pt}

\noindent Exactness in $\check{H}^{p}(S^{p}_{X})$ is due to the fact that a non-integral class vanishes in $\check{H}^{p}(\tilde{S}^{p}_{X})$ if and only if it can be realized by constant transition functions and zero potentials, i.e.\ if it is flat. Exactness in $\check{H}^{p}(\tilde{S}^{p}_{X})$ is due to the fact that the twisting class vanishes if and only if the non-integral Deligne class is the projection (up to constant functions) of a non-twisted one. Exactness in $H^{p+1}(X; \UU(1))$ is due to the fact that the $\underline{\UU}(1)$-cohomology class of $\zeta$ is trivialized by the transition functions $g$ (in particular, $\tilde{t}$ is not surjective in general). Again the morphism $\color{blue} \check{H}^{p}(S^{p}_{X}) \to \check{H}^{p}(\tilde{S}^{p}_{X})$ in sequence \eqref{ExSeqTildeS1} can be generalized to $\check{H}^{p}_{\zeta}(S^{p}_{X}) \to \check{H}^{p}(\tilde{S}^{p}_{X})$ for any (constant) twisting cocycle $\zeta$, the image being contained in $\tilde{t}^{-1}[\zeta]$.

Actually, the sheaf cohomology of \eqref{ComplexS1Twidle} has been introduced only as a preliminary step towards the partially non-abelian setting. In fact, since the quotient by $\UU(1)$ cuts the flat part, the curvature is the only meaningful information, hence $\check{H}^{p}(\tilde{S}^{p}_{X}) \simeq \Omega^{p+1}_{cl}(X)$.

\paragraph{\textbf{Abelian framework -- Relative topological version.}} Given a smooth closed embedding $\rho \colon Y \hookrightarrow X$, we have seen that the corresponding relative Cech cohomology of the sheaf $\underline{\UU}(1)$ is by definition the one of the cone complex of $\rho^{!} \colon \underline{\UU}(1)_{X} \to \rho_{*}\underline{\UU}(1)_{Y}$. In the non-integral setting, it is natural to consider $\tilde{\rho}^{!} \colon \underline{\UU}(1)_{X}/\UU(1)_{X} \to \rho_{*}(\underline{\UU}(1)_{Y}/\UU(1)_{Y})$, but also the intermediate one $\hat{\rho}^{!} \colon \underline{\UU}(1)_{X} \to \rho_{*}(\underline{\UU}(1)_{Y}/\UU(1)_{Y})$. The natural projections $\underline{\UU}(1) \to \underline{\UU}(1)/\UU(1)$, both in $X$ and $Y$, induce the following diagram:
\begin{equation}\label{DiagMorphAbRelTop}
\xymatrix{
	\underline{\UU}(1)_{X} \ar[rr]^{\rho^{!}} \ar@{=}[d] & & \rho_{*}\underline{\UU}(1)_{Y} \ar[d] \\
	\underline{\UU}(1)_{X} \ar[rr]^{\hat{\rho}^{!}} \ar[d] & & \rho_{*}(\underline{\UU}(1)_{Y}/\UU(1)_{Y}) \ar@{=}[d] \\
	\underline{\UU}(1)_{X}/\UU(1)_{X} \ar[rr]^{\tilde{\rho}^{!}} & & \rho_{*}(\underline{\UU}(1)_{Y}/\UU(1)_{Y}),
}\end{equation}
that induces the morphisms in cohomology
\begin{equation}\label{MorphAbRelTopCohom}
\xymatrix{
	\check{H}^{p}(\rho^{!}) \ar[r] & \check{H}^{p}(\hat{\rho}^{!}) \ar[r] & \check{H}^{p}(\tilde{\rho}^{!}).
}\end{equation}
From the short exact sequence $0 \to \rho^{!}_{\fl} \to \rho^{!} \to \tilde{\rho}^{!} \to 0$, where $\rho^{!}_{\fl} \colon \UU(1)_{X} \to \rho_{*}\UU(1)_{Y}$, we obtain the following long exact sequence, that is the relative version of \eqref{DiagramT}:
\begin{equation}\label{DiagramTRel}
\xymatrix{
	\cdots \ar[r] & \check{H}^{p}(\rho^{!}) \ar[r] & \check{H}^{p}(\tilde{\rho}^{!}) \ar[r]^(.38){\tilde{t}} & H^{p+1}(\rho; \UU(1)) \ar[r] & \cdots.
}\end{equation}
The image of the map $\tilde{t}$ is a class $[\zeta] = [(\zeta_{X}, \zeta_{Y})]$, whose meaning is the following one. Given a class $[(g, h)] \in \check{H}^{p}(\tilde{\rho}^{!})$, we have that $\check{\delta}^{p}g = \zeta_{X}$ (i.e.\ $\zeta_{X}$ is the twisting cocycle of $g$), and $h$ is a class such that $\rho^{*}g = \check{\delta}^{p-1}h \cdot \zeta_{Y}$, i.e.\ $\zeta_{Y}$ is the constant correction of the relative cocycle condition in $Y$. The pair $(\zeta_{X}, \zeta_{Y})$ is a relative cocycle.

With respect to $\hat{\rho}^{!}$, from the short exact sequence $0 \to \rho_{*}(\UU(1)_{Y}) \to \rho^{!} \to \hat{\rho}^{!} \to 0$, we obtain the following long exact sequence:
\begin{equation}\label{DiagramTRel2}
\xymatrix{
	\cdots \ar[r] & \check{H}^{p}(\rho^{!}) \ar[r] & \check{H}^{p}(\hat{\rho}^{!}) \ar[r]^(.38){\hat{t}} & H^{p}(Y; \UU(1)) \ar[r] & \cdots.
}\end{equation}
The image of the map $\hat{t}$ is a class $[\zeta_{Y}]$, whose meaning is the following one. Given a class $[(g, h)] \in \check{H}^{p}(\hat{\rho}^{!})$, we have that $\check{\delta}^{p}g = 0$ (i.e.\ $g$ is not twisted), and $h$ is a class such that $\rho^{*}g = \check{\delta}^{p-1}h \cdot \zeta_{Y}$, i.e.\ $\zeta_{Y}$ is the constant correction of the relative cocycle condition in $Y$ and it turns out to be a cocycle, as it is easy to verify by direct computation too.

We have a natural morphism of exact sequences from $0 \to \rho_{*}(\UU(1)_{Y}) \to \rho^{!} \to \hat{\rho}^{!} \to 0$ to $0 \to \rho^{!}_{\fl} \to \rho^{!} \to \tilde{\rho}^{!} \to 0$, inducing the following morphism of long exact sequences from \eqref{DiagramTRel2} to \eqref{DiagramTRel}:
\begin{equation}\label{DiagramTRelMorph}
\xymatrix{
	\cdots \ar[r] & \check{H}^{p}(\rho^{!}) \ar[r] \ar@{=}[d] & \check{H}^{p}(\hat{\rho}^{!}) \ar[r]^(.38){\hat{t}} \ar[d] & H^{p}(Y; \UU(1)) \ar[r] \ar[d] & \cdots \\
	\cdots \ar[r] & \check{H}^{p}(\rho^{!}) \ar[r] & \check{H}^{p}(\tilde{\rho}^{!}) \ar[r]^(.38){\tilde{t}} & H^{p+1}(\rho; \UU(1)) \ar[r] & \cdots.}\end{equation}
The central vertical arrow is the map appearing in \eqref{MorphAbRelTopCohom} and the one on the right is the Bockstein map of the long exact sequence induced by $\rho$ in singular cohomology with $\UU(1)$-coefficients.

\paragraph{\textbf{Abelian framework -- Relative differential version.}} Given a smooth closed embedding $\rho \colon Y \hookrightarrow X$, we have seen that the corresponding relative Deligne cohomology is by definition the one of the cone complex of $\rho^{!} \colon S^{p}_{X} \to \rho_{*}S^{p-1}_{Y}$. In the non-integral setting, it is natural to consider $\tilde{\rho}^{!} \colon \tilde{S}^{p}_{X} \to \rho_{*}\tilde{S}^{p-1}_{Y}$ and intermediate morphism $\hat{\rho}^{!} \colon S^{p}_{X} \to \rho_{*}\tilde{S}^{p-1}_{Y}$, both defined by a diagram similar to \eqref{RelativeDeligneCplx}. We get the following differential refinement of diagram \eqref{DiagMorphAbRelTop}:
\begin{equation}\label{DiagMorphAbRelDiff}
\xymatrix{
	S^{p}_{X} \ar[rr]^{\rho^{!}} \ar@{=}[d] & & \rho_{*}S^{p}_{X} \ar[d] \\
	S^{p}_{X} \ar[rr]^{\hat{\rho}^{!}} \ar[d] & & \rho_{*}\tilde{S}^{p}_{Y} \ar@{=}[d] \\
	\tilde{S}^{p}_{X} \ar[rr]^{\tilde{\rho}^{!}} & & \rho_{*}\tilde{S}^{p}_{Y},
}\end{equation}
that induces the morphisms in cohomology $\check{H}^{p}(\rho^{!}) \to \check{H}^{p}(\hat{\rho}^{!}) \to \check{H}^{p}(\tilde{\rho}^{!})$, refining \eqref{MorphAbRelTopCohom}. Composing the projections $S^{p} \to \underline{\UU}(1)$ and $\tilde{S}^{p} \to \underline{\UU}(1)/\UU(1)$, both in $X$ and $Y$, with the maps $\tilde{t}$ and $\hat{t}$ in \eqref{DiagramTRel} and \eqref{DiagramTRel2}, we get the maps $\tilde{t} \colon \check{H}^{p}(\tilde{\rho}^{!}) \to H^{p+1}(\rho; \UU(1))$ and $\hat{t} \colon \check{H}^{p}(\hat{\rho}^{!}) \to H^{p}(Y; \UU(1))$, fitting in the following relative version of \eqref{ExSeqTildeS1}:

\begin{tiny}
\begin{equation}\label{ExSeqTildeS1Hat}
	\xymatrix{
	& & & & H^{p}(X; \UU(1)) \ar[d] \ar[dl] \\
	0 \ar[r] & H^{p-1}(Y; \UU(1)) \ar[r] \ar[d] & {\color{red} \check{H}^{p}(\rho^{!})} \ar@[red][r] \ar@{=}[d] & {\color{red} \check{H}^{p}(\hat{\rho}^{!})} \ar@[red][r]^(.4){\color{red} \hat{t}} \ar[d] & {\color{red} H^{p}(Y; \UU(1))} \ar[r] \ar[d] & \Tor \, H^{p+1}(Y; \Z) \ar[r] \ar[d] & 0 \\
	0 \ar[r] & H^{p}(\rho; \UU(1)) \ar[r] & {\color{blue} \check{H}^{p}(\rho^{!})} \ar@[blue][r] & {\color{blue} \check{H}^{p}(\tilde{\rho}^{!})} \ar@[blue][r]^(.4){\color{blue} \tilde{t}} \ar[d] & {\color{blue} H^{p+1}(\rho; \UU(1))} \ar[r] \ar[dl] & \Tor \, H^{p+2}(\rho; \Z) \ar[r] & 0 \\
	& & & H^{p+1}(X; \UU(1)).
}
\end{equation}
\end{tiny}

\paragraph{\textbf{Partially non-abelian framework -- Absolute topological version.}} In order to extend the previous picture to any non-integral vector bundle (not necessarily of rank $1$), we consider the sheaf quotient $\underline{\UU}(r)/\UU(1)$. We get the following Cech complex, analogous to \eqref{ComplexUr}:

\vspace{-10pt}
\begin{small}
\begin{equation}\label{ComplexUrU1Flat}
\xymatrix{
	\check{C}^{0}(\underline{\UU}(r)/\UU(1)) \ar@{-->}[r]^{\check{\delta}^{0}} & \check{C}^{1}(\underline{\UU}(r)/\UU(1)) \ar[r]^{\check{\delta}^{1}} & \check{C}^{2}(\underline{\UU}(r)/\UU(1)) \ar[r]^{\check{\delta}^{2}} & \check{C}^{3}(\underline{\UU}(1)/\UU(1)) \ar[r]^(.65){\check{\delta}^{3}} & \cdots.
}\end{equation}
\end{small}
\vspace{-10pt}

\noindent The group $\check{H}^{1}(\underline{\UU}(r)/\UU(1))$ encodes non-integral vector bundles up to constant functions. We have the natural projection $\underline{\UU}(r) \to \underline{\UU}(r)/\UU(1)$, inducing the corresponding push-forward in cohomology, and we define the map $t \colon \check{H}^{1}(\underline{\UU}(r)/\UU(1)) \to H^{2}(X; \UU(1))$, $[\{g_{ij}\}] \mapsto [\{\zeta_{ijk} := g_{ki}g_{jk}g_{ij}\}]$. We get the exact sequence, that generalizes \eqref{DiagramT} with $p = 1$:
\begin{equation}\label{ExSeqUrU1Flat}
	\xymatrix{
	\check{H}^{1}(\underline{\UU}(r)) \ar[r] & \check{H}^{1}(\underline{\UU}(r)/\UU(1)) \ar[r]^{t} & H^{2}(X; \UU(1)).
}\end{equation}
In this non-abelian framework, we can also consider twisted vector bundles, without imposing that the twisting cocycle is constant. In this case we have to consider the sheaf quotient $\underline{\UU}(r)/\underline{\UU}(1)$ (that vanishes in the abelian context). We get the following Cech complex, analogous to \eqref{ComplexUrU1Flat}:

\vspace{-10pt}
\begin{small}
\begin{equation}\label{ComplexUrU1}
\xymatrix{
	\check{C}^{0}(\underline{\UU}(r)/\underline{\UU}(1)) \ar@{-->}[r]^{\check{\delta}^{0}} & \check{C}^{1}(\underline{\UU}(r)/\underline{\UU}(1)) \ar[r]^{\check{\delta}^{1}} & \check{C}^{2}(\underline{\UU}(r)/\underline{\UU}(1)) \ar[r]^{\check{\delta}^{2}} & \check{C}^{3}(\underline{\UU}(1)/\underline{\UU}(1)) \ar[r]^(.65){\check{\delta}^{3}} & \cdots.
}\end{equation}
\end{small}
\vspace{-8pt}

\noindent The group $\check{H}^{1}(\underline{\UU}(r)/\underline{\UU}(1))$ encodes twisted vector bundles up to $\underline{\UU}(1)$-valued transition functions. We define the map $\bar{t} \colon \check{H}^{1}(\underline{\UU}(r)/\underline{\UU}(1)) \to H^{3}(X; \Z)$, $[\{g_{ij}\}] \mapsto [\{\zeta_{ijk} := g_{ki}g_{jk}g_{ij}\}]$ and we get the exact sequence:
\begin{equation}\label{ExSeqUrU1}
	\xymatrix{
	\check{H}^{1}(\underline{\UU}(r)) \ar[r] & \check{H}^{1}(\underline{\UU}(r)/\underline{\UU}(1)) \ar[r]^(.55){\bar{t}} & H^{3}(X; \Z).
}\end{equation}
There is a natural morphism of exact sequences from \eqref{ExSeqUrU1Flat} to \eqref{ExSeqUrU1}, the right vertical arrow being the Bockstein map induced by the sequence $0 \to \Z \to \R \to \UU(1) \to 0$ and the central one being induced by the projection $\underline{\UU}(r)/\UU(1) \to \underline{\UU}(r)/\underline{\UU}(1)$. Moreover, as in the abelian framwework, the morphism $\check{H}^{1}(\underline{\UU}(r)) \to \check{H}^{1}(\underline{\UU}(r)/\underline{\UU}(1))$ in sequence \eqref{ExSeqUrU1} is the untwisted version of $\check{H}^{1}_{\zeta}(\underline{\UU}(r)) \to \check{H}^{1}(\underline{\UU}(r)/\underline{\UU}(1))$ for any twisting cocycle $\zeta$. In this case the image is contained in $\bar{t}^{-1}[\zeta]$, that, by exactness, is a coset of the image of $\check{H}^{1}(\underline{\UU}(r))$.

\paragraph{\textbf{Partially non-abelian framework -- Absolute differential version.}} In order to endow the bundle with a connection, we consider the following complex, analogous to \eqref{ComplexS1Twidle}:
\begin{equation}\label{ComplexS1rTwidleU1Flat}
	\xymatrix{
	\tilde{S}^{1,r}_{X} := \underline{\UU}(r)/\UU(1) \ar@{-->}[r]^(.65){\tilde{d}} & \Omega^{1}_{i\uu(r)}.
}\end{equation}
We get the Cech double complex analogous to \eqref{DoubleComplexUr}. The group $\check{H}^{1}(\tilde{S}^{1,r}_{X})$ corresponds to the set of non-integral rank-$r$ vector bundles with connection, up to constant $\UU(1)$-valued transition functions. We have the natural projection $S^{1,r}_{X} \to \tilde{S}^{1,r}_{X}$, inducing the corresponding push-forward in cohomology, and we define the map $\tilde{t} \colon \check{H}^{1}(\tilde{S}^{1,r}_{X}) \to H^{2}(X; \UU(1))$, $[\{g_{ij}, A_{i}\}] \mapsto [\{\zeta_{ijk} := g_{ki}g_{jk}g_{ij}\}]$. We get the following exact sequence, that generalizes \eqref{ExSeqTildeS1} and refines \eqref{ExSeqUrU1Flat}:
\begin{equation}\label{ExSeqS1rU1Flat}
	\xymatrix{
	0 \ar[r] & H^{1}(X; \UU(1)) \ar[r]^(.55){i} & {\color{blue} \check{H}^{1}(S^{1,r}_{X})} \ar@[blue][r]^{\color{blue} p} & {\color{blue} \check{H}^{1}(\tilde{S}^{1,r}_{X})} \ar@[blue][r]^(.42){\color{blue} \tilde{t}} & {\color{blue} H^{2}(X; \UU(1))}.
}\end{equation}
We briefly justify the exactness of this sequence. The map $i$ is the embedding of $H^{1}(X; \UU(1))$ in $\check{H}^{1}(S^{1,r}_{X})$ that sends a flat line bundle with connection $L$ to the direct sum $L \oplus \cdots \oplus L$ iterated $r$ times, hence it is injective. The kernel of $p$ consists of the set of rank-$r$ vector bundles with connection that are trivial when applying the quotient by $\UU(1)$ as the centre of $\UU(r)$, hence it coincides with the image of $i$. The kernel of $\tilde{t}$ is formed by vector bundles with connection that are twisted with respect to a $\UU(1)$-cocycle representing a trivial cohomology class, i.e.\ by twisted bundles that are equivalent to a non-twisted one up to the action of $\UU(1)$.

If we consider twisted vector bundles, without imposing that the twisting cocycle is constant, we consider the following complex:
\begin{equation}\label{ComplexS1rTwidleU1}
	\xymatrix{
	\tilde{\tilde{S}}^{1,r}_{X} := \underline{\UU}(r)/\underline{\UU}(1) \ar@{-->}[r]^(.58){\tilde{d}} & \Omega^{1}_{i\uu(r)}/\Omega^{1}_{\R}.
}\end{equation}
We get the Cech double complex analogous to \eqref{DoubleComplexUr}, such that $\check{H}^{1}(\tilde{\tilde{S}}^{1,r}_{X})$ corresponds to the set of twisted vector bundles with connection, up to (non-twisted) lined bundles with connection. We get the following exact sequence, that generalizes \eqref{ExSeqTildeS1} and refines \eqref{ExSeqUrU1}:
\begin{equation}\label{ExSeqS1rU1}
	\xymatrix{
	0 \ar[r] & \check{H}^{1}(S^{1}_{X}) \ar[r] & {\color{blue} \check{H}^{1}(S^{1,r}_{X})} \ar@[blue][r] & {\color{blue} \check{H}^{1}(\tilde{\tilde{S}}^{1,r}_{X})} \ar@[blue][r]^(.47){\color{blue} \tilde{\tilde{t}}} & {\color{blue} H^{3}(X; \Z)}.
}\end{equation}
We have a natural morphism of exact sequences from \eqref{ExSeqS1rU1Flat} to \eqref{ExSeqS1rU1}, the right vertical arrow being the Bockstein map induced by the sequence $0 \to \Z \to \R \to \UU(1) \to 0$ and the central one being induced by the projection $\tilde{S}^{1,r}_{X} \to \tilde{\tilde{S}}^{1,r}_{X}$. If we include 2-forms in the codomain of $\tilde{\tilde{t}}$, we can refine $H^{3}(X; \Z)$ to $\check{H}^{2}(S^{2}_{X})$ in \eqref{ExSeqS1rU1}. Again the morphism $\check{H}^{1}(S^{1,r}_{X}) \to \check{H}^{1}(\tilde{\tilde{S}}^{1,r}_{X})$ in sequence \eqref{ExSeqS1rU1} is the untwisted version of $\check{H}^{1}_{(\zeta, \Lambda)}(S^{1,r}_{X}) \to \check{H}^{1}(\tilde{\tilde{S}}^{1,r}_{X})$ for any $(\zeta, \Lambda)$.

\paragraph{\textbf{Partially non-abelian framework -- Relative topological version.}} The picture is similar to the abelian one, but considering also the quotient by $\underline{\UU}(1)$. We have seen that the non-twisted relative cohomology is the one of $\rho^{!} \colon \underline{\UU}(1)_{X} \to \rho_{*}\underline{\UU}(r)_{Y}$. The five twisted versions are:
\begin{itemize}
	\item $\tilde{\rho}^{!} \colon \underline{\UU}(1)_{X}/\UU(1)_{X} \to \rho_{*}(\underline{\UU}(r)_{Y}/\UU(1)_{Y})$;
	\item $\hat{\rho}^{!} \colon \underline{\UU}(1)_{X} \to \rho_{*}(\underline{\UU}(r)_{Y}/\UU(1)_{Y})$;
	\item $\tilde{\tilde{\rho}}^{!} \colon 0 = \underline{\UU}(1)_{X}/\underline{\UU}(1)_{X} \overset{\!0}\to \rho_{*}(\underline{\UU}(r)_{Y}/\underline{\UU}(1)_{Y})$;
	\item $\hat{\tilde{\rho}}^{!} \colon \underline{\UU}(1)_{X}/\UU(1)_{X} \overset{\!0}\to \rho_{*}(\underline{\UU}(r)_{Y}/\underline{\UU}(1)_{Y})$;
	\item $\hat{\hat{\rho}}^{!} \colon \underline{\UU}(1)_{X} \overset{\!0}\to \rho_{*}(\underline{\UU}(r)_{Y}/\underline{\UU}(1)_{Y})$.
\end{itemize}
Diagram \eqref{DiagMorphAbRelTop}, which is essentially $\rho^{!} \to \hat{\rho}^{!} \to \tilde{\rho}^{!}$, now can be enriched as follows:
\begin{equation}\label{DiagMorphNonAbRelTop}
	\xymatrix{
	& \rho^{!} \ar[dl] \ar[dr] \\
	\hat{\rho}^{!} \ar[rr] \ar[d] & & \tilde{\rho}^{!} \ar[d] \\
	\hat{\hat{\rho}}^{!} \ar[r] & \hat{\tilde{\rho}} \ar[r] & \tilde{\tilde{\rho}}^{!}.
}\end{equation}
The corresponding $t$-maps provide the twisting class, that is a cohomology class of the complex formed by the denominators. Therefore, we get the following extension of diagram \eqref{DiagramTRel2}:

\begin{scriptsize}
\begin{equation}\label{DiagramTRel2NonAb}
\xymatrix{
	& \check{H}^{2}(\rho^{!}) \ar[dl] \ar[dr] & & {\color{green!70!black} H^{2}(Y; \UU(1))} \ar@[green][rr] \ar@[green][d] & & {\color{green!70!black} H^{3}(\rho; \UU(1))} \ar@[green][d] \\
	\check{H}^{2}(\hat{\rho}^{!}) \ar[rr] \ar[d] \ar@[green]@{.>}[urrr]^{{\color{green!70!black}\hat{t}}} & & \check{H}^{2}(\tilde{\rho}^{!}) \ar[d] \ar@[green]@{.>}[urrr]^{{\color{green!70!black}\tilde{t}}} & {\color{green!70!black} H^{3}(Y; \Z)} \ar@[green][r] & {\color{green!70!black} H^{3}(\UU(1)_{X} \to \rho^{*}\underline{\UU}(1)_{Y})} \ar@[green][r] & {\color{green!70!black} H^{4}(\rho; \Z)} \\
	\check{H}^{2}(\hat{\hat{\rho}}^{!}) \ar[r] \ar@[green]@{.>}[urrr]^{{\color{green!70!black}\hat{\hat{{t}}}}} & \check{H}^{2}(\hat{\tilde{\rho}}) \ar[r] \ar@[green]@{.>}[urrr]^{{\color{green!70!black}\hat{\tilde{{t}}}}} & \check{H}^{2}(\tilde{\tilde{\rho}}^{!}). \ar@[green]@{.>}[urrr]_{{\color{green!70!black}\tilde{\tilde{{t}}}}}
}\end{equation}
\end{scriptsize}
 
\paragraph{\textbf{Partially non-abelian framework -- Relative differential version.}} Again the picture is similar to the abelian one, but considering also the quotient by $\underline{\UU}(1)$. We have seen that the non-twisted relative cohomology is the one of $\rho^{!} \colon S^{2}_{X} \to \rho_{*}S^{1,r}_{Y}$. The five twisted versions are:
\begin{itemize}
	\item $\tilde{\rho}^{!} \colon \tilde{S}^{2}_{X} \to \rho_{*}\tilde{S}^{1,r}_{Y}$;
	\item $\hat{\rho}^{!} \colon S^{2}_{X} \to \rho_{*}\tilde{S}^{1,r}_{Y}$;
	\item $\tilde{\tilde{\rho}}^{!} \colon 0 \overset{\!0}\to \rho_{*}\tilde{\tilde{S}}^{1,r}_{Y}$;
	\item $\hat{\tilde{\rho}}^{!} \colon \tilde{S}^{2}_{X} \overset{\!0}\to \rho_{*}\tilde{\tilde{S}}^{1,r}_{Y}$;
	\item $\hat{\hat{\rho}}^{!} \colon S^{2}_{X} \overset{\!0}\to \rho_{*}\tilde{\tilde{S}}^{1,r}_{Y}$.
\end{itemize}
Diagrams \eqref{DiagMorphNonAbRelTop} and \eqref{DiagramTRel2NonAb} holds in the differentials setting too, composing the $t$-maps in \eqref{DiagramTRel2NonAb} with the projections to the underlying topological class.
 
\paragraph{\textbf{Partially non-abelian framework -- Parallel classes.}} The same constructions can be applied to parallel classes too. We leave the detailed to the interested reader, since they will not be necessary in the following.

\paragraph{\textbf{Finer quotients.}} As we have seen in definition \ref{DefNonIntVBZ2}, we can consider the quotients $\underline{\UU}(r)/\Gamma_{k}$ or $\underline{\UU}(r)/\Gamma$, that are intermediate between $\underline{\UU}(r)$ and $\underline{\UU}(r)/\UU(1)$. All of the previous diagrams and sequences can be enriched with such sheaves, both in the topological and differential frameworks. We consider directly the differential non-abelian setting for $\Z_{2}$-twistings, in order not to make the exposition too long. We set:
	\[\xymatrix{
	\tilde{S}^{p}_{[2] \, X} := \underline{\UU}(1)/\Gamma_{2} \ar[r]^(.72){\tilde{d}} & \Omega^{1}_{\R} \ar[r]^{d} & \cdots \ar[r]^{d} & \Omega^{p}_{\R}
} \quad\qquad \xymatrix{
	\tilde{S}^{1,r}_{[2] \, X} := \underline{\UU}(r)/\Gamma_{2} \ar@{-->}[r]^(.65){\tilde{d}} & \Omega^{1}_{i\uu(r)}.
}
\]
We remark that $\Z_{2} = \{0, 1\}$ (with additive notation) and $\Gamma_{2} = \{\pm 1\}$ (with multiplicative notation). We get the projections $S^{p}_{X} \to \tilde{S}^{p}_{[2] \, X} \to \tilde{S}^{p}_{X}$ and $S^{1,r}_{X} \to \tilde{S}^{1,r}_{[2] \, X} \to \tilde{S}^{1,r}_{X}$. Moreover, we set:
\begin{equation}\label{Z2Maps}
	\hat{\rho}^{!}_{[2]} \colon S^{2}_{X} \to \tilde{S}^{1,r}_{[2] \, Y} \qquad\qquad \tilde{\rho}^{!}_{[2]} \colon \tilde{S}^{2}_{[2] \, X} \to \tilde{S}^{1,r}_{[2] \, Y}.
\end{equation}
Of course we have other intermediate possibilities that we are neglecting. In diagram \eqref{DiagMorphNonAbRelTop} the upper triangle can be enriched as follows:
\begin{equation}\label{DiagMorphNonAbRelTopZ2}
	\xymatrix{
	& & \rho^{!} \ar[dl] \ar[dr] \\
	& \hat{\rho}^{!}_{[2]} \ar[rr] \ar[dl] & & \tilde{\rho}^{!}_{[2]} \ar[dr] \\
	\hat{\rho}^{!} \ar[rrrr] & & & & \tilde{\rho}^{!}.
}\end{equation}
Diagram \eqref{DiagramTRel2NonAb} gets enriched coherently, in the differential version as well. In particular, we have the maps:
\begin{equation}\label{TMapsZ2}
	\hat{t}_{[2]} \colon \check{H}^{2}(\hat{\rho}^{!}_{[2]}) \to H^{2}(Y; \Z_{2}) \qquad\qquad \tilde{t}_{[2]} \colon \check{H}^{2}(\tilde{\rho}^{!}_{[2]}) \to H^{3}(\rho; \Z_{2}).
\end{equation}

\subsection{Long exact sequences}

In the abelian setting, the long exact sequence in Deligne cohomology, associated to $\rho \colon Y \hookrightarrow X$, is made by segments of the form $\cdots \to \check{H}^{\bullet-1}(S^{p-1}_{Y}) \to \check{H}^{\bullet}(\rho^{!}) \to \check{H}^{\bullet}(S^{p}_{X}) \to \check{H}^{\bullet}(S^{p-1}_{Y}) \to \cdots$, where $\rho^{!} \colon S^{p}_{X} \to \rho_{*}S^{p-1}_{Y}$. Considering remark \ref{CohDeligneNotP}, that holds in the relative case as well, we get the sequence of the flat part until degree $p-1$ in $X$, and we get the sequence of (topological) singular cohomology from degree $p$ in $Y$ on. Hence, the complete exact sequence is the following one:
\begin{equation}\label{LongExactRho1}
\begin{tikzcd}
\cdots \arrow[r]& H^{p-1}(\rho; \UU(1))\arrow[r]
\arrow[d, phantom, ""{coordinate, name=Z}]
& H^{p-1}(X; \UU(1)) \arrow[r]& \check{H}^{p-1}(S^{p-1}_{Y})\arrow[dll,rounded corners=8pt,curvarr=Z]& 
\\
&\check{H}^{p}(\rho^{!}) \arrow[r]\arrow[d, phantom, ""{coordinate, name=W}] & \check{H}^{p}(S^{p}_{X})\arrow[r]
& H^{p}(Y; \Z)\arrow[dll,rounded corners=8pt,curvarr=W]& 
\\
& H^{p+1}(\rho; \Z) \arrow[r]& H^{p+1}(X; \Z)\arrow[r]& H^{p+1}(Y; \Z)\arrow[r]&\cdots.
\end{tikzcd}
\end{equation}
We can construct an analogous sequence with parallel classes, but we will not need it in the present paper. If we consider $\tilde{\rho}^{!} \colon \tilde{S}^{p}_{X} \to \rho_{*}\tilde{S}^{p-1}_{Y}$, then the flat part vanishes at the quotient, hence integral cohomology is replaced by the real one. We get the following sequence:
\begin{equation}\label{LongExactRho2}
\begin{tikzcd}
\cdots \arrow[r]& 0\arrow[r]
\arrow[d, phantom, ""{coordinate, name=Z}]
& 0 \arrow[r]& \check{H}^{p-1}(\tilde{S}^{p-1}_{Y})\arrow[dll,rounded corners=8pt,curvarr=Z]& 
\\
&\check{H}^{p}(\tilde{\rho}^{!}) \arrow[r]\arrow[d, phantom, ""{coordinate, name=W}] & \check{H}^{p}(\tilde{S}^{p}_{X})\arrow[r]
& H^{p}(Y; \R)\arrow[dll,rounded corners=8pt,curvarr=W]& 
\\
& H^{p+1}(\rho; \R) \arrow[r]& H^{p+1}(X; \R)\arrow[r]& H^{p+1}(Y; \R)\arrow[r]&\cdots.
\end{tikzcd}
\end{equation}
The natural projection morphisms $S^{p}_{X} \to \tilde{S}^{p}_{X}$ and $S^{p-1}_{Y} \to \tilde{S}^{p-1}_{Y}$, inducing $\rho^{!} \to \tilde{\rho}^{!}$, induce a morphism of exact sequences from \eqref{LongExactRho1} to \eqref{LongExactRho2}. The reader can construct the analogous sequence induced by $\hat{\rho}^{!} \colon S^{p}_{X} \to \rho_{*}\tilde{S}^{p-1}_{Y}$.

Considering the partially non-abelian generalization, from the five maps of diagram \eqref{DiagMorphNonAbRelTop} we get the long exact sequence analogous to \eqref{LongExactRho1} and \eqref{LongExactRho2} (with $p=2$), whose most relevant segments are the following ones:
\begin{align}
	& \xymatrix{
	\cdots \ar[r] & H^{1}(X; \UU(1)) \ar[r] & \check{H}^{1}(S^{1,r}_{Y}) \ar[r] & \check{H}^{2}(\rho^{!}) \ar[r] & \check{H}^{2}(S^{2}_{X}) \ar[r] & \cdots
} \label{LongExactRhoNonAb1} \\
	& \xymatrix{
	\cdots \ar[r] & 0 \ar[r] & \check{H}^{1}(\tilde{S}^{1,r}_{Y}) \ar[r] & \check{H}^{2}(\tilde{\rho}^{!}) \ar[r] & \check{H}^{2}(\tilde{S}^{2}_{X}) \ar[r] & \cdots
} \label{LongExactRhoNonAb2} \\
	& \xymatrix{
	\cdots \ar[r] & H^{1}(X; \UU(1)) \ar[r]^(.55){0} & \check{H}^{1}(\tilde{S}^{1,r}_{Y}) \ar[r] & \check{H}^{2}(\hat{\rho}^{!}) \ar[r] & \check{H}^{2}(S^{2}_{X}) \ar[r] & \cdots
} \label{LongExactRhoNonAb3} \\
	& \xymatrix{
	\cdots \ar[r] & 0 \ar[r] & \check{H}^{1}(\tilde{\tilde{S}}^{1,r}_{Y}) \ar[r]^{\simeq} & \check{H}^{2}(\tilde{\tilde{\rho}}^{!}) \ar[r] & 0 \ar[r] & \cdots
} \label{LongExactRhoNonAb4} \\
	& \xymatrix{
	\cdots \ar[r] & 0 \ar[r] & \check{H}^{1}(\hat{\tilde{S}}^{1,r}_{Y}) \ar[r] & \check{H}^{2}(\hat{\tilde{\rho}}^{!}) \ar[r] & \check{H}^{2}(\tilde{S}^{2}_{X}) \ar[r] & \cdots
} \label{LongExactRhoNonAb5} \\
	& \xymatrix{
	\cdots \ar[r] & H^{1}(X; \UU(1)) \ar[r]^(.55){0} & \check{H}^{1}(\tilde{\tilde{S}}^{1,r}_{Y}) \ar[r] & \check{H}^{2}(\hat{\hat{\rho}}^{!}) \ar[r] & \check{H}^{2}(S^{2}_{X}) \ar[r] & \cdots
}. \label{LongExactRhoNonAb6}
\end{align}
The morphisms of diagram \eqref{DiagMorphNonAbRelTop} induce the corresponding morphisms of exact sequences. From the maps \eqref{Z2Maps}, we get:
\begin{align}
	& \xymatrix{
	\cdots \ar[r] & H^{1}(X; \UU(1)/\Z_{2}) \ar[r] & \check{H}^{1}(\tilde{S}^{1,r}_{[2] \, Y}) \ar[r] & \check{H}^{2}(\tilde{\rho}^{!}_{[2]}) \ar[r] & \check{H}^{2}(\tilde{S}^{2}_{[2] \, X}) \ar[r] & \cdots.
} \label{LongExactRhoNonAb7} \\
	& \xymatrix{
	\cdots \ar[r] & H^{1}(X; \UU(1)) \ar[r] & \check{H}^{1}(S^{1,r}_{[2] \, Y}) \ar[r] & \check{H}^{2}(\hat{\rho}^{!}_{[2]}) \ar[r] & \check{H}^{2}(S^{2}_{X}) \ar[r] & \cdots
}. \label{LongExactRhoNonAb8}
\end{align}

\subsection{Classification of gauge theories}\label{ClassGauge}

Given a smooth embedding $\rho \colon Y \hookrightarrow X$, we call $\id_{Y}$ the identity map of $Y$ and we consider the following natural morphisms of complexes of sheaves on $X$:
\begin{equation}\label{DiagMorphClassGauge}
	\rho^{!} \to \hat{\rho}^{!}_{[2]} \to \rho_{*}\tilde{\id}_{Y}^{!} \to \rho_{*}\tilde{\tilde{\id}}_{Y}^{!}.
\end{equation}
This is equivalent to considering the following diagram:

\begin{small}
\[\xymatrix{
	S^{2}_{X} \ar[rr]^{\rho^{!}} \ar[d] & & \rho_{*}S^{1,r}_{Y} \ar[d] \\
	S^{2}_{X} \ar[rr]^{\hat{\rho}^{!}_{[2]}} \ar[d] & & \rho_{*}\tilde{S}^{1,r}_{[2] \, Y} \ar[d] \\
	\rho_{*}\tilde{S}^{2}_{Y} \ar[rr]^{\rho_{*}\tilde{\id}_{Y}^{!}} \ar[d] & & \rho_{*}\tilde{S}^{1,r}_{Y} \ar[d] \\
	\rho_{*}\tilde{\tilde{S}}^{2}_{Y} \ar[rr]^{\rho_{*}\tilde{\tilde{\id}}_{Y}^{!}} & & \rho_{*}\tilde{\tilde{S}}^{1,r}_{Y}.
}\]
\end{small}

\noindent The first and the last vertical morphisms are the ones appearing in diagrams \eqref{DiagMorphNonAbRelTop} and \eqref{DiagMorphNonAbRelTopZ2}, with respect to $\rho$ and $\id_{Y}$ respectively. The central one is formed by $\hat{\rho}^{!} \colon S^{2}_{X} \to \rho_{*}\tilde{S}^{2}_{Y}$ on the left and the natural projection $\tilde{S}^{1,r}_{[2] \, Y} \to \tilde{S}^{1,r}_{Y}$ on the right. Considering the exact sequences \eqref{LongExactRhoNonAb1}, \eqref{LongExactRhoNonAb2}, \eqref{LongExactRhoNonAb8} and \eqref{LongExactRhoNonAb4} respectively, and considering the $t$-maps in \eqref{TMapsZ2}, \eqref{ExSeqS1rU1Flat} and \eqref{ExSeqS1rU1} respectively, from \eqref{DiagMorphClassGauge} we get the following diagram in cohomology, that is the key to classify the possible gauge theories on a stack of D-branes:

\begin{footnotesize}
\begin{equation}\label{DiagMorphClassGauge2}
\xymatrix{
	\cdots \ar[r] & H^{1}(X; \UU(1)) \ar[r] \ar[d] & \check{H}^{1}(S^{1,r}_{Y}) \ar[r] \ar[d] & {\color{magenta} \check{H}^{2}(\rho^{!})} \ar[r] \ar[d] & \check{H}^{2}(S^{2}_{X}) \ar[r] \ar[d] & \cdots & {\color{green!70!black} H^{2}(Y; \Z_{2}) } \ar@[green][d] \\
	\cdots \ar[r] & H^{1}(X; \UU(1)) \ar[r] \ar[d] & \check{H}^{1}(\tilde{S}^{1,r}_{[2] \, Y}) \ar[r] \ar[d] & {\color{orange} \check{H}^{2}(\hat{\rho}^{!}_{[2]})} \ar[r] \ar[d] \ar@[green]@{.>}[urrr]^{{\color{green!70!black} \hat{t}_{[2]}}} & \check{H}^{2}(S^{2}_{X}) \ar[r] \ar[d] & \cdots & {\color{green!70!black} H^{2}(Y; \UU(1)) } \ar@[green][d] \\
	\cdots \ar[r] & 0 \ar[r] \ar[d] & \check{H}^{1}(\tilde{S}^{1,r}_{Y}) \ar[r] \ar[d] \ar@[green]@{.>}[urrrr]^{{\color{green!70!black} \tilde{t}}} & {\color{cyan!75!black} \check{H}^{2}(\tilde{\id}_{Y}^{!})} \ar[r] \ar[d] & \check{H}^{2}(\tilde{S}^{2}_{Y}) \ar[r] \ar[d] & \cdots & {\color{green!70!black} H^{3}(Y; \Z) } \\
	\cdots \ar[r] & 0 \ar[r] & \check{H}^{1}(\tilde{\tilde{S}}^{1,r}_{Y}) \ar[r] \ar@[green]@{.>}[urrrr]^{{\color{green!70!black} \tilde{\tilde{t}}}} & {\color{purple!50!blue} \check{H}^{2}(\tilde{\tilde{\id}}_{Y}^{!})} \ar[r] & 0 \ar[r] & \cdots.
}\end{equation}
\end{footnotesize}

\noindent Let us call $\B := [(\zeta, \Lambda, B)]$ the $B$-field class on $X$, with curvature $H = dB$, and $\A := [(g, A)]$ the $A$-field class on $Y$, i.e.\ the $(\xi, \Theta)$-twisted bundle with connection, that is the gauge theory to be classified. In order to choose a sign convention coherent with formula \eqref{FWCondition} in chapter \ref{TwKChar}, we consider the dual bundle $\A^{*} := [(\bar{g}, -\bar{A})]$, which is $-(\xi, \Theta)$-twisted.\footnote{We could choose other coherent conventions, considering $\A$ instead of $\A^{*}$. For example, we could replace $\rho^{!}$ by $-\rho^{!}$ in the definition of the mapping cone (see \eqref{ConeCplx}), or we could add a minus sign to a term of formula \eqref{DefChi} in chapter \ref{TwKChar}, so that formula \eqref{FWCondition} would change coherently. We could also change the convention in the definition of twisting for a vector bundle, introducing a minus sign when necessary. All of such choices are equivalent.} Moreover, we call $\Ss$ the spin$^{c}$-gerbe on $Y$, which is a flat gerbe classified by the second Stiefel-Whitney class $w_{2}(Y) \in H^{2}(Y; \Z_{2})$.\footnote{As a (complex) gerbe, the flat holonomy of $\Ss$ is classified by the image of $w_{2}(Y)$ through the inclusion $\Z_{2} \hookrightarrow \UU(1)$. It follows that its first Chern class is $W_{3}(Y)$. Anyway, such a gerbe is the complexification of a real one, classified by $w_{2}$ itself, hence we can choose its transition functions in $\Gamma_{2}$. In this way we can choose the more refined function $\hat{\rho}^{!}_{[2]}$, instead of $\hat{\rho}^{!}$, in diagram \eqref{DiagMorphClassGauge2}.} In order for the open string action to be well-defined, $\A^{*}$ has to trivialize topologically the tensor product $\rho^{*}\B \otimes \Ss$, hence $\rho^{*}(\zeta, \Lambda) + (\epsilon, 0) = -(\xi, \Theta)$, where $\epsilon$ is a $\Gamma_{2}$-cocycle representing $w_{2}(Y)$. Equivalently:
\begin{equation}\label{FWConditionSA}
	\rho^{*}(\zeta, \Lambda) + (\xi, \Theta) = (\epsilon, 0),
\end{equation}
i.e.\ $\rho^{*}\zeta \cdot \xi = \epsilon$ and $\Lambda + \Theta = 0$. Considering the coboundary $\check{\pmb{D}}^{2}$ of the complex \eqref{ComplexConeUr}, this means that
\begin{equation}\label{CocycleCondGeneral}
	\check{\pmb{D}}^{2}(\zeta, \Lambda, B, \bar{g}, -\bar{A}) = (0, 0, 0, \epsilon, 0).
\end{equation}
Since $\epsilon$ vanishes quotienting out by $\Z_{2}$, in general $[(\zeta, \Lambda, B, \bar{g}, -\bar{A})] \in {\color{orange} \check{H}^{2}(\hat{\rho}^{!}_{[2]})}$, the latter group being part of diagram \eqref{DiagMorphClassGauge2}. More precisely, $[(\zeta, \Lambda, B, \bar{g}, -\bar{A})] \in \hat{t}_{[2]}^{-1}(w_{2}(Y))$, the latter being a coset of $\Ker(\hat{t}_{[2]})$. Such a coset is the most general classification of the possible $A$-field and $B$-field configurations in type II superstring theory (cfr.\ \cite{BFS} and \cite{FR2}). If $w_{2}(Y) = 0$, then we can choose $\epsilon = 1$ and lift the class to $[(\zeta, \Lambda, B, \bar{g}, -\bar{A})] \in {\color{magenta} \check{H}^{2}(\rho^{!})}$, the lift being part of the physical datum, since it is not unique in general.

\paragraph{\textbf{Freed-Witten anomaly.}} Fixing $X$, $Y$ and $\B$, condition \eqref{CocycleCondGeneral} can be realized by some configurations of $\A$ if and only if the Freed-Witten anomaly vanishes \cite{FW, Kapustin}, i.e.\ $\rho^{*}[\zeta] + [\xi] = W_{3}(Y)$ in $H^{3}(Y; \Z)$. We denote such a condition by (FW). Let us show in detail the equivalence \eqref{CocycleCondGeneral} $\Leftrightarrow$ (FW). From \eqref{CocycleCondGeneral} we easily get (FW) by taking the first Chern class on both sides. Conversely, assuming (FW), we choose any representatives $(\zeta, \Lambda)$ and $\epsilon$ and, up to multiplying any fixed $\xi$ by a coboundary, we set $(\xi, \Theta) := (\epsilon, 0) - \rho^{*}(\zeta, \Lambda)$. Since $[\xi]$ is the twisting class of a vector bundle by construction, it has finite order, hence there exist some $(\xi, \Theta)$-twisted bundles on $Y$ independently of $\Theta$, i.e.\ there exist some admissible $A$-field configurations.

Since $[\xi]$ is part of $\A$ and the only constraint about it consists in being a finite-order class, it follows that, fixing $X$, $Y$ and $\B$, (FW) can be realized if and only if $\rho^{*}[\zeta]$ has finite order (equivalently, $\rho^{*}H$ is exact). In general, the rank $r$ of the $A$-filed depends on the configuration, since, given $[\xi]$, not every rank is realizable. If we have a single D-brane, i.e.\ we impose $r = 1$, then $[\xi] = 0$ in $H^{3}(Y; \Z)$, hence the only possibility is $\rho^{*}[\zeta] = W_{3}(Y)$, which is a much stronger condition than $\rho^{*}[\zeta]$ being torsion. This is the original condition described in the seminal paper \cite{FW}, that has been refined to the partially non-abelian setting (i.e.\ to stacks of D-branes) in \cite{Kapustin}.

\paragraph{\textbf{The classification.}} Now we can classify the possible gauge theories on the world-volume, analysing diagram \eqref{DiagMorphClassGauge2}. \vspace{3pt}

\texttt{Case I: Vanishing $B$-field and spin world-volume.} If the $B$-field vanishes on the whole space-time $X$, i.e.\ $[(\zeta, \Lambda, B)] = 0$, and $w_{2}(Y) = 0$, then $[(\zeta, \Lambda, B, \bar{g}, -\bar{A})] \in {\color{magenta} \check{H}^{2}(\rho^{!})}$ and, in particular, it belongs to the kernel of the map ${\color{magenta} \check{H}^{2}(\rho^{!})} \to \check{H}^{2}(S^{2}_{X})$. By exactness of the first line of \eqref{DiagMorphClassGauge2}, it can be lifted to a class $[(g, A)] \in \check{H}^{1}(S^{1,r}_{Y})$, that is unique up to the image of $H^{1}(X; \UU(1)) \to \check{H}^{1}(S^{1,r}_{Y})$, i.e.\ up to the restriction to $Y$ of a \emph{flat} line bundle on $X$. Such an ambiguity is the ``residual gauge freedom'', whose physical meaning has been described in detail in \cite{BFS, FR2, FR4}. \vspace{3pt}

\texttt{Case II: Vanishing $B$-field.} If the $B$-field vanishes on the whole space-time $X$, but we have no hypotheses on the world-volume, then $[(\zeta, \Lambda, B, \bar{g}, -\bar{A})] \in {\color{orange} \check{H}^{2}(\hat{\rho}^{!}_{[2]})}$ and, in particular, it belongs to the kernel of the map ${\color{orange} \check{H}^{2}(\hat{\rho}^{!}_{[2]})} \to \check{H}^{2}(S^{2}_{X})$. By exactness of the second line of \eqref{DiagMorphClassGauge2}, it can be lifted to a class $[(g, A)] \in \check{H}^{1}(\tilde{S}^{1,r}_{[2] \, Y})$, that is unique up to the image of $H^{1}(X; \UU(1)) \to \check{H}^{1}(\tilde{S}^{1,r}_{[2] \, Y})$. This means that we get a $\Z_{2}$-non-integral vector bundle with connection, twisted by $w_{2}(Y)$, up to the restriction of a \emph{flat} line bundle on $X$ and up to a \emph{real} line bundle on $Y$, the latter ambiguity being the action of $H^{1}(X; \Z_{2})$ on $\Z_{2}$-twisted bundles. This case has not been considered explicitly in \cite{BFS, FR2, FR4}. \vspace{3pt}

\texttt{Case III: Flat $B$-field on $Y$.} If $\rho^{*}H = 0$, projecting $[(\zeta, \Lambda, B, \bar{g}, -\bar{A})] \in {\color{orange} \check{H}^{2}(\hat{\rho}^{!}_{[2]})}$ to ${\color{cyan!75!black} \check{H}^{2}(\tilde{\id}_{Y}^{!})}$ in diagram \eqref{DiagMorphClassGauge}, by exactness of the third line we get a unique lift $[(g, A)] \in \check{H}^{1}(\tilde{S}^{1,r}_{Y})$, i.e.\ we get a non-integral vector bundle on $Y$, that is canonically defined up to a flat line bundle on $Y$ (i.e.\ up to the action of $H^{1}(Y; \R/\Z)$; see comments after definition \ref{DefNonIntVB}). By construction and because of the chosen sign convention, we have that $[(g, A)] \in \tilde{t}^{-1}(w'_{2}(Y) \cdot \rho^{*}\B^{-1})$, where $w'_{2}(Y) \in H^{2}(Y; \UU(1))$ is the image of $w_{2}(Y)$ trough the embedding $\Z_{2} \hookrightarrow \UU(1)$. \vspace{3pt}

\texttt{Case IV: No hypotheses on the $B$-field.} In a completely generic configuration, we have to project $[(\zeta, \Lambda, B, \bar{g}, -\bar{A})] \in {\color{orange} \check{H}^{2}(\hat{\rho}^{!}_{[2]})}$ to ${\color{purple!50!blue} \check{H}^{2}(\tilde{\tilde{\id}}_{Y}^{!})}$ and, by exactness, we get a unique class $[(g, A)] \in \check{H}^{1}(\tilde{\tilde{S}}^{1,r}_{Y})$, i.e.\ a twisted vector bundle up to any line bundle with connection, belonging $\tilde{\tilde{t}}^{-1}(W_{3}(Y) - \rho^{*}[\zeta])$. This means that the bundle is $(\xi, \Theta)$-twisted, where $[(\xi, \Theta)] \in \check{H}^{2}(S^{1}_{Y}) \simeq H^{3}(Y; \Z)$ coincides with $W_{3}(Y) - \rho^{*}[\zeta]$ because of the Freed-Witten anomaly.

\paragraph{\textbf{The classification for simply-connected world-volumes.}} If the world-volume is simply connected, many of the ambiguities in the previous classification disappear, hence the classification becomes simpler and more natural. In particular, any non-integral vector bundle with connection is canonically defined from the $\UU(1)$-twisting class, hence ordinary and $\Z_{2}$-twisted vector bundles are just particular cases. With the cohomological language, the maps $\check{H}^{1}(S^{1,r}_{Y}) \hookrightarrow \check{H}^{1}(\tilde{S}^{1,r}_{[2] \, Y}) \hookrightarrow \check{H}^{1}(\tilde{S}^{1,r}_{Y})$ become embeddings. Moreover, the upper-left maps from $H^{1}(X; \UU(1))$ in diagram \eqref{DiagMorphClassGauge2} become $0$-maps (independently of the fact that $H^{1}(X; \UU(1))$ vanishes or not), since the image is a (vanishing) flat line bundle on $Y$. Even the map ${\color{magenta} \check{H}^{2}(\rho^{!})} \hookrightarrow {\color{orange} \check{H}^{2}(\hat{\rho}^{!}_{[2]})}$ becomes an embedding, as the reader can verify by diagram chasing in \eqref{DiagMorphClassGauge2}. Therefore, diagram \eqref{DiagMorphClassGauge2} becomes the following one:

\begin{footnotesize}
\begin{equation}\label{DiagMorphClassGauge2SC}
\xymatrix{
	0 \ar[r] \ar[d] & \check{H}^{1}(S^{1,r}_{Y}) \ar[r] \ar@{^(->}[d] & {\color{magenta} \check{H}^{2}(\rho^{!})} \ar[r] \ar@{^(->}[d] & \check{H}^{2}(S^{2}_{X}) \ar[r] \ar[d] & \cdots & {\color{green!70!black} H^{2}(Y; \Z_{2}) } \ar@[green][d] \\
	0 \ar[r] \ar[d] & \check{H}^{1}(\tilde{S}^{1,r}_{[2] \, Y}) \ar[r] \ar@{^(->}[d] & {\color{orange} \check{H}^{2}(\hat{\rho}^{!}_{[2]})} \ar[r] \ar[d] \ar@[green]@{.>}[urrr]^{{\color{green!70!black} \hat{t}_{[2]}}} & \check{H}^{2}(S^{2}_{X}) \ar[r] \ar[d] & \cdots & {\color{green!70!black} H^{2}(Y; \UU(1)) } \ar@[green][d] \\
	0 \ar[r] \ar[d] & \check{H}^{1}(\tilde{S}^{1,r}_{Y}) \ar[r] \ar[d] \ar@[green]@{.>}[urrrr]^{{\color{green!70!black} \tilde{t}}} & {\color{cyan!75!black} \check{H}^{2}(\tilde{\id}_{Y}^{!})} \ar[r] \ar[d] & \check{H}^{2}(\tilde{S}^{2}_{Y}) \ar[r] \ar[d] & \cdots & {\color{green!70!black} H^{3}(Y; \Z) } \\
	0 \ar[r] & \check{H}^{1}(\tilde{\tilde{S}}^{1,r}_{Y}) \ar[r] \ar@[green]@{.>}[urrrr]^{{\color{green!70!black} \tilde{\tilde{t}}}} & {\color{purple!50!blue} \check{H}^{2}(\tilde{\tilde{\id}}_{Y}^{!})} \ar[r] & 0 \ar[r] & \cdots.
}\end{equation}
\end{footnotesize}

Hence, case I of the previous classification is just a particular case of II, therefore we can think of $[(\zeta, \Lambda, B, \bar{g}, -\bar{A})] \in {\color{orange} \check{H}^{2}(\hat{\rho}^{!}_{[2]})}$, without analysing $w_{2}(Y)$. Moreover, case II becomes a particular case of III as well, since, in spite of ${\color{orange} \check{H}^{2}(\hat{\rho}^{!}_{[2]})} \to {\color{cyan!75!black} \check{H}^{2}(\tilde{\id}_{Y}^{!})}$ not being an embedding in general (since it forgets the behaviour of $\B$ outside $Y$), the eventual lift to $\check{H}^{1}(\tilde{S}^{1,r}_{[2] \, Y})$ in the second line is a particular case of the eventual lift to $\check{H}^{1}(\tilde{S}^{1,r}_{Y})$ in the third line. Therefore, the only meaningful information is whether $\rho^{*}H$ vanishes or not, leading to the following two possibilities: \vspace{3pt}

\texttt{Case III (including I and II): Flat $B$-field on $Y$.} Projecting the class $[(\zeta, \Lambda,$ $B, \bar{g}, -\bar{A})] \in {\color{orange} \check{H}^{2}(\hat{\rho}^{!}_{[2]})}$ to ${\color{cyan!75!black} \check{H}^{2}(\tilde{\id}_{Y}^{!})}$ in diagram \eqref{DiagMorphClassGauge2SC}, by exactness of the third line we get a unique lift $[(g, A)] \in \check{H}^{1}(\tilde{S}^{1,r}_{Y})$, i.e.\ we get a non-integral vector bundle on $Y$, that is canonically defined. By construction $[(g, A)] \in \tilde{t}^{-1}(w'_{2}(Y) \cdot \rho^{*}\B^{-1})$. \vspace{3pt}

\texttt{Case IV: No hypotheses on the $B$-field.} We project $[(\zeta, \Lambda, B, \bar{g}, -\bar{A})] \in {\color{orange} \check{H}^{2}(\hat{\rho}^{!}_{[2]})}$ to ${\color{purple!50!blue} \check{H}^{2}(\tilde{\tilde{\id}}_{Y}^{!})}$ and, by exactness, we get a unique class $[(g, A)] \in \check{H}^{1}(\tilde{\tilde{S}}^{1,r}_{Y})$, i.e.\ a twisted vector bundle up to any line bundle with connection, belonging $\tilde{\tilde{t}}^{-1}(W_{3}(Y) - \rho^{*}[\zeta])$.

\begin{Rmk}\label{RmkSCBTilde} \emph{The two cases of the classification for simply-connected world-volumes can be unified through the following statement, that will be important in order to define the Wess-Zumino action:}
\begin{quote}
	\emph{$(\star)$ Fixing a 2-form $\tilde{B} \in \Omega^{2}_{\R, Y}$, such that $d\tilde{B} = \rho^{*}H$, we get canonically a non-integral vector bundle with connection $[(E, \nabla)]$ on $Y$ (up to isomorphism), twisted with respect to the class $[\xi] \in H^{2}(Y; \R/\Z)$ such that
	\begin{equation}\label{FWConditionSC}
		\rho^{*}[(\zeta, \Lambda, B)] + [(\xi, 0, \tilde{B})] = w'_{2}(Y).
	\end{equation}
If we replace $\tilde{B}$ by $\tilde{B'}$, then $[(E, \nabla)]$ is replaced by $[(E, \nabla)] \otimes [(L, \nabla_{L})]$, where $[(L, \nabla_{L})]$ is the unique (up to isomorphism) non-integral line bundle with connection whose curvature is $\tilde{B}' - \tilde{B}$.}
\end{quote}
\emph{This is the general case IV, but, when $\rho^{*}H = 0$, we have the canonical choice $\tilde{B} = 0$, so that we obtain a well-defined twisted vector bundle with connection, as stated in case III. In order to prove $(\star)$, we observe that, since $[\xi]$ is a torsion class, we can always choose $\xi$ constant and $\Theta = 0$, i.e.\ we can represent $w'_{2}(Y) - \rho^{*}[(\zeta, \Lambda, B)]$ in the form $[(\xi, 0, \tilde{B})]$. In this way we get a $(\xi, 0)$-twisted bundle $[(E, \nabla)]$ (i.e.\ a non-integral vector bundle) and a fixed form $\tilde{B}$. Changing representative to $(\xi', 0, \tilde{B'})$, through \emph{any} cochain $(\eta, \lambda)$ such that $\check{D}^{1}(\eta, \lambda) = (\xi'\xi^{-1}, 0, \tilde{B}' - \tilde{B}$), not necessarily with $\eta$ constant, is equivalent to multiplying $[(E, \nabla)]$ by the $[\xi'\xi^{-1}]$-non-integral line bundle $[(\eta, \lambda)]$. Since $Y$ is simply-connected, the latter is completely determined by its curvature $\tilde{B}' - \tilde{B}$. In particular, if $\tilde{B'} = \tilde{B}$, then $[(E, \nabla)]$ does not change, hence it is completely determined by $\tilde{B}$.}
\end{Rmk}

Let us prove the statement $(\star)$ with the intrinsic language of homological algebra. We consider the map $(\hat{\id}_{Y})^{!}_{[2] \, \fl} \colon (S^{2}_{Y})_{\fl} \to \rho_{*}S^{1,r}_{[2] \, Y}$, described by the following complex, intermediate between \eqref{RelativeDeligneCplxRankRFlat} and \eqref{RelativeDeligneCplxRankRParFlat}, but with $X = Y$ and quotienting out by $\Z_{2}$ in the image:
\begin{equation}\label{RelativeDeligneCplxRankRFlat2}
\xymatrix{
	\UU(1)_{Y} \ar[r] \ar[d]^{\rho^{!,0}} & 0 \ar[r] \ar[d] & \Omega^{2}_{Y, \R} \ar[d] \\
	\rho_{*}\underline{\UU}(r)_{Y}/\Gamma_{2} \ar[r]^{\tilde{d}} & \rho_{*}\Omega^{1}_{Y, i\uu(r)} \ar[r] & 0.
}\end{equation}
Considering the complex $(\hat{\id}_{Y})^{!}_{[2]} \colon S^{2}_{Y} \to \rho_{*}S^{1,r}_{[2] \, Y}$, we get the short exact sequence of complexes $0 \to (\hat{\id}_{Y})^{!}_{[2] \, \fl} \to (\hat{\id}_{Y})^{!}_{[2]} \to \tilde{S}^{1}_{Y} \to 0$. We have a natural map $\hat{\rho}^{!}_{[2]} \to \rho_{*}(\hat{\id}_{Y})^{!}_{[2]}$, inducing ${\color{orange} \check{H}^{2}(\hat{\rho}^{!}_{[2]})} \to \check{H}^{2}((\hat{\id}_{Y})^{!}_{[2]})$, therefore we get the image of $[(\zeta, \Lambda, B, \bar{g}, -\bar{A})]$ in $\check{H}^{2}((\hat{\id}_{Y})^{!}_{[2]})$ (with no hypotheses on the $B$-field).  Since $\check{H}^{2}(\tilde{S}^{1}_{Y}) \simeq H^{2}(Y; \R)$, by the Freed-Witten anomaly the image of $[(\zeta, \Lambda, B, \bar{g}, -\bar{A})]$ in $\check{H}^{2}(\tilde{S}^{1}_{Y})$ vanishes, hence by exactness, we get a lift to $\check{H}^{2}((\hat{\id}_{Y})^{!}_{[2] \, \fl})$, the latter being formed precisely by a $\xi$-twisted non-integral vector bundle and a global 2-form $\tilde{B}$. The lift is unique up to the action of $\check{H}^{1}(\tilde{S}^{1}_{Y}) \simeq \Omega^{2}_{Y; \R}$, where the latter isomorphism identifies a non-integral line bundle with its curvature $F$ (since $Y$ is simply connected, the curvature is the only meaningful information). Since a non-zero curvature $F$ shifts $\tilde{B}$ to $\tilde{B} + F$, it follows that, fixing $\tilde{B}$, the lift is unique.

We conclude observing that, fixing $\tilde{B}$, we can write the (relative) curvature of $[(\zeta, \Lambda, B, \bar{g},$ $-\bar{A})] \in {\color{orange} \check{H}^{2}(\hat{\rho}^{!}_{[2]})}$ as $(H, r\tilde{B} - \Tr\, F)$, where $F$ is the field strength of the corresponding bundle $[(E, \nabla)]$. Therefore, fixing $\tilde{B}$ is equivalent to fixing $\Tr \, F$, and this choice determines the twisted vector bundle up to isomorphism, since, $Y$ being simply-connected, it leaves no freedom for the action of $\check{H}^{1}(S^{1}_{Y})$.

\paragraph{\textbf{Final remarks.}} Up to know we considered the case of vanishing $B$-field in the whole $X$, corresponding to the kernel of ${\color{magenta} \check{H}^{2}(\rho^{!})} \to \check{H}^{2}(S^{2}_{X})$ or of ${\color{orange} \check{H}^{2}(\hat{\rho}^{!}_{[2]})} \to \check{H}^{2}(S^{2}_{X})$ in diagram \eqref{DiagMorphClassGauge}, and the case of flat $B$-field on $Y$, corresponding to the kernel of ${\color{cyan!75!black} \check{H}^{2}(\tilde{\id}_{Y}^{!})} \to \check{H}^{2}(\tilde{S}^{2}_{Y})$. The reader may inquire why we did not consider the intermediate cases of flat $B$-field on $X$ or of vanishing $B$-field on $Y$. The reason is that we get no new information with respect to case III, that include both. In fact, the case $H = 0$ corresponds to the map $\tilde{\rho}^{!}$, whose exact sequence is \eqref{LongExactRhoNonAb2}. By exactness, we get a unique lift $[(g, A)] \in \check{H}^{1}(\tilde{S}^{1,r}_{Y})$, exactly as in the more general case III. Similarly, the case of vanishing $B$-field on $Y$ corresponds to the map $\hat{\id}_{Y}^{!}$, whose exact sequence is \eqref{LongExactRhoNonAb3}. Again we get a unique lift $[(g, A)] \in \check{H}^{1}(\tilde{S}^{1,r}_{Y})$. Moreover, it is worth to mention the case in which the $B$-field is flat on $Y$ and its holonomy is exactly $w'_{2}(Y)$. In this case we get a non-twisted vector bundle on $Y$, but anyway it is meaningful only up to a flat line bundle on $Y$, since we fit in case III, with a class belonging to $\Ker(\tilde{t})$.

The last remark is about the world-volume being spin$^{c}$, since $W_{3}(Y)$ appears in Freed-Witten anomaly. Actually the most interesting cases of the previous classification deal with non-integral bundles, that's why $w_{2}(Y)$ is more interesting than $W_{3}(Y)$. The latter is partially meaningful only in case IV, since we are forced to deal with generic twisted bundles. In this case, the condition $W_{3}(Y) = 0$ is equivalent to the fact that $[(g, A)] \in \Ker(\tilde{\tilde{t}})$, i.e.\ we get a non-twisted vector bundle, but anyway up to any twisted line bundle.

%%%%%%%%%%%%%%%%%%%%%%%%%%%%%%%%%%%%%%%%%%%%%%%%%%%%%%%%%%%%%%%%%%%%%%%%%%%%%%%%%%%%%%%%%%%%%%%

\section{Twisted differential K-theory}\label{DiffTwistedKTh}

It is well known that $K$-theory is a better tool than ordinary cohomology in order to classify D-brane charges \cite{Evslin, MM, OS, FR4, FS}. When the $B$-field is vanishing on the whole space-time and the world-volume is spin, we use ordinary $K$-theory, otherwise we must consider its twisted version \cite{CW, CMW, BM2, Kapustin}. Since there are various models in the literature, we summarize the ones that we are going to use in this paper, focusing on the features that will be important for our purposes.

This section is supposed to be mainly expository, but we think it is necessary in order to provide an organic presentation of differential twisted K-theory with the language of the present paper, paying particular attention to the dependence on the twisting cocycle. Subsection \ref{TopFiniteOrder} is based mainly on \cite{Karoubi}, but giving more emphasis to the structure of twisted cohomology theory only through the language of twisted vector bundles. Subsection \ref{FLModelTwisted} is based mainly on \cite{Park}, but we complete the construction to the differential cohomology groups of any degree, considering product and integration, and we link it more explicitly to relative Deligne cohomology. Subsection \ref{DiffTwistedKGen} is based on \cite{CMW}, except for \ref{DepCocycleGTwC} and \ref{EqTorsTwSec}. The former deals with the dependence on the cocycle in the differential setting. Surely its content is well known to the experts too, but we could not find a similar discussion in literature and it is very relevant for our purposes. In \ref{EqTorsTwSec} we sketch the construction of the isomorphism between the two models of differential twisted K-theory we are dealing with, deferring the details to a future paper. We conclude this section with subsection \ref{TopTwistedDeligne}, in which we introduce a simple but useful definition, that will be applied to define the topological charge of a D-brane.

\subsection{Topological twisted K-theory -- Finite order}\label{TopFiniteOrder}

Any general definition of twisted $K$-theory involves some infinite-dimensional geometric objects, like projective Hilbert bundles. Nevertheless, this is not necessary when the twisting class has finite order \cite{Karoubi}, as we are going to summarize.

The direct sum of $\zeta$-twisted vector bundles is defined as $(\{E_{i}\}, \{\varphi_{ij}\}) \oplus (\{F_{i}\}, \{\psi_{ij}\}) := (\{E_{i} \oplus F_{i}\}, \{\varphi_{ij} \oplus \psi_{ij}\})$. The set $\VB_{\zeta}(X)$, endowed with this operation, is a commutative semi-group, hence we can define the corresponding Grothendieck group, that we call \emph{$\zeta$-twisted K-theory group} of $X$ and we denote by $K_{\zeta}(X)$. This is the $0$-degree group. In order to construct twisted K-theory of any degree, we consider the $n$-dimensional torus $\mathbb{T}^{n} := (S^{1})^{n}$ and the natural embeddings $\iota_{j} \colon \mathbb{T}^{n-1} \times X \hookrightarrow \mathbb{T}^{n} \times X$, for $j \in \{1, \ldots, n\}$, defined supposing that $S^{1}$ has a marked point (e.g.\ $1 \in S^{1} \subset \mathbb{C}$). Fixing a good cover of $S^{1}$, we easily get a good cover of $\mathbb{T}^{n} \times X$ by cartesian product; moreover, the cocycle $\zeta$ on $X$ determines a cocycle on $\mathbb{T}^{n} \times X$, that we also denote by $\zeta$, by pull-back with respect to the projection $\mathbb{T}^{n} \times X \to X$. Then we set:
\begin{equation}\label{TwistedDegreeN}
	K^{-n}_{\zeta}X := \Ker(\iota_{1}^{*}) \cap \ldots \cap \Ker(\iota_{n}^{*}) \subset K_{\zeta}(\mathbb{T}^{n} \times X).
\end{equation}
This construction concerns non-positive degrees. Before extending it to positive degrees, we need to introduce the Bott periodicity, hence we need products. Given a $\zeta$-twisted bundle $(\{E_{i}\}, \{\varphi_{ij}\})$ and an $\eta$-twisted bundle $(\{F_{i}\}, \{\psi_{ij}\})$, we set $(\{E_{i}\}, \{\varphi_{ij}\}) \otimes (\{F_{i}\}, \{\psi_{ij}\}) := (\{E_{i} \otimes F_{i}\}, \{\varphi_{ij} \otimes \psi_{ij}\})$. The result is a $(\zeta\eta)$-twisted bundle. Therefore, we define the product $\VB_{\zeta}(X) \otimes \VB_{\eta}(Y) \to \VB_{\zeta\eta}(X \times Y)$, $[E] \cdot [F] := \pi_{X}^{*}[E] \cdot \pi_{Y}^{*}[F]$, where $\pi_{X} \colon X \times Y \to X$ and $\pi_{Y} \colon X \times Y \to Y$ are the projections. Considering the corresponding $K$-theory classes, we get $K_{\zeta}(X) \otimes K_{\eta}(Y) \to K_{\zeta\eta}(X \times Y)$. From definition \eqref{TwistedDegreeN} and the natural homeomorphism $(\mathbb{T}^{n} \times X) \times (\mathbb{T}^{n} \times Y) \approx \mathbb{T}^{n+m} \times X \times Y$, we get $K^{-n}_{\zeta}(X) \otimes K^{-m}_{\eta}(Y) \to K^{-n-m}_{\zeta\eta}(X \times Y)$. Composing with the pull-back via the diagonal map, we get $K^{-n}_{\zeta}(X) \otimes K^{-m}_{\eta}(X) \to K^{-n-m}_{\zeta\eta}(X)$. In particular, this shows that $K^{\bullet}_{\zeta}(X)$ is a graded module over $K^{\bullet}(X)$. Now we can define the Bott periodicity. We consider the dual of the tautological line bundle of $\PPP^{1}(\C) \simeq S^{2}$; its pull-back via the projection $\mathbb{T}^{2} \to S^{2}$ is a line bundle $\eta$ on the torus, such that $\eta - 1$ is a generator of $K^{-2}(pt) \simeq \tilde{K}(\mathbb{T}^{2}) \simeq \Z$. We get the morphism $B \colon K^{\bullet}_{\zeta}(X) \to K^{\bullet-2}_{\zeta}(X)$, $\alpha \mapsto (\eta - 1)\alpha$, which is a group isomorphism. This shows that only the parity of the (non-positive) degree is meaningful up to canonical isomorphism, hence, for $n \geq 0$, we set $K^{n}_{\zeta}(X) := K^{-n}_{\zeta}(X)$.

\paragraph{\textbf{Dependence on the cocycle.}} If $\zeta$ and $\xi$ are cohomologous and we fix $\eta$ such that $\xi = \zeta \cdot \check{\delta}^{1}\eta$, the isomorphism \eqref{IsoPhiEta} extends to the corresponding Grothendieck groups, defining $\Phi_{\eta} \colon K_{\zeta}(X) \overset{\!\simeq}\longrightarrow K_{\xi}(X)$. This shows that the isomorphism class of the group $K_{\zeta}(X)$ only depends on $[\zeta]$ in a non-canonical way, the set of isomorphisms of the form $\Phi_{\eta}$ being a torsor over $\check{H}^{1}(\U, \underline{\UU}(1)) \simeq H^{2}(X, \Z)$. In particular, if $H^{2}(X, \Z) = 0$, then $K_{[\zeta]}(X)$ is canonically defined and does not depend on the cover. In general, only the quotient up to the action of $H^{2}(X, \Z)$, that we denote by $K_{[\zeta]}(X)$ anyway, depends on $[\zeta]$ in a canonical way.

\paragraph{\textbf{Non-integral vector bundles and K-theory.}} Of course we are free to choose $\zeta$ constant, getting the Grothendieck group of non-integral vector bundles twisted by $\zeta \in \check{Z}^{2}(\U, \UU(1))$. All of the previous considerations keep on holding, except for the fact that the set of isomorphisms of the form $\Phi_{\eta}$ is a torsor over the image of the natural map $H^{1}(\U, \UU(1)) \to H^{1}(\U, \underline{\UU}(1))$, canonically isomorphic to $\Tor\,H^{2}(X; \Z)$. Therefore, if $\Tor\,H^{2}(X; \Z) = 0$ (in particular, if $X$ is simply connected), then $K_{[\zeta]}(X)$ is canonically defined with $[\zeta] \in H^{2}(X; \R/\Z)$.

\SkipTocEntry \subsubsection{Relative version}\label{RelativeVersion}

Given a pair of spaces $(X, Y)$, such that $\U\vert_{Y}$ is a good cover, we define the relative $\zeta$-twisted K-theory group $K_{\zeta}(X, Y)$ as follows. We call $\Ll_{\zeta}(X, Y)$ the set of triples $(E, F, \alpha)$, where $E$ and $F$ are $\zeta$-twisted vector bundles on $X$ and $\alpha \colon E\vert_{Y} \to F\vert_{Y}$ is an isomorphism. There exists a natural operation of direct sum in $\Ll_{\zeta}(X, Y)$, defined by $(E, F, \alpha) \oplus (E', F', \alpha') := (E \oplus E', F \oplus F', \alpha \oplus \alpha')$. Two triples $(E, F, \alpha)$ and $(E', F', \alpha')$ are \emph{isomorphic} if there exist two isomorphisms $f \colon E \to E'$ and $g \colon F \to F'$ such that $\alpha' \circ f\vert_{Y} = g\vert_{Y} \circ \alpha$. We use the notation $(E, F, \alpha) \simeq (E', F', \alpha')$. Moreover, a triple is called \emph{elementary} if it is of the form $(E, E, \id)$. We introduce the following equivalence relation in $\Ll_{\zeta}(X, Y)$: the triple $(E, F, \alpha)$ is equivalent to $(E', F', \alpha')$ if there exist two elementary triples $(G, G, \id)$ and $(H, H, \id)$ such that $(E, F, \alpha) \oplus (G, G, \id) \simeq (E', F', \alpha') \oplus (H, H, \id)$. We call $K_{\zeta}(X, Y)$ the quotient of $\Ll_{\zeta}(X, Y)$ by this relation. We obtain an abelian group, whose zero-element is the class $[(E, E, \id)]$, for any $E$, and such that $-[(E, F, \alpha)] = [(F, E, \alpha^{-1})]$. When $Y = \emptyset$, we recover the usual group $K_{\zeta}(X)$ by identifying $[(E, F, \emptyset)]$ with $[E] - [F]$.

The definition of the relative groups of any degree is analogous to \eqref{TwistedDegreeN}. In particular, we consider the $n$-dimensional torus $\mathbb{T}^{n} := (S^{1})^{n}$ and the natural embeddings $\iota_{j} \colon (\mathbb{T}^{n-1} \times X, \mathbb{T}^{n-1} \times Y) \hookrightarrow (\mathbb{T}^{n} \times X, \mathbb{T}^{n} \times Y)$, for $j \in \{1, \ldots, n\}$, and we set:
\begin{equation}\label{TwistedRelDegreeN}
	K^{-n}_{\zeta}(X, Y) := \Ker(\iota_{1}^{*}) \cap \ldots \cap \Ker(\iota_{n}^{*}) \subset K_{\zeta}(\mathbb{T}^{n} \times X, \mathbb{T}^{n} \times Y).
\end{equation}
There is a natural product $K^{\bullet}_{\zeta}(X) \times K^{\bullet}_{\xi}(X, Y) \to K^{\bullet}_{\zeta\xi}(X, Y)$, defined by $E \cdot [(F, G, \alpha)] := [(E \otimes F, E \otimes G, \id \otimes \alpha)]$. In particular, the Bott periodicity morphism $B \colon K^{\bullet}_{\zeta}(X, Y) \to K^{\bullet-2}_{\zeta}(X, Y)$, $\alpha \mapsto (\eta - 1)\alpha$, is well-defined. This shows that only the parity of the (non-positive) degree is meaningful up to canonical isomorphism, hence, for $n \geq 0$, we set $K^{n}_{\zeta}(X, Y) := K^{-n}_{\zeta}(X, Y)$. With these definitions we get a twisted cohomology theory on the category of pairs $(X, Y)$ such that $X$ is compact and $Y \subset X$ is closed.\footnote{Such a theory can be extended to pairs having the same homotopy type of a pair $(X, Y)$, with $X$ compact and $Y$ closed, but we omit the details.} Moreover, the isomorphism \eqref{IsoPhiEtaBdl} easily extends to the relative setting, applying it to both $E$ and $F$ in the triple $(E, F, \alpha)$, hence the isomorphism class of $K^{\bullet}_{\zeta}(X, Y)$ only depends on $[\zeta]$. In particular, if $H^{2}(X; \Z) = 0$, then $K^{\bullet}_{[\zeta]}(X, Y)$ is canonically defined. Choosing $\zeta$ constant, we get a canonical group $K^{\bullet}_{[\zeta]}(X, Y)$, with $[\zeta] \in H^{2}(X; \R/\Z)$, when $\Tor \, H^{2}(X; \Z) = 0$.

\SkipTocEntry \subsubsection{Compact support}\label{CptSupp}

When $X$ is locally compact, we are going to define the compactly-supported $\zeta$-twisted K-theory groups $K^{\bullet}_{\zeta, \cpt}(X)$. Given a compact subset $K \subset X$, we set $X \dblsetminus K := \overline{X \setminus K}$, i.e.\ the complement of the interior of $K$ in $X$. We say that $K$ is \emph{$\U$-compact} if $\U\vert_{X \dblsetminus K}$ is a good cover. We denote by $\K_{X, \U}$ the directed set formed by the $\U$-compact subsets of $X$, the partial ordering being given by set inclusion. We assume that $\U$ is refined enough so that the union of the $\U$-compact subsets is the whole $X$. We think of $\K_{X, \U}$ as a category, whose objects are the $\U$-compact subsets of $X$ and whose morphisms are defined as follows: the set $\Hom_{\K_{X, \U}}(K, H)$ contains one element if $K \subset H$ and it is empty otherwise. Calling $\Top_{2}$ the category of pairs of topological spaces, there is a natural contravariant functor $\Cc_{X} \colon \K_{X, \U} \to \Top_{2}$, assigning to an object $K$ the pair $(X, X \dblsetminus K)$ and to a morphism $K \hookrightarrow H$ the natural inclusion $(X, X \dblsetminus H) \hookrightarrow (X, X \dblsetminus K)$. Calling $\A_{\Z}$ the category of graded abelian groups, we define $K^{\bullet}_{\zeta, \cpt}(X)$ as the colimit of the composition functor $K^{\bullet}_{\zeta} \circ \Cc_{X} \colon \K_{X, \U} \to \A_{\Z}$:
\begin{equation}
  K^{\bullet}_{\zeta, \cpt}(X) := \colim \bigl( \, K^{\bullet}_{\zeta} \circ \Cc_{X} \colon \K_{X, \U} \to \A_{\Z} \, \bigr).
\end{equation}
Since $K^{\bullet}_{\zeta}$ and $\Cc_{X}$ are both contravariant, the composition is covariant. Concretely, an element of $K^{\bullet}_{\zeta, \cpt}(X)$ is an equivalence class $[\alpha]$, represented by a class $\alpha \in K^{\bullet}_{\zeta}(X, X \dblsetminus K)$, $K$ being a $\U$-compact subset of $X$. The colimit is taken over the groups $K^{\bullet}_{\zeta}(X, X \dblsetminus K)$, where, if $K \subset H$, the corresponding morphism in the direct system is the pull-back $i_{KH}^{*} \colon K^{\bullet}_{\zeta}(X, X \dblsetminus K) \to K^{\bullet}_{\zeta}(X, X \dblsetminus H)$.

We have defined the compactly-supported groups associated to a space $X$. We can make $K^{\bullet}_{\zeta, \cpt}$ a covariant functor, defining its behaviour on \emph{open embeddings}. In fact, let us fix an open embedding $\iota \colon Y \hookrightarrow X$, such that the good cover of $X$ restricts to a good cover of $Y$. For any $\U$-compact subset $K \subset Y$, from the embedding of pairs $\iota_{K} \colon (Y, Y \dblsetminus K) \hookrightarrow (X, X \dblsetminus \iota(K))$, we get the induced morphism $\iota_{K}^{*} \colon K^{\bullet}_{\zeta}(X, X \dblsetminus \iota(K)) \to K^{\bullet}_{\zeta}(Y, Y \dblsetminus K)$, which is an excision isomorphism. If $K \subset H$, the following diagram commutes:
\[\xymatrix{
  K^{\bullet}_{\zeta}(Y, Y \dblsetminus K) \ar[rr]^{(\iota_{K}^{*})^{-1}} \ar[d]^{i_{KH}^{*}} & & K^{\bullet}_{\zeta}(X, X \dblsetminus \iota(K)) \ar[d]^{i_{KH}^{*}} \\
  K^{\bullet}_{\zeta}(Y, Y \dblsetminus H) \ar[rr]^{(\iota_{H}^{*})^{-1}} & & K^{\bullet}_{\zeta}(X, X \dblsetminus \iota(H))
}\]
therefore we get an induced morphism between the colimits, i.e., $\iota_{*} \colon K^{\bullet}_{\zeta,\cpt}(Y) \to K^{\bullet}_{\zeta,\cpt}(X)$, as desired. We also have the following natural product
\begin{equation}\label{ProdCptMod}
	K^{\bullet}_{\zeta}(X) \times K^{\bullet}_{\xi,\cpt}(X) \to K^{\bullet}_{\zeta\xi,\cpt}(X),
\end{equation}
defined by $\alpha \cdot [\beta] := [\alpha \cdot \beta]$, the product $\alpha \cdot \beta$ being an instance of $K^{\bullet}_{\zeta}(X) \times K^{\bullet}_{\xi}(X, X \dblsetminus K) \to K^{\bullet}_{\zeta\xi}(X, X \dblsetminus K)$. Finally, the following canonical isomorphism will be useful later on:
\begin{equation}\label{IsoCptS1}
	\int_{\R} \colon K^{\bullet}_{\zeta, \cpt}(\R \times X) \overset{\!\simeq}\longrightarrow K^{\bullet-1}_{\zeta, \cpt}(X),
\end{equation}
where $\int_{\R} := i_{*}$, for $i \colon \R \times X \hookrightarrow S^{1} \times X$ the open embedding induced by $\R \hookrightarrow \R^{+} \approx S^{1}$. In order to show that \eqref{IsoCptS1} is an isomorphism, on the l.h.s.\ we take the direct limit of $K^{\bullet}_{\zeta}((\R \times X), (\R \times X) \dblsetminus \iota(K))$ over the cofinal subset of $\K_{\R \times X, \U}$ formed by the elements $K := I_{n} \times K'$, with $I_{n} := [-n, n] \subset \R$ and $K' \subset X$. The direct limit is the group of compactly-supported classes in $S^{1} \times X$ relative to $\{\infty\} \times X$.\footnote{Since the compact support in $S^{1} \times X$ only concerns $X$, we can define \emph{relative} classes with respect to the subspace $\{\infty\}$ of $S^{1}$, considering pairs of the form $(S^{1} \times K', \{\infty\} \times K')$ and applying the colimit. A similar consideration holds about the notion of pull-back, that we use in the following paragraph.} Considering the embedding $i_{\infty} \colon X \hookrightarrow X \times S^{1}$, $x \mapsto (x, \infty)$, such a group is the kernel of $i_{\infty}^{*} \colon K^{\bullet}_{\zeta, \cpt}(S^{1} \times X) \to K^{\bullet}_{\zeta, \cpt}(X)$, which is exactly $K^{\bullet-1}_{\zeta, \cpt}(X)$ by definition \eqref{TwistedDegreeN}.

The isomorphism \eqref{IsoPhiEtaBdl}, extended to relative twisted K-theory, induces an isomorphism between the compactly-supported groups. It follows that the isomorphism class of $K^{\bullet}_{\zeta, \cpt}(X)$ only depends on $[\zeta]$. In particular, if $H^{2}(X; \Z) = 0$, then $K^{\bullet}_{[\zeta], \cpt}(X)$ is canonically defined. If $\Tor\,H^{2}(X; \Z) = 0$ and we choose $\zeta$ constant, then $K^{\bullet}_{[\zeta], \cpt}(X)$ is well defined.

\SkipTocEntry \subsubsection{$S^{1}$-Integration}

We are going to define the integration map
\begin{equation}\label{S1Integration}
	\int_{S^{1}} \colon K^{\bullet+1}_{\zeta}(S^{1} \times X) \to K^{\bullet}_{\zeta}(X),
\end{equation}
calling $\zeta$ both the twisting cocycle on $X$ and its pull-back on $S^{1} \times X$. Let us consider the embedding $\iota_{1} \colon X \hookrightarrow S^{1} \times X$, defined through a marked point of $S^{1}$, and the projection $\pi_{1} \colon S^{1} \times X \to X$. We set $\int_{S^{1}} \colon K_{\zeta}(S^{1} \times X) \to K^{-1}_{\zeta}(X)$, $\alpha \mapsto \alpha - \pi_{1}^{*}i_{1}^{*}\alpha$. Since $\pi_{1} \circ i_{1} = \id_{X}$, it follows that $\pi_{1}^{*}i_{1}^{*}\alpha \in \Ker(i_{1}^{*})$, such a kernel being $K^{-1}_{\zeta}(X)$ by definition \eqref{TwistedDegreeN}. Now, since $K^{-1}_{\zeta}(S^{1} \times X) \subset K_{\zeta}(S^{1} \times S^{1} \times X)$ and $K^{-2}_{\zeta}(X) \subset K^{-1}_{\zeta}(S^{1} \times X)$, we define $\int_{S^{1}} \colon K^{-1}_{\zeta}(S^{1} \times X) \to K^{-2}_{\zeta}(X)$ as the restriction of $\int_{S^{1}} \colon K_{\zeta}(S^{1} \times S^{1} \times X) \to K^{-1}_{\zeta}(S^{1} \times X)$. The construction can be iterated and it can be extended to positive degrees by the Bott periodicity.

\SkipTocEntry \subsubsection{Thom isomorphism}\label{SecThomIso}

We recall some basic facts on spin geometry in order to fix the notation within the framework of twisted bundles. Given a real orientable vector bundle $\pi \colon E \to X$ of rank $2r$, with fixed metric and orientation, the good cover $\U = \{U_{i}\}_{i \in I}$ on $X$ induces the good cover $\pi^{*}\U := \{E_{i} := \pi^{-1}(U_{i})\}_{i \in I}$ on $E$. Let us consider the frame bundle $p \colon \SO(E) \to X$ and the corresponding restrictions $p_{i} \colon \SO(E_{i}) \to U_{i}$. Since $U_{i}$ is contractible, we can choose a spin lift $p'_{i} \colon \Spin(E_{i}) \to U_{i}$ for each $i$. Moreover, we fix principal bundle isomorphisms $\varphi'_{ij} \colon \Spin(E_{i})\vert_{U_{ij}} \to \Spin(E_{j})\vert_{U_{ij}}$, lifting the identity $\SO(E_{i})\vert_{U_{ij}} = \SO(E_{j})\vert_{U_{ij}}$. It follows that $\varphi'_{ki} \circ \varphi'_{jk} \circ \varphi'_{ij} = \epsilon_{ijk} = \pm 1$. We get the class $w_{2}(E) := [\epsilon_{ijk}] \in H^{2}(X; \Z_{2})$, that vanishes if and only if there exists a global spin lift. Moreover, we call $\rho \colon \Spin(2r) \to \UU(2^{r})$ the natural unitary representation of $\Spin(2r)$, acting on the vector space $S := \C^{2^{r}}$, that splits in the two irreducible chirality representations $S = S_{+} \oplus S_{-}$. From each local spin lift $\Spin(E_{i})$, we get the associated vector bundle $S(E_{i}) := \Spin(E_{i}) \times_{\rho} S$ of rank $2^{r}$ (the bundle of spinors), with the chirality splitting $S(E_{i}) = S_{+}(E_{i}) \oplus S_{-}(E_{i})$. We get the $\epsilon$-twisted vector bundle
\begin{equation}\label{SETwisted}
	S(E) := (\{S(E_{i})\}, \{\varphi''_{ij}\}),
\end{equation}
where $\varphi''_{ij} := \varphi'_{ij} \times_{\rho} 1 \colon \Spin(E_{i}) \times_{\rho} S \to \Spin(E_{j}) \times_{\rho} S$. We also have the global splitting of $\epsilon$-twisted vector bundles $S(E) = S_{+}(E) \oplus S_{-}(E)$, where $S_{\pm}(E) := (\{S_{\pm}(E_{i})\}, \{\varphi''_{ij}\vert_{S_{\pm}(E_{i})}\})$.

Let us consider the projection $\pi \colon E \to X$ and the pull-back $\pi^{*}S(E) = \pi^{*}S_{+}(E) \oplus \pi^{*}S_{-}(E)$ on $E$. We define the twisted-bundle morphism $\mu \colon \pi^{*}S_{+}(E) \to \pi^{*}S_{-}(E)$ as follows. For any fixed point $e \in E_{x}$, with $x \in U_{i}$, the morphism $\mu$ acts between the fibres $(S_{+}(E_{i}))_{x}$ and $(S_{-}(E_{i}))_{x}$, both contained in $(S(E_{i}))_{x}$, as the Clifford multiplication by $e \in \CCl(E_{x}) = \CCl((E_{i})_{x})$.\footnote{We recall that $E_{i} = E\vert_{U_{i}}$, hence $E_{x} = (E_{i})_{x}$ for any $x \in U_{i}$.} It is easy to verify that $\mu$ is actually a morphism of twisted bundles and that it is an isomorphism on the closure of the complement of the disk bundle $D_{E}$ of $E$. Refining $\pi^{*}\U$ on $E$ in a suitable way,\footnote{For example, we can fix a trivialization $E_{i} \simeq U_{i} \times \R^{2n}$ and consider the cover formed by the sets $U_{i} \times B$, where $B$ is an open ball in $\R^{2n}$.} we get a good cover $\V = \{V_{j}\}_{j \in J}$ such that $D_{E}$ is $\V$-compact and the union of the $\V$-compact sets is the whole $E$. In this way we get a class $\tilde{u} := [\pi^{*}S_{+}(E), \pi^{*}S_{-}(E), \mu] \in K_{\epsilon}(E, E \dblsetminus D_{E})$, representing a compactly-supported class $u \in K_{\epsilon, \cpt}(E)$, the latter being a (twisted) Thom class. Of course, when $w_{2}(E) = 0$, we can choose $\epsilon = 1$ and we get an ordinary Thom class.

Up to now the rank of $E$ has been taken even by hypothesis. If $\rk(E)$ is odd, we consider a Thom class in $E \oplus 1 \to X$, where $1 = X \times \R$ is the trivial real line bundle.\footnote{Since $w_{2}(E \oplus 1) = w_{2}(E)$ and $W_{3}(E \oplus 1) = W_{3}(E)$, nothing changes about the characteristic classes involved.} We get $u \in K_{\epsilon, \cpt}(E \oplus 1) = K_{\epsilon, \cpt}(E \times \R) \simeq K^{-1}_{\epsilon, \cpt}(E)$, the last isomorphism being \eqref{IsoCptS1}. It follows that, if $\rk \, E = n$ (even or odd), then $u \in K^{n}_{\epsilon, \cpt}(E)$.

The Thom isomorphism is defined in the following way, choosing a refinement map $\phi \colon J \to I$ from $\pi^{*}\U$ to $\V$ and using the product \eqref{ProdCptMod}:
\begin{equation}\label{ThomIso}
\begin{split}
	T \colon & K_{\zeta}^{\bullet}(X) \overset{\!\simeq}\longrightarrow K^{\bullet+n}_{\phi^{*}(\zeta\epsilon),\cpt}(E) \\
	& \alpha \mapsto \pi^{*}\alpha \cdot u.
\end{split}
\end{equation}
Of course we are identifying $\zeta$ and $\epsilon$ in $X$ with $\pi^{*}\zeta$ and $\pi^{*}\epsilon$ in $E$ as twisting cocycles. In general \eqref{ThomIso} depends on $\zeta$, $\epsilon$ and $\phi$. Nevertheless, it becomes canonical when $H^{2}(X; \Z) = 0$. In fact, in this case, $H^{2}(E; \Z) = 0$ as well, since $E$ retracts by deformation on $X$. It follows that both $K_{\zeta}^{\bullet}(X)$ and $K^{\bullet+n}_{\zeta\epsilon,\cpt}(E)$ only depends on the cohomology class of their twisting cocycle, hence the isomorphism \eqref{ThomIso} can be written intrinsically as follows:
\begin{equation}\label{ThomIso2}
\begin{split}
	T \colon & K_{[\zeta]}^{\bullet}(X) \overset{\!\simeq}\longrightarrow K^{\bullet+n}_{[\zeta] + W_{3}(E),\cpt}(E) \\
	& \alpha \mapsto \pi^{*}\alpha \cdot u.
\end{split}
\end{equation}
Since $\epsilon$ is constant, if $\zeta$ is constant too we get the a canonical isomorphism similar to \eqref{ThomIso2} on any manifold such that $\Tor\,H^{2}(X; \Z) = 0$, replacing $W_{3}(E)$ by $w_{2}(E)$. The Thom isomorphism, stated with the present language, is coherent with the the one constructed in \cite{CW}. We defer the details of the comparison to a future paper.

\paragraph{\textbf{Thom isomorphism and spin$^{c}$ structures.}} Let us suppose that $W_{3}(E) = 0$, but $w_{2}(E) \neq 0$. In this case we cannot choose $\epsilon = 1$, but we should recover the ordinary Thom isomorphism anyway, trough a spin$^{c}$-structure of $E$ or $E \oplus 1$. Let us show how. Again we summarize some basic facts about spin$^{c}$ geometry in order to fix the notation. By definition $\Spin^{c}(2r) := \Spin(2r) \times_{\Z_{2}} \UU(1)$, where the symbol `$\times_{\Z_{2}}$' indicates that $(-s, \lambda) \simeq (s, -\lambda)$. From the cover $\xi \colon \Spin(2r) \to \SO(2r)$, we get the natural two-sheet cover $\xi^{c} \colon \Spin^{c}(2r) \to \SO(2r) \times \UU(1)$, $(s, \lambda) \mapsto (\xi(s), \lambda^{2})$. In order to construct a (local) spin$^{c}$-lift of $E$, we fix a complex Hermitian line bundle $\tilde{\pi} \colon L \to X$ as a part of the initial datum. We get the corresponding unitary frame bundle $\tilde{p} \colon \UU(L) \to X$. For each open set $U_{i}$, we choose a spin lift $p'_{i} \colon \Spin(E_{i}) \to U_{i}$ as above, with the two-sheet cover $\xi_{i} \colon \Spin(E_{i}) \to \SO(E_{i})$. Moreover, we choose a lift $\tilde{p}'_{i} \colon \UU'(L_{i}) \to U_{i}$, where $L_{i} := L\vert_{U_{i}}$, with a two-sheet cover $\tilde{\xi}_{i} \colon \UU'(L_{i}) \to \UU(L_{i})$ compatible with $\tilde{\xi} \colon \UU(1) \to \UU(1)$, $\lambda \mapsto \lambda^{2}$. We get the $\Spin^{c}$-lift $\pi^{c}_{i} \colon \Spin^{c}(E_{i}) := \Spin(E_{i}) \times_{\Z_{2}, U_{i}} \UU'(L_{i}) \to U_{i}$, with the two-sheet cover $\xi^{c}_{i} := \xi_{i} \times \tilde{\xi}_{i} \colon \Spin^{c}(E_{i}) \to \SO(E_{i}) \times_{U_{i}} \UU(L_{i})$. From the representation $\rho \colon \Spin(2r) \to \UU(2^{r})$, acting on $S := \C^{2^{r}}$, we easily get the representation $\rho^{c} \colon \Spin^{c}(2r) \to \UU(2^{r})$, $\rho^{c}(s, \lambda)(v) := \lambda\rho(s)(v)$. The chirality splitting $S = S_{+} \oplus S_{-}$ is preserved, hence we get the associated vector bundle $S^{c}(E_{i}) := \Spin^{c}(E_{i}) \times_{\rho^{c}} S$ or rank $2^{r}$, called bundle of complex spinors, with the chirality splitting $S^{c}(E_{i}) = S^{c}_{+}(E_{i}) \oplus S^{c}_{-}(E_{i})$.

As before, we fix principal bundle isomorphisms $\varphi'_{ij} \colon \Spin(E_{i})\vert_{U_{ij}} \to \Spin(E_{j})\vert_{U_{ij}}$, lifting the identity $\SO(E_{i})\vert_{U_{ij}} = \SO(E_{j})\vert_{U_{ij}}$. It follows that $\varphi'_{ki} \circ \varphi'_{jk} \circ \varphi'_{ij} = \epsilon_{ijk}$. Moreover, we fix principal bundle isomorphisms $\psi'_{ij} \colon \UU'(L_{i})\vert_{U_{ij}} \to \UU'(L_{j})\vert_{U_{ij}}$, lifting the identity $\UU(L_{i})\vert_{U_{ij}} = \UU(L_{j})\vert_{U_{ij}}$. It follows that $\psi'_{ki} \circ \psi'_{jk} \circ \psi'_{ij} = \theta_{ijk} = \pm 1$. We get the principal bundle isomorphisms $\varphi^{c}_{ij} := \varphi'_{ij} \times \psi'_{ij} \colon \Spin^{c}(E_{i})\vert_{U_{ij}} \to \Spin^{c}(E_{j})\vert_{U_{ij}}$, lifting the identity $(\SO(E_{i}) \times \UU(L_{i}))\vert_{U_{ij}} = (\SO(E_{j}) \times \UU(L_{j}))\vert_{U_{ij}}$. It follows that $\varphi^{c}_{ki} \circ \varphi^{c}_{jk} \circ \varphi^{c}_{ij} = \epsilon_{ijk}\theta_{ijk}$. We can construct a global bundle $\Spin^{c}(E)$ if and only if it is possible to choose these data in such a way that $\epsilon_{ijk}\theta_{ijk} = 1$, that is equivalent to $\theta_{ijk} = \epsilon_{ijk}$. Fixing a set of local unitary sections $t_{i} \colon U_{i} \to L\vert_{U_{i}}$, we get the set of transition functions $h_{ij} \colon U_{ij} \to \UU(1)$ such that $t_{i} = h_{ij}t_{j}$. We lift the sections $t_{i}$ to $t'_{i} \colon U_{i} \to \UU'(L_{i})$, so that we get transition functions $h'_{ij} \colon U_{ij} \to \UU(1)$ such that $\psi'_{ij}(t'_{i}) = h'_{ij}t'_{j}$. It follows that $h'_{ki}h'_{jk}h'_{ij} = \theta_{ijk}$, hence there exists a global spin$^{c}$-lift of $E$ if and only if it is possible to find a cochain $\{h'_{ij}\} \in \check{C}^{2}(\U; \underline{\UU}(1))$ such that
\begin{equation}\label{GlobalSpinC}
	h'_{ki}h'_{jk}h'_{ij} = \epsilon_{ijk} \cdot I.
\end{equation}
This is equivalent to the triviality of $[\epsilon]$ as a $\underline{\UU}(1)$-cocyle, that is equivalent to $W_{3}(E) = 0$, since $\check{H}^{2}(X, \underline{\UU}(1)) \simeq H^{3}(X; \Z)$.

Let us suppose that $W_{3}(E) = 0$ and let us show how to recover the Thom isomorphism in ordinary K-theory from \eqref{ThomIso}. Choosing $\zeta = \epsilon$, we get the isomorphism $T \colon K_{\epsilon}^{\bullet}(X) \to K^{\bullet+n}_{\cpt}(E)$. Since $W_{3}(E) = 0$ and since $W_{3}(E)$ is the twisting (integral) class represented by $\epsilon$, it follows that $K_{\epsilon}^{\bullet}(X) \simeq K^{\bullet}(X)$ in a non-canonical way. In order to find an isomorphism of the form \eqref{IsoPhiEta}, we must fix a trivialization of $\epsilon$ in $\underline{\UU}(1)$. If we call $\{h'_{ij}\}$ such a trivialization, we get \eqref{GlobalSpinC}. This means that the choice of a spin$^{c}$ structure is equivalent to the choice of an isomorphism $\Phi_{h'} \colon K_{\epsilon}^{\bullet}(X) \overset{\!\simeq}\longrightarrow K^{\bullet}(X)$. The composition between $\Phi_{h'}^{-1}$ and \eqref{ThomIso}, the latter in the form $T \colon K_{\epsilon}^{\bullet}(X) \to K^{\bullet+n}_{\cpt}(E)$, is the ordinary Thom isomorphism, with respect to the Thom class induced by the chosen spin$^{c}$-structure.\footnote{If we consider the associated bundles $L'_{i} := \UU'(L_{i}) \times_{1} \C$, where `$1$' denotes the fundamental representation of $\UU(1)$, and the isomorphisms $\psi''_{ij} := \psi'_{ij} \times 1 \colon L'_{i}\vert_{U_{ij}} \to L'_{j}\vert_{U_{ij}}$,  we get the $\epsilon$-twisted line bundle $\sqrt{L} := (\{L'_{i}\}, \{\psi''_{ij}\})$, such that $\sqrt{L} \otimes \sqrt{L} = L$. With this language, the isomorphism $\Phi_{h'}$ can be also written in the form $\Phi_{\sqrt{L}} \colon K_{\epsilon}^{\bullet}(X) \overset{\!\simeq}\longrightarrow K^{\bullet}(X)$, $\alpha \mapsto \alpha \otimes \sqrt{L}$.} The choice of $\epsilon$ as a representative of $w_{2}(E)$ is immaterial (even if $\Tor_{(2)}H^{2}(X; \Z)$ does not vanish), since, choosing another representative $\epsilon'$ and a cochain $\upsilon$ such that $\epsilon' = \epsilon \cdot \check{\delta}^{1}\upsilon$, the identifications $\varphi'_{ij} \mapsto \varphi'_{ij}\upsilon_{ij}$ and $\psi'_{ij} \mapsto \psi'_{ij}\upsilon_{ij}$ determine the same isomorphism $\varphi^{c}_{ij}$, hence the two $\Z_{2}$-ambiguities cut, inducing the same ordinary Thom isomorphism.

\subsection{Freed-Lott model of differential twisted K-theory}\label{FLModelTwisted}

We are going to describe differential twisted K-theory as the Grothendieck group of a suitable semi-group, as in the topological setting, following the construction shown in \cite{FL} about ordinary differential K-theory and in \cite{Park} in the twisted case. Moreover, we complete the construction of \cite{Park} to any degree, considering product and integration, and we link it more explicitly to relative Deligne cohomology.

\SkipTocEntry \subsubsection{Preliminary notions}

We briefly review the notions of Chern character and Cheeger-Simons class in the twisted framework, since they will be applied to define differential K-theory in this context.

\paragraph{\textbf{Chern character of a twisted connection.}} We have seen in section \ref{ReviewCoh} that, given a class $\hat{\alpha} \in \hat{H}^{p}(\rho)$, its curvature is a relative form $(\omega, \eta) \in \Omega^{p}(\rho)$, where $\omega$ is the curvature of the class on $X$ and $\eta$ is the global potential representing the pull-back on $Y$. In the partially non-abelian setting, it is natural to define the curvature of $[(\zeta, \Lambda, B, g, A)] \in \hat{H}^{3, r}(\rho)$ as the relative 3-form $(H, rB_{i} - \Tr \, F_{i})$, the first component being the field strength on the space-time $X$ and the second one being the field strength on the world-volume $Y$. Coherently, the class is parallel when the second component vanishes. Actually, as for ordinary connections, we can extract higher-degree gauge-invariant forms as well, starting from the following simple lemma.
\begin{Lemma} Given a Deligne cocycle $(\zeta, \Lambda, B)$ and a $(\zeta, \Lambda)$-twisted bundle with connection $(\{E_{i}\}, \{\varphi_{ij}\}, \{\nabla_{i}\})$, represented by $(g, A)$, we have that:
\begin{equation}\label{CurvatureCC}
	F_{j} - B_{j}I_{r} = g_{ij}^{-1}(F_{i} - B_{i}I_{r})g_{ij},
\end{equation}
where we recall that $F_{i} := dA_{i} + 2\pi i A_{i} \wedge A_{i}$.
\end{Lemma}
The proof can be realized by direct computation. We remark that formula \eqref{CurvatureCC} holds for any $B$ completing the twisting cocycle $(\zeta, \Lambda)$, no matter if the connection is $(\zeta, \Lambda, B)$-twisted or not. It easily follows that $e^{F_{j} - B_{j}I_{r}} = g_{ij}^{-1}e^{F_{i} - B_{i}I_{r}}g_{ij}$, therefore we get the globally-defined even form
\begin{equation}\label{DefChNabla}
	\ch(\nabla) := \Tr \, e^{F_{i} - B_{i}I_{r}}.
\end{equation}
If we set $d_{H} := d + H \wedge$, we have that
\begin{equation}\label{ChHClosed}
	d_{H}\ch(\nabla) = 0,
\end{equation}
i.e.\ $\ch(\nabla)$ is a $d_{H}$-closed even form. It is immediate to verify that $\ch(\nabla \oplus \nabla') = \ch(\nabla) + \ch(\nabla')$ and that $\ch(f^{*}\nabla) = f^{*}\ch(\nabla)$ for any smooth function $f \colon Y \to X$.

Fixing a $(\zeta, \Lambda)$-twisted connection $\nabla$, the Chern character $\ch(\nabla)$ depends on the chosen cocycle $(\zeta, \Lambda, B)$ completing $(\zeta, \Lambda)$. When we need to specify the completion, we use the notation $\ch_{(\zeta, \Lambda, B)}(\nabla)$. Fixing $B$, any other choice is of the form $(\zeta, \Lambda, B + \tilde{B})$, where $\tilde{B}$ is a global 2-form. It immediately follows from definition \eqref{DefChNabla} that
\begin{equation}\label{ChChangesB}
	\ch_{(\zeta, \Lambda, B + \tilde{B})}(\nabla) = e^{-\tilde{B}} \wedge \ch_{(\zeta, \Lambda, B)}(\nabla).
\end{equation}
In the case of a non-integral vector bundle (i.e.\ $\Lambda = 0$), we have the canonical choice $B = 0$, leading to the usual formula $\ch(\nabla) = \Tr \, e^{F_{i}}$. Considering the definition \eqref{ChernClasses} of Chern classes, we get the usual relation $\ch(\nabla) = r + c_{1}(\nabla) + \frac{1}{2}\bigl(c_{1}(\nabla)^{2} - 2c_{2}(\nabla)\bigr) + \cdots$.

Given a closed embedding $\rho \colon Y \hookrightarrow X$, we get the following map:
\begin{equation}\label{ChNonAbelinaCoh}
\begin{split}
	\ch \colon & \hat{H}^{3, r}(\rho) \to \Omega^{\ev}_{H\textnormal{-}\cl}(Y) \\
	& [(\zeta, \Lambda, B, g, A)] \mapsto \Tr \, e^{F_{i} - \rho^{*}B_{i}I_{r}}.
\end{split}
\end{equation}
The image of $[(\zeta, \Lambda, B, g, A)]$ through \eqref{ChNonAbelinaCoh} is the Chern character of the $\rho^{*}(\zeta, \Lambda)$-connection $\nabla$ represented by $(g, A)$, with respect to the cocycle $\rho^{*}(\zeta, \Lambda, B)$.

\paragraph{\textbf{Cheeger-Simons class.}} Let us consider two $(\zeta, \Lambda)$-twisted connections $\nabla = \{\nabla_{i}\}$ and $\nabla' = \{\nabla'_{i}\}$ on the same bundle $E = (\{E_{i}\}, \{\varphi_{ij}\})$ on $X$. Fixing a completion $(\zeta, \Lambda, B)$, so that the Chern character \eqref{DefChNabla} is well-defined, we get the Cheeger-Simons class $\CS(\nabla, \nabla') \in \Omega^{\odd}(X)/\IIm(d_{H})$, defined as follows. On the space $I \times X$ we have the good cover $\pi^{*}\U$ and the twisted bundle $\pi_{I}^{*}E$, where $\pi_{I} \colon I \times X \to X$ is the natural projection. We endow each vector bundle $\pi_{I}^{*}E_{i}$ with the connection $\tilde{\nabla}_{i}$, that interpolates between $\nabla_{i}$ and $\nabla'_{i}$. More precisely, for any sections $s$ of $\pi_{I}^{*}E_{i}$ and $(a, V)$ of the tangent bundle $T(I \times X)$, we set $(\tilde{\nabla}_{i})_{(a,V)} s := (1-t)(\nabla_{V}s) + t(\nabla'_{V}s) + a\partial_{t}s$, where $t$ is the coordinate of $I$. We obtain the twisted connection $\tilde{\nabla} = \{\tilde{\nabla}_{i}\}$ on $\pi_{I}^{*}E$ such that, if $\nabla$ and $\nabla'$ are represented respectively by $(\{g_{ij}\}, \{A_{i}\})$ and $(\{g_{ij}\}, \{A'_{i}\})$ with respect to the local sections $\{s_{i}\}$ of $E$, then $\tilde{\nabla}$ is represented by $(\{\pi_{I}^{*}g_{ij}\}, \{(1-t)(\pi_{I}^{*}A_{i}) + t(\pi_{I}^{*}A'_{i})\})$ with respect to the sections $\{\pi_{I}^{*}s_{i}\}$. We set:
\begin{equation}
	\CS(\nabla, \nabla') := \int_{I} \ch(\tilde{\nabla}) \mod \IIm(d_{H}).
\end{equation}
The following relation holds:
\begin{equation}\label{CSProp0}
	\ch(\nabla) - \ch(\nabla') = d_{H}\CS(\nabla, \nabla').
\end{equation}

\paragraph{\textbf{Chern character in twisted K-theory.}} Equations \eqref{ChHClosed} and \eqref{CSProp0} imply that $\ch(E) \in H^{\ev}_{\dR, H}(X)$ is well-defined, since it does not depend on the choice of the connection. We get the natural transformation:
\begin{equation}\label{ChTwistedK}
\begin{split}
	\ch \colon & K^{0}_{\zeta}(X) \to H^{\ev}_{\dR, H}(X) \\
	& E - F \mapsto \ch(E) - \ch(F),
\end{split}
\end{equation}
where $H$ is the curvature of any Deligne cocycle $(\zeta, \Lambda, B)$. From definition \eqref{TwistedDegreeN} we easily get:
\begin{equation}\label{ChTwistedK1}
\begin{split}
	\ch^{-n} \colon & K^{-n}_{\zeta}(X) \to H^{n \,\textnormal{mod}\, 2}_{\dR, H}(X) \\
	& \alpha \mapsto \int_{\mathbb{T}^{n}} \ch \, \alpha,
\end{split}
\end{equation}
where $\ch\,\alpha$ is computed thinking of $\alpha \in K^{0}_{\zeta}(\mathbb{T}^{n} \times X)$.

Some comments are in order about the choice of the cocycle $(\zeta, \Lambda, B)$. Since $[\zeta]$ is necessarily torsion, then $H$ is exact and $H^{\ev}_{\dR, H}(X) \simeq H^{\ev}_{\dR}(X)$. More precisely, we can always choose a \emph{flat} cocycle $(\zeta, \Lambda, B_{0})$, i.e.\ $H_{0} = dB_{0} = 0$, so that the codomain of \eqref{ChTwistedK} is ordinary even de-Rham cohomology. Any other cocycle completing $(\zeta, \Lambda)$ is of the form $(\zeta, \Lambda, B_{0} + \tilde{B})$, where $\tilde{B}$ is a global 2-form, hence $H = d\tilde{B}$ and we have the isomorphism $H^{\ev}_{\dR}(X) \simeq H^{\ev}_{\dR, d\tilde{B}}(X)$, $[\omega] \mapsto [e^{-\tilde{B}} \wedge \omega]$, coherently with formula \eqref{ChChangesB}. Therefore, if we are free to fix any $B$, we can always choose ordinary de-Rham cohomology in \eqref{ChTwistedK}, but this is not possible if the class $[(\zeta, \Lambda, B)]$ is fixed, therefore $H$ too (e.g.\ formula \eqref{ChNonAbelinaCoh}).

\SkipTocEntry \subsubsection{Twisted differential K-theory}

The following definition enriches a (topological) vector bundle with differential information.
\begin{Def}\label{DefTwistedDiffVB} A \emph{$(\zeta, \Lambda, B)$-twisted differential vector bundle} on $X$ is a triple $(E, \nabla, \omega)$ where:
\begin{itemize}
	\item $E$ is a $\zeta$-twisted vector bundle on $X$;
	\item $\nabla$ is a $(\zeta, \Lambda)$-twisted connection on $E$;
	\item $\omega \in \Omega^{\odd}(X)/\IIm(d_{H})$, where $H$ is the curvature of $(\zeta, \Lambda, B)$.
\end{itemize}
The \emph{direct sum} between differential vector bundles is defined as:
	\[(E, \nabla, \omega) \oplus (E', \nabla', \omega') := (E \oplus E', \nabla \oplus \nabla', \omega + \omega').
\]
An \emph{isomorphism of differential vector bundles} $\Phi \colon (E, \nabla, \omega) \to (E', \nabla', \omega')$ is an isomorphism of twisted vector bundles $\Phi \colon E \to E'$ such that:
\begin{equation}\label{IsomorphismQuadruples}
	\omega - \omega' \in \CS(\nabla, \Phi^{*}\nabla').
\end{equation}
\end{Def}
The isomorphism classes of differential vector bundles form an abelian semi-group, that we call $(\DiffVect_{(\zeta, \Lambda, B)}(X), \oplus)$. Moreover, the following map is well-defined and it is a surjective semi-group homomorphism:
\begin{equation}\label{MapI}
\begin{split}
	I \colon & \DiffVect_{(\zeta, \Lambda, B)}(X) \rightarrow \VB_{\zeta}(X) \\
	&I[(E, \nabla, \omega)] := [E].
\end{split}
\end{equation}
Since $(\DiffVect_{(\zeta, \Lambda, B)}(X), \oplus)$ is an abelian semi-group, it is natural to consider its Grothendieck group.
\begin{Def} The \emph{$(\zeta, \Lambda, B)$-twisted differential K-theory group} of $X$ is the group:
\begin{equation}\label{DiffKGroup}
\hat{K}_{(\zeta, \Lambda, B)}(X) := K(\DiffVect_{(\zeta, \Lambda, B)}(X), \oplus).
\end{equation}
\end{Def}
By definition an element of $\hat{K}(X)$ is a difference $[(E, \nabla, \omega)] - [(E', \nabla', \omega')]$, where $[(E, \nabla, \omega)]$ is the class of $(E, \nabla, \omega)$ up to the stable equivalence relation. It follows that, if $(E, \nabla, \omega)$ is equivalent to $(E', \nabla', \omega')$, then the two twisted bundles $E$ and $E'$ represent the same K-theory class, therefore the following map is well-defined and it is a surjective group homomorphism:
\begin{equation}\label{MapIK}
\begin{split}
	I \colon & \hat{K}_{(\zeta, \Lambda, B)}(X) \to K_{\zeta}(X) \\
	& I[(E, \nabla, \omega)] := [E].
\end{split}
\end{equation}

\paragraph{\textbf{Axioms of (twisted) differential cohomology.}} We are looking for a differential extension of twisted K-theory, therefore we need to define the following natural transformations:\footnote{Since the operator $d_{H}$ is defined only among even or odd forms, when we use the notation $\Omega^{\bullet}(X)$ we assume that only the parity of the degree is meaningful. In the non-twisted setting, this corresponds to considering forms with value in the K-theory of the point, as in the usual framework of differential cohomology.}
\begin{itemize}
	\item $I \colon \hat{K}^{\bullet}_{(\zeta, \Lambda, B)}(X) \to K^{\bullet}_{\zeta}(X)$;
	\item $R \colon \hat{K}^{\bullet}_{(\zeta, \Lambda, B)}(X) \to \Omega^{\bullet}_{H\textnormal{-}\cl}(X)$, called \emph{curvature};
	\item $a \colon \Omega^{\bullet-1}(X)/\IIm(d_{H}) \to \hat{K}^{\bullet}_{(\zeta, \Lambda, B)}(X)$,
\end{itemize}
satisfying the following axioms, analogous to $R1$-$R3$ of \cite[p.\ 4]{BS2}:
\begin{itemize}
	\item[R1.] $R \circ a = d_{H}$;
	\item[R2.] the following diagram is commutative:
		\[\xymatrix{
		\hat{K}^{\bullet}_{(\zeta, \Lambda, B)}(X) \ar[rr]^{I} \ar[d]_{R} & & K^{\bullet}_{\zeta}(X) \ar[d]^{\ch} \\
		\Omega^{\bullet}_{H\textnormal{-}\cl}(X) \ar[rr]^{\dR} & & H^{\bullet}_{\dR, H}(X); \\
	}\]
	\item[R3.] the following sequence is exact:
		\[\xymatrix{
		0 \ar[r] & \Omega^{\bullet-1}_{\ch, H}(X) \ar[r] & \Omega^{\bullet-1}(X) \ar[r]^(.45){a} & \hat{K}^{\bullet}_{(\zeta, \Lambda, B)}(X) \ar[r]^(.58){I} & K^{\bullet}_{\zeta}(X) \ar[r] & 0,
	}\]
	where $\Omega^{\bullet-1}_{\ch, H}(X)$ denotes the forms representing a class belonging to the image of $\ch^{\bullet-1} \colon K^{\bullet-1}_{\zeta}(X) \to H^{\bullet-1}_{\dR, H}(X)$.
\end{itemize}
We fulfil these axioms, only in degree $0$ up to now, through the following definition.
\begin{Def} Given a differential K-theory class $[(E, \nabla, \omega)]$ we define:
\begin{itemize}
	\item the natural transformation $I$ through \eqref{MapIK}, i.e.\ $I[(E, \nabla, \omega)] := [E]$;
	\item the curvature as the form $R[(E, \nabla, \omega)] := \ch(\nabla) - d_{H}\omega$;
	\item $a(\omega) := [(E, \nabla, 0)] - [(E, \nabla, \omega)]$ for any $(\zeta, \Lambda)$-twisted bundle $(E, \nabla)$.
\end{itemize}
\end{Def}

\paragraph{\textbf{Extension to any degree.}} We begin defining $\hat{K}^{-n}_{(\zeta, \Lambda, B)}(X)$ for every $n \geq 0$. In particular, we use a general property of differential cohomology \cite[Lemma 2.16]{FRRB} as the definition. We use the same notations of formula \eqref{TwistedDegreeN} about the torus $\mathbb{T}^{n}$ and the embeddings $i_{1}, \ldots, i_{n}$.
\begin{Def}\label{KMinusN} The group $\hat{K}^{-n}_{(\zeta, \Lambda, B)}(X)$ is the subgroup of $\hat{K}_{(\zeta, \Lambda, B)}(\mathbb{T}^{n} \times X)$ whose elements are the classes $\hat{\alpha}$ such that:
\begin{itemize}
	\item for every $j = 1, \ldots, n$:
	\begin{equation}\label{KnKer}
		\hat{\alpha} \in \Ker(i_{j}^{*} \colon \hat{K}_{(\zeta, \Lambda, B)}(\mathbb{T}^{n} \times X) \to \hat{K}_{(\zeta, \Lambda, B)}(\mathbb{T}^{n-1} \times X));
	\end{equation}
	\item there exists $\rho \in \Omega^{-n}(X)$ such that, calling $\pi_{\mathbb{T}^{n}} \colon \mathbb{T}^{n} \times X \to X$ the projection, we have:
	\begin{equation}\label{KnCurvature}
		R(\hat{\alpha}) = dt_{1} \wedge \ldots \wedge dt_{n} \wedge \pi_{\mathbb{T}^{n}}^{*}\rho.
	\end{equation}
\end{itemize}
\end{Def}
It follows from \eqref{KnKer} and \eqref{TwistedDegreeN} that the following natural transformation is well-defined:
\begin{equation}\label{MapIKn}
\begin{split}
	I^{-n} \colon & \hat{K}^{-n}_{(\zeta, \Lambda, B)}(X) \to K^{-n}_{\zeta}(X) \\
	& I^{-n}([E, \nabla, \omega] - [F, \tilde{\nabla}, \xi]) := [E] - [F].
\end{split}
\end{equation}
Moreover, we can define the curvature of a class in $\hat{K}^{-n}_{(\zeta, \Lambda, B)}(X)$ as the form $\rho$ appearing in \eqref{KnCurvature}, or, equivalently:
\begin{equation}\label{CurvatureKn}
\begin{split}
	R^{-n} \colon & \hat{K}^{-n}_{(\zeta, \Lambda, B)}(X) \rightarrow \Omega^{-n}_{H\textnormal{-}\cl}(X) \\
	& R^{-n}(\hat{\alpha}) := \int_{\mathbb{T}^{n}} R(\hat{\alpha}).
\end{split}
\end{equation}
Finally, we define the natural transformation:
\begin{equation}\label{MapA}
\begin{split}
	a^{-n} \colon & \Omega^{-n-1}(X)/\IIm(d_{H}) \to \hat{K}^{-n}_{(\zeta, \Lambda, B)}(X) \\
	& a^{-n}(\omega) := [(E, \nabla, 0)] - [(E, \nabla, (-1)^{n} dt_{1} \wedge \ldots \wedge dt_{n} \wedge \pi_{\mathbb{T}^{n}}^{*}\omega)]
\end{split}
\end{equation}
for any $(E, \nabla)$. It is quite easy to prove that these definitions fulfil the axioms $R1$-$R3$ in any negative degree.

\paragraph{\textbf{Product.}} We have the following natural product:

\vspace{-10pt}
\begin{small}
\[\begin{split}
	\otimes \colon & \hat{K}_{(\zeta, \Lambda, B)}(X) \times \hat{K}_{(\zeta', \Lambda', B')}(X) \to \hat{K}_{(\zeta, \Lambda, B)+(\zeta', \Lambda', B')}(X) \\
	& (E, \nabla, \omega) \otimes (F, \tilde{\nabla}, \rho) := (E \otimes F, \nabla \otimes \tilde{\nabla}, \omega \wedge (\ch \tilde{\nabla} - d_{H'}\rho) + (\ch \nabla - d_{H}\omega) \wedge \rho + \omega \wedge d_{H'}\rho).
\end{split}\]
\end{small}
\vspace{-5pt}

\noindent In particular, we get a module structure on $\hat{K}_{(\zeta, \Lambda, B)}(X)$ over $\hat{K}(X)$, such that $I$ and $R$ are multiplicative and $a(\omega) \cdot \hat{\alpha} = a(\omega \wedge R(\hat{\alpha}))$, as required by the axioms of multiplicativity in differential cohomology. We can also define in a similar way the exterior product
	\[\boxtimes \colon \hat{K}_{(\zeta, \Lambda, B)}(X) \times \hat{K}_{(\zeta', \Lambda', B')}(Y) \to \hat{K}_{(\zeta, \Lambda, B)+(\zeta', \Lambda', B')}(X \times Y),
\]
applying the pull-backs through the projections $\pi_{X} \colon X \times Y \to X$ and $\pi_{Y} \colon X \times Y \to Y$ to the corresponding terms. Calling $\Delta \colon X \to X \times X$ the diagonal embedding, one has $(E, \nabla, \omega) \otimes (F, \tilde{\nabla}, \rho) = \Delta^{*}((E, \nabla, \omega) \boxtimes (F, \tilde{\nabla}, \rho))$. The exterior product can be extended to any non-positive degree:
\begin{equation}\label{ExtProd}
	\boxtimes \colon \hat{K}^{-n}_{(\zeta, \Lambda, B)}(X) \times \hat{K}^{-m}_{(\zeta', \Lambda', B')}(Y) \to \hat{K}^{-n-m}_{(\zeta, \Lambda, B)+(\zeta', \Lambda', B')}(X \times Y).
\end{equation}
In fact, given $\hat{\alpha} \in \hat{K}^{-n}_{(\zeta, \Lambda, B)}(X) \subset \hat{K}_{(\zeta, \Lambda, B)}(\mathbb{T}^{n} \times X)$ and $\hat{\beta} \in \hat{K}^{-m}_{(\zeta', \Lambda', B')}(X) \subset \hat{K}_{(\zeta', \Lambda', B')}(\mathbb{T}^{m} \times X)$, we consider $\hat{\alpha} \boxtimes_{0} \hat{\beta} \in \hat{K}_{(\zeta, \Lambda, B)+(\zeta', \Lambda', B')}(\mathbb{T}^{n} \times X \times \mathbb{T}^{m} \times Y)$, where $\boxtimes_{0}$ is the exterior product in degree $0$. Then we consider the natural diffeomorphism $\varphi_{n,m} \colon \mathbb{T}^{n+m} \times X \times Y \to \mathbb{T}^{n} \times X \times \mathbb{T}^{m} \times Y$ and we set $\hat{\alpha} \boxtimes \hat{\beta} := (-1)^{nm}\varphi_{n,m}^{*}(\hat{\alpha} \boxtimes_{0} \hat{\beta})$. The factor $(-1)^{nm}$ is necessary to make the first Chern class and the curvature multiplicative.\footnote{For example, considering the curvature, we have $R(\hat{\alpha} \boxtimes_{0} \hat{\beta}) = dt_{1} \wedge \ldots \wedge dt_{n} \wedge \pi_{\mathbb{T}^{n}}^{*}R^{-n}(\hat{\alpha}) \wedge dt_{n+1} \wedge \ldots \wedge dt_{n+m} \wedge \pi_{\mathbb{T}^{m}}^{*}R^{-m}(\hat{\beta}) = (-1)^{nm}dt_{1} \wedge \ldots \wedge dt_{n+m} \wedge \pi_{\mathbb{T}^{n+m}}^{*}(\pi_{X}^{*}R^{-n}(\hat{\alpha}) \wedge \pi_{Y}^{*}R^{-m}(\hat{\beta}))$, therefore we need to multiply by $(-1)^{nm}$ to get $R^{-n-m}(\hat{\alpha} \boxtimes \hat{\beta}) = \pi_{X}^{*}R^{-n}(\hat{\alpha}) \wedge \pi_{Y}^{*}R^{-m}(\hat{\beta})$.} Then, via the diagonal embedding $\Delta \colon \mathbb{T}^{n+m} \times X \to \mathbb{T}^{n+m} \times X \times X$, we get the interior product
\begin{equation}\label{IntProd}
	\otimes \colon \hat{K}^{-n}_{(\zeta, \Lambda, B)}(X) \times \hat{K}^{-m}_{(\zeta', \Lambda', B')}(X) \to \hat{K}^{-n-m}_{(\zeta, \Lambda, B)+(\zeta', \Lambda', B')}(X).
\end{equation}
The reader can verify by direct computation that $a^{-n}(\omega) \cdot \hat{\alpha} = a^{-n-m}(\omega \wedge R^{-m}(\hat{\alpha}))$.

\paragraph{\textbf{$S^{1}$-integration.}} We have to define the map:
\begin{equation}\label{IntS10}
	\int_{S^{1}} \colon \hat{K}^{-n}_{(\zeta, \Lambda, B)}(S^{1} \times X) \to \hat{K}^{-n-1}_{(\zeta, \Lambda, B)}(X)
\end{equation}
satisfying the following axioms:
\begin{itemize}
	\item[I1.] Calling $t \colon S^{1} \to S^{1}$ the conjugation $e^{i\theta} \mapsto e^{-i\theta}$ and considering the map $t \times \id \colon S^{1} \times X \to S^{1} \times X$, we have $\int_{S^{1}} (t \times \id)^{*}\hat{\alpha} = -\int_{S^{1}} \hat{\alpha}$.
	\item[I2.] $\int_{S^{1}} \circ \pi_{1}^{*} = 0$, where $\pi_{1} \colon S^{1} \times X \to X$ is the natural projection.
	\item[I3.] The integration map \eqref{IntS10} commutes with $I$, $R$ and $a$.
\end{itemize}
Considering the projection $\pi_{1} \colon S^{1} \times X \to X$ and the embedding $i_{1} \colon X \to S^{1} \times X$, the latter being defined fixing a marked point on $S^{1}$, topologically we have the canonical isomorphism:
\begin{equation}\label{SplittingS1}
\begin{split}
	& K^{-n}_{\zeta}(S^{1} \times X) \simeq K_{\zeta}^{-n-1}(X) \oplus K^{-n}_{\zeta}(X) \\
	& \alpha \simeq (\alpha - \pi_{1}^{*}i_{1}^{*}\alpha) \oplus i_{1}^{*}\alpha,
\end{split}
\end{equation}
the inverse morphism being $\alpha \oplus \beta \mapsto \alpha + \pi_{1}^{*}\beta$. Topological $S^{1}$-integration corresponds to the projection to the first term of \eqref{SplittingS1}. Given any differential class $\hat{\alpha} \in \hat{K}^{-n}_{(\zeta, \Lambda, B)}(S^{1} \times X)$, we get from \eqref{SplittingS1} that $I^{-n}(\hat{\alpha}) = \beta + \pi_{1}^{*}\gamma$, where $\beta \in K^{-n-1}_{\zeta}(S^{1} \times X)$ and $\gamma \in K^{-n}_{\zeta}(X)$, in a unique way. Refining $\beta$ and $\gamma$ to any differential classes $\hat{\beta} \in \hat{K}^{-n-1}_{(\zeta, \Lambda, B)}(S^{1} \times X)$ and $\hat{\gamma} \in \hat{K}^{-n}_{(\zeta, \Lambda, B)}(X)$, we have that:
\begin{equation}\label{DecompAlphaS1}
	\hat{\alpha} = \hat{\beta} + \pi_{1}^{*}\hat{\gamma} + a^{-n}(\rho)
\end{equation}
for a suitable form $\rho \in \Omega^{-n}(S^{1} \times X)$. We define the integral of $\hat{\beta}$ as $\hat{\beta}$ itself, i.e.\ the $S^{1}$-integration restricts to the identity on $\hat{K}^{-n-1}_{(\zeta, \Lambda, B)}(X)$. The integral of $\pi_{1}^{*}\hat{\gamma}$ vanishes because of axiom $I2$ and the integral of $a^{-n}(\rho)$ is determined by axiom $I3$, in particular by the commutativity between $a$ and $S^{1}$-integration. Therefore, starting from the decomposition \eqref{DecompAlphaS1}, we set:
\begin{equation}\label{IntS1Def}
	\int_{S^{1}} \hat{\alpha} := \hat{\beta} + a^{-n-1}\biggl(\int_{S^{1}}\rho\biggr).
\end{equation}
Such a definition does not depend on the decomposition \eqref{DecompAlphaS1} and it fulfils axioms $I1$-$I3$.

\paragraph{\textbf{Bott periodicity and positive degrees.}} Bott periodicity can be extended to the differential framework as follows. Let us consider the generator $\eta - 1 \in \tilde{K}(\mathbb{T}^{2}) \simeq \mathbb{Z}$ leading to topological Bott periodicity. There exists a unique differential refinement $\hat{\eta}-1$ such that $R(\hat{\eta} - 1) = dt_{1} \wedge dt_{2}$ and $i_{1}^{*}(\hat{\eta} - 1) = i_{2}^{*}(\hat{\eta} - 1) = 0$, where $i_{1}, i_{2} \colon \mathbb{T} \to \mathbb{T}^{2}$ are the natural embeddings: such a class is the unique element of $\hat{K}^{-2}(pt)$ with curvature $1$ and first Chern class $1$. Then the map
\begin{equation}\label{DiffBott}
\begin{split}
	\hat{B}^{-n} \colon & \hat{K}^{-n}_{(\zeta, \Lambda, B)}(X) \overset{\!\simeq}\longrightarrow \hat{K}^{-n-2}_{(\zeta, \Lambda, B)}(X) \\
	& \hat{\alpha} \mapsto (\hat{\eta} - 1) \boxtimes \hat{\alpha}
\end{split}
\end{equation}
is an isomorphism too. Thus, for any $n > 0$, we set $\hat{K}^{n}_{(\zeta, \Lambda, B)}(X) := \hat{K}^{-n}_{(\zeta, \Lambda, B)}(X)$ and we easily extend to positive degrees the functors $I$, $R$ and $a$, together with product and $S^{1}$-integration.

\paragraph{\textbf{Parallel and compactly-supported versions.}} We just need the definition of \emph{parallel} relative classes, since it will be used in order to define the compactly-supported version. It is defined exactly as in the topological case (see section \ref{RelativeVersion}), replacing the set $\Ll_{\zeta}(X, Y)$ by the set $\hat{\Ll}_{(\zeta, \Lambda, B)}(X, Y)$. The latter is formed by triples $\bigl( (E, \nabla_{E}, \omega_{E}), (F, \nabla_{F}, \omega_{F}), \alpha \bigr)$, where $(E, \nabla_{E}, \omega_{E})$ and $(F, \nabla_{F}, \omega_{F})$ are $(\zeta, \Lambda)$-twisted vector bundles on $X$, such that $\omega_{E}$ and $\omega_{F}$ vanish on $Y$, and $\alpha \colon E\vert_{Y} \to F\vert_{Y}$ is an isomorphism such that $\alpha^{*}\nabla_{F} = \nabla_{E}$. We define in an analogous way the notions of \emph{isomorphism of triples} and \emph{elementary triple}, getting the corresponding set $\hat{K}_{(\zeta, \Lambda, B)}(X, Y)$. With this definition, we can construct the compactly-supported version exactly as in topological framework (see section \ref{CptSupp}), since parallel classes satisfy excision as well \cite[Lemma 2.15]{FRRB}. We get the group $\hat{K}_{(\zeta, \Lambda, B), \cpt}(X)$.

\SkipTocEntry \subsubsection{Dependence on the cocycle}\label{DepCocycleFL}

We have defined $\hat{K}^{\bullet}_{(\zeta, \Lambda, B)}(X)$ as a twisted differential cohomology theory, fixing the cocycle $(\zeta, \Lambda, B)$. If we fix a cohomologous cocycle $(\zeta', \Lambda', B')$ and a cochain $(\eta, \lambda)$, such that $(\zeta', \Lambda', B') = (\zeta, \Lambda, B) + \check{D}^{1}(\eta, \lambda)$ in $S^{2}_{X}$, then we get the equivalence of differential cohomology theories
\begin{equation}\label{PsiKDiff}
	\Psi_{(\eta, \lambda)} \colon \hat{K}^{\bullet}_{(\zeta, \Lambda, B)}(X) \to \hat{K}^{\bullet}_{(\zeta', \Lambda', B')}(X),
\end{equation}
defined as follows. It is enough to define the image of $[(E, \nabla, \omega)] \in \hat{K}^{\bullet}_{(\zeta, \Lambda, B)}(X)$, since the image of a formal difference follows by additivity. We have in particular that $(\zeta', \Lambda') = (\zeta, \Lambda) + \check{D}^{1}(\eta, \lambda)$ in $S^{1}_{X}$, thus we apply the isomorphism \eqref{IsoPhiEtaDiffVB} to $(E, \nabla)$, fixing $\omega$. This means that \eqref{PsiKDiff} is defined by $[(E = (\{E_{i}\}, \{\varphi_{ij}\}), \nabla = \{\nabla_{i}\}, \omega)] \mapsto [(E' := (\{E_{i}\}, \{\varphi_{ij}\eta_{ij}\}), \nabla' = \{\nabla_{i} + \lambda_{i} \cdot I\}, \omega)]$. This isomorphism does not change the curvature, since the image of the field strength $F_{i}$ of $\nabla$ is $F'_{i} := F_{i} + d\lambda_{i} \cdot I_{r} = F_{i} + (B_{i} - B'_{i}) \cdot I_{r}$, therefore $F'_{i} + B'_{i} \cdot I_{r} = F_{i} + B_{i} \cdot I_{r}$. Moreover, it induces the isomorphism \eqref{IsoPhiEtaBdl} through the underlying vector bundles. The set of isomorphisms of the form \eqref{PsiKDiff} is a torsor over the flat part of $\check{H}^{1}(S^{1}_{X})$, i.e.\ over $H^{1}(X; \R/\Z) \simeq \Hom(H_{1}(X; \Z); \R/\Z)$. It follows that, if $H_{1}(X; \Z) = 0$, then $\hat{K}^{\bullet}_{[(\zeta, \Lambda, B)]}(X)$ is canonically defined, with a canonical curvature. Similar considerations hold about parallel and compactly-supported versions. We will see how to define the transformation $I$ canonically in section \ref{TopTwistedDeligne}.

\paragraph{\textbf{Dependence on $B$.}} We remark that, given two cocycles of the form $(\zeta, \Lambda, B)$ and $(\zeta, \Lambda, B')$, we have a canonical isomorphism between the corresponding differential K-theory groups, even if they are not cohomologous (in particular, even if $H \neq H'$). In fact, necessarily $B' = B + \tilde{B}$, with $\tilde{B}$ a global 2-form on $X$, hence we get:
\begin{equation}\label{PsiKDiff2}
\begin{split}
	\Psi_{\tilde{B}} \colon & \hat{K}^{\bullet}_{(\zeta, \Lambda, B)}(X) \to \hat{K}^{\bullet}_{(\zeta, \Lambda, B + \tilde{B})}(X) \\
	& [(E, \nabla, \omega)] \mapsto [(E, \nabla, \omega \wedge e^{-\tilde{B}})].
\end{split}
\end{equation}
In this case the underlying vector bundle remains unchanged, but the curvature changes by a factor $e^{-\tilde{B}}$. In fact, applying formula \eqref{ChChangesB}:
	\[\ch_{(\zeta, \Lambda, B)}(\nabla) - d_{H}\omega \mapsto \ch_{(\zeta, \Lambda, B')}(\nabla) - d_{H'}(\omega \wedge e^{-\tilde{B}}) = (\ch_{(\zeta, \Lambda, B)}(\nabla) - d_{H}\omega) \wedge e^{-\tilde{B}}.
\]
It follows from \eqref{PsiKDiff2} that we can define $\hat{K}^{\bullet}_{(\zeta, \Lambda)}(X)$ canonically for a fixed cocycle $(\zeta, \Lambda) \in \check{H}^{2}(S^{1}_{X})$. When $[(\zeta, \Lambda)] = [(\zeta', \Lambda')]$, a cochain $(\eta, \lambda)$, trivializing the difference, induces an isomorphism analogous to \eqref{PsiKDiff}, and the set of such isomorphisms is a torsor over $\hat{H}^{2}(X)$. When $[(\zeta, \Lambda, B)] = [(\zeta, \Lambda, B')]$, i.e.\ when $\tilde{B}$ has integral periods, we can choose $(\eta, \lambda)$ such that $\check{D}^{1}(\eta, \lambda) = (1, 0, \tilde{B})$, but $\Psi_{(\eta, \lambda)} \neq \Psi_{\tilde{B}}$, as it is clear from the behaviour of the curvature. We do not analyse in detail the isomorphism \eqref{PsiKDiff2}, since it is not relevant for our purposes.

\subsection{Generic twisting class}\label{DiffTwistedKGen}

Up to now we considered a twisting class of finite order. Considering the $B$-field, this finite-order hypothesis corresponds to the exactness of the $H$-flux. This is always true on a D-brane world-volume, because of the Freed-Witten anomaly, but not on the whole space-time. That's why we need to briefly review the general case, in which some infinite-dimensional tools are necessary, and explicitly relate it to the language that we have used up to now, analysing in particular the dependence on the cocycle. We choose the model described in \cite{CMW}. In a future paper, it would be interesting to compare it with the more recent model in \cite{GL}, that could also be used as a reference here.

\SkipTocEntry \subsubsection{Topological twisted K-theory}\label{TwistedGen}

We fix a separable Hilbert space $\HH$. We can easily generalize definition \ref{DefTwistedBdl} as follows.
\begin{Def}\label{DefTwistedHilbBdl} Given a cocycle $\zeta := \{\zeta_{ijk}\} \in \check{Z}^{2}(\U, \underline{\UU}(1))$, a \emph{$\zeta$-twisted Hilbert bundle} with fibre $\HH$ on $X$ is a collection of trivial Hilbert bundles $\pi_{i} \colon E_{i} \to U_{i}$ with fibre $\HH$ and of Hilbert bundle isomorphisms $\varphi_{ij} \colon E_{i}\vert_{U_{ij}} \to E_{j}\vert_{U_{ij}}$, such that $\varphi_{ki} \varphi_{jk} \varphi_{ij} = \zeta_{ijk} \cdot \id$.
\end{Def}
The corresponding definition of (iso)morphism coincides with \ref{MorphTwistedVB}. For every $\zeta \in \check{Z}^{2}(\U, \underline{\UU}(1))$, not necessarily of finite order in cohomology, there exists a $\zeta$-twisted Hilbert bundle \cite{AS}, the main difference with respect to the finite-dimensional setting being that any two $\zeta$-twisted Hilbert bundles (for a fixed $\zeta$) are isomorphic \cite{Karoubi}.

\paragraph{\textbf{Projective Hilbert bundles.}} Given a twisted bundle $E = (\{E_{i}\}, \{\varphi_{ij}\})$, projecting each fibre $(E_{i})_{x} \setminus \{0\}$ to the corresponding projective space, we get a well-defined (non-twisted) projective bundle, that we denote by $\PPP(E)$. It follows from local triviality that every projective bundle can be obtained in this way up to isomorphism, therefore we get a surjective map from isomorphism classes of twisted bundles to isomorphism classes of projective bundles. In the finite-dimensional case such a map is not injective for a fixed $\zeta$ (for example, every line bundle projects to the trivial one). On the contrary, in the infinite-dimensional case, the unique isomorphism class of $\zeta$-twisted Hilbert bundles induces a unique isomorphism class of projective bundles. Moreover, fixing $\zeta$ and $\zeta' := \zeta \cdot \check{\delta}^{1}\eta$, let us consider the bijection
\begin{equation}
\begin{split}
	\Phi_{\eta} \colon & \widetilde{\VB}_{\zeta}(X) \overset{\!\simeq}\longrightarrow \widetilde{\VB}_{\zeta'}(X) \\
	& E = (\{E_{i}\}, \{\varphi_{ij}\}) \mapsto \Phi_{\eta}(E) := (\{E_{i}\}, \{\varphi_{ij}\eta_{ij}\}),
\end{split}
\end{equation}
where $\widetilde{\VB}_{\zeta}(X)$ denotes the set of $\zeta$-twisted Hilbert bundles on $X$ (not quotiented out up to isomorphism). Since $\PPP(E) = \PPP(\Phi_{\eta}(E))$, the isomorphism class of $\PPP(E)$ only depends on $[\zeta] \in \check{H}^{2}(\U, \underline{\UU}(1)) \simeq H^{3}(X; \Z)$ (see \cite{AS}). It follows that $H^{3}(X; \Z)$ classifies projective Hilbert bundles on $X$.

 If $\check{\delta}^{1}{\eta} = 1$, then, since any two $\zeta$-twisted bundles are isomorphic, there exists an isomorphism $f = \{f_{i}\} \colon E \to \Phi_{\eta}(E)$. This means that $f_{i} \colon E_{i} \to E_{i}$ and $\varphi_{ij}\eta_{ij}f_{i} = f_{i}\varphi_{ij}$, hence $f$ induces an automorphism $\bar{f} \colon \PPP(E) \to \PPP(E)$. Let us see that any automorphism $\bar{f}$ can be realized in this way from suitable $\eta$ and $f$. In fact, by local triviality, we can lift $\bar{f}$ to $f_{i} \colon E_{i} \to E_{i}$ for each $i$. Since the family $\{f_{i}\}$ glues to $\bar{f}$, there exists $\eta_{ij}$ such $f_{j}\varphi_{ij} = \varphi_{ij}f_{i}\eta_{ij}$. The latter condition necessarily implies $\check{\delta}^{1}\eta = 1$. Moreover, the only freedom we had in constructing the cocycle $\eta$ was the choice of the lifts $f_{i}$. Any other choice is of the form $f_{i}\xi_{i}$, that replaces $\eta$ by $\eta \cdot \check{\delta}^{0}\xi$. Therefore, the following map is well-defined:
\begin{equation}\label{PhiAutH2}
\begin{split}
	\Phi \colon & \Aut(\PPP(E)) \to H^{2}(X, \Z) \\
	& \bar{f} \mapsto [\{\eta_{ij}\}].
\end{split}
\end{equation}
It is easy to prove that it is a group homomorphism. Moreover, it follows from the previous construction that $\bar{f} \in \Aut(\PPP(E))$ lifts to an automorphisms of $E$ if and only if $\Phi(\bar{f}) = 0$, therefore $\Phi(\bar{f})$ can be thought of as the obstruction to the existence such a lift. This remark leads quite easily to the following lemma, that also follows from the fact that $\PU(\HH)$ is an Eilenberg-MacLane space $K(\Z, 2)$ \cite{AS}.
\begin{Lemma}\label{LemmaPhiAutH2} The morphism \eqref{PhiAutH2} is surjective. Moreover, its kernel is the connected component of the identity of $\Aut(\PPP(E))$, therefore $\Phi$ induces a canonical bijection between the connected components of $\Aut(\PPP(E))$ and $H^{2}(X, \Z)$.
\end{Lemma}

\paragraph{\textbf{Definition of twisted K-theory.}} We fix a cocycle $\zeta \in \check{Z}^{2}(\U, \underline{\UU}(1))$ and a $\zeta$-twisted Hilbert bundle $E = (\{E_{i}\}, \{\varphi_{ij}\})$, inducing the corresponding projective bundle $\PPP(E)$. We denote by $P_{\PPP(E)}$ the bundle of projective reference frames of $\PPP(E)$ and by $\Fred(\HH)$ the space of Fredholm operators acting on $\HH$. We have a natural adjoint action of $\PU(\HH)$ on $\Fred(\HH)$ by conjugation, that we denote by $\rho \colon \PU(\HH) \to C^{0}(\Fred(\HH))$, hence we construct the associated $\Fred(\HH)$-bundle $F_{\PPP(E)} := P_{\PPP(E)} \times_{\rho} \Fred(\HH)$. We denote by $\Gamma(F_{\PPP(E)})$ its set of global sections and by $\bar{\Gamma}(F_{\PPP(E)})$ the corresponding quotient with respect to homotopy of sections. The latter carries a natural abelian group structure, induced by composition of Fredholm operators.

\begin{Def}\label{DefKuX} The \emph{twisted K-theory group} $K_{\zeta}(X)$ is defined as the abelian group $\bar{\Gamma}(F_{\PPP(E)})$ for any $\zeta$-twisted Hilbert bundle $E$.
\end{Def}

Since the space of bounded invertible operators in $\HH$ is contractible (like $\UU(\HH)$), a section of $F_{\PPP(E)}$, which is point-wise invertible, is always homotopic to the identity. Therefore, if a section is point-wise invertible in a subset of $X$, we consider it trivial on such a subset. This fact justifies the following definition.

\begin{Def}\label{DefKuXCpt} A section of $F_{\PPP(E)}$ is called \emph{compactly supported} if it is point-wise invertible in the complement of a compact subset of $X$. We denote by $\Gamma_{\cpt}(F_{\PPP(E)})$ and $\bar{\Gamma}_{\cpt}(F_{\PPP(E)})$ respectively the space of compactly-supported sections of $F_{\PPP(E)}$ and its quotient up to compactly-supported homotopy. We define the \emph{compactly supported twisted K-theory group} $K_{\zeta, \cpt}(X)$ as the abelian group $\bar{\Gamma}_{\cpt}(F_{\PPP(E)})$ for any $\zeta$-twisted Hilbert bundle $E$.\footnote{When $X$ is compact, definitions \ref{DefKuX} and \ref{DefKuXCpt} are equivalent. Actually, we will never apply definition \ref{DefKuX} when $X$ is not compact, hence it would be sufficient to state definition \ref{DefKuXCpt} for every (locally compact) space.}
\end{Def}

\paragraph{\textbf{Dependence on the cocycle.}} Definitions \ref{DefKuX} and \ref{DefKuXCpt} seem to depend on $E$, not only on $\zeta$. Nevertheless, fixing two $\zeta$-twisted bundles $E$ and $E'$, an isomorphism $f \colon E \to E'$ is unique up to an automorphism of $E$. It follows from lemma \ref{LemmaPhiAutH2} that the induced isomorphism $\bar{f} \colon \PPP(E) \to \PPP(E')$ is unique up to an automorphism of $\PPP(E)$ connected to the identity, the latter inducing the identity on $\bar{\Gamma}(F_{\PPP(E)})$ and $\bar{\Gamma}_{\cpt}(F_{\PPP(E)})$. Hence, $K_{\zeta}(X)$ and $K_{\zeta, \cpt}(X)$ are canonically defined.\footnote{This is the advantage of starting from a twisted bundle $E$ instead of directly considering a projective Hilbert bundle. Otherwise, we would have an intrinsic ambiguity even fixing the cocycle $\zeta$.} On the contrary, the definition is not canonical if we only fix the cohomology class $[\zeta]$. In fact, let us consider a $\zeta$-twisted bundle $E$ and a $\zeta'$-twisted bundle $E'$, such that $\zeta' = \zeta \cdot \check{\delta}^{1}\eta$. We have the isomorphism
\begin{equation}\label{IsoTwistedKGeneral}
	\Phi_{\eta} \colon K_{\zeta}(X) \overset{\!\simeq}\longrightarrow K_{\zeta'}(X),
\end{equation}
analogous to the one induced by \eqref{IsoPhiEta} in the finite-order setting, defined as follows. We fix an isomorphism $\bar{f} \colon \PPP(E) \to \PPP(E')$, belonging to the inverse image of $[\eta]$ through \eqref{PhiAutH2}, and we apply the induced one between the corresponding K-theory groups. This is equivalent to inducing the identity between $\bar{\Gamma}(F_{\PPP(E)})$ and $\bar{\Gamma}(F_{\PPP(\Phi_{\eta}(E))})$, that represent respectively $K_{\zeta}(X)$ and $K_{\zeta'}(X)$.

The isomorphism \eqref{IsoTwistedKGeneral} depends on $\eta$ up to coboundaries. Equivalently, the set of isomorphisms of the form \eqref{IsoTwistedKGeneral} is a torsor over $H^{2}(X; \Z)$, hence, if $\zeta = \zeta'$, we get an action of $H^{2}(X; \Z)$ on $K_{\zeta}(X)$. Only the quotient up to such an action is well-defined. Of course, if $H^{2}(X; \Z) = 0$, then we have the canonical group $K_{[\zeta]}(X)$, as in the finite-order setting. Analogous considerations hold about compactly-supported K-theory.

\SkipTocEntry \subsubsection{Differential extension}

We follow \cite{CMW} to define differential twisted K-theory with no hypotheses on the torsion class. We keep on denoting by $\HH$ a separable infinite-dimensional Hilbert space. Moreover, given two such Hilbert spaces $\HH_{1}$ and $\HH_{2}$, for any $p \in [1, +\infty)$ we denote by $\LL^{p}(\HH_{1}, \HH_{2})$ the corresponding $p$-Schatten class, i.e.\ the space of boundend linear operators $A \colon \HH_{1} \to \HH_{2}$ such that
	\[\norm{A}_{\LL^{p}} := \left( \Tr\bigl((A^{\dagger}A)^{\frac{p}{2}}\bigr) \right)^{\frac{1}{p}} < \infty.
\]
We set $\LL^{p}(\HH) := \LL^{p}(\HH, \HH)$.

\paragraph{\textbf{Schatten Grassmannians.}} We set $\hat{\HH} := \HH \oplus \HH$ and we denote by $\HH_{+}$ and $\HH_{-}$ respectively the first and the second component of the direct sum. It follows that $\hat{\HH}$ is $\Z_{2}$-graded, the corresponding involution being $\epsilon \colon \hat{\HH} \to \hat{\HH}$ such that $\HH_{\pm}$ is the $\pm 1$-eigenspace. Given a closed subspace $V \subset \hat{\HH}$, we get the corresponding decomposition $\hat{\HH} = V \oplus V^{\bot}$ and the corresponding self-adjoint involution $\epsilon_{V}$. We define the space $\Gr^{p}(\hat{\HH}, \epsilon)$ as the subspace of the Grassmannian of $\hat{\HH}$ such that $V \in \Gr^{p}(\hat{\HH}, \epsilon)$ if and only if $\epsilon_{V} = \epsilon + A$, with $A \in \LL^{p}(\hat{\HH})$. The space $\Gr^{p}(\hat{\HH}, \epsilon)$, identified with a subspace of bounded linear operators on $\hat{\HH}$ through $\epsilon_{V}$, is a Banach manifold and it is smoothly homotopically equivalent to $\Fred(\HH)$. In particular, its homotopy type does not depend on $p$.

In order to construct an explicit homotopy equivalence, we consider the group $\GL_{p}(\hat{\HH}, \epsilon)$, formed by bounded invertible operators on $\hat{\HH}$ of the form:
	\[A = \begin{bmatrix} a & b \\ c & d \end{bmatrix}, \quad b \in \LL^{2p}(\HH_{-}, \HH_{+}), \; c \in \LL^{2p}(\HH_{+}, \HH_{-}).
\]
It follows from this definition that $a \in \Fred(\HH_{+})$. From now on we denote $\Gr^{p}(\hat{\HH}, \epsilon)$ and $\GL_{p}(\hat{\HH}, \epsilon)$ respectively by $\Gr^{p}$ and $\GL_{p}$. We have a natural transitive action $\GL_{p} \times \Gr^{p} \to \Gr^{p}$, $(A, V) \mapsto A(V)$. In particular, a closed subspace $V \subset \hat{\HH}$ belongs to $\Gr^{p}$ if and only if there exists $A \in \GL_{p}$ such that $A(\HH_{+}) = V$. We have the following natural maps, that turn out to be homotopy equivalences:
\begin{equation}\label{HomEquivH}
\begin{matrix}
	\Fred(\HH) & \overset{\psi}\longleftarrow & \GL_{p} & \overset{\varphi}\longrightarrow & \Gr^{p} \\
	a & \mapsfrom & A = \begin{bmatrix} a & b \\ c & d \end{bmatrix} & \mapsto & A(\HH_{+}) = \{(ax, cx): x \in \HH_{+}\}.
\end{matrix}
\end{equation}
The group $\PU(\HH)$ acts by conjugation on each of the previous spaces. We denote such actions respectively by $\rho \colon \PU(\HH) \to C^{0}(\Fred(\HH))$, $\rho'' \colon \PU(\HH) \to C^{0}(\GL_{p})$ and $\rho' \colon \PU(\HH) \to C^{0}(\Gr^{p})$. In $\rho'$ and $\rho''$ we are applying the diagonal embedding $\PU(\HH) \hookrightarrow \PU(\hat{\HH})$. We get the three bundles
	\[F_{\PPP(E)} := P_{\PPP(E)} \times_{\rho} \Fred(\HH) \qquad F''_{\PPP(E)} := P_{\PPP(E)} \times_{\rho''} \GL_{p} \qquad F'_{\PPP(E)} := P_{\PPP(E)} \times_{\rho'} \Gr^{p}.
\]
It is straightforward to verify that the homotopy equivalences $\varphi$ and $\psi$ in \eqref{HomEquivH} commute with the actions of $\PU(\HH)$, therefore we get the induced maps:
\begin{equation}\label{HomEquivBdles}
\begin{matrix}
	F_{\PPP(E)} & \overset{\psi_{*}}\longleftarrow & F''_{\PPP(E)} & \overset{\varphi_{*}}\longrightarrow & F'_{\PPP(E)} \\
	[p, \psi(A)] & \mapsfrom & [p, A] & \mapsto & [p, \varphi(A)].
\end{matrix}
\end{equation}
As above, we denote by $\Gamma(F_{\PPP(E)})$ the space of sections of $F_{\PPP(E)}$ and by $\bar{\Gamma}(F_{\PPP(E)})$ the quotient up to homotopy. We use the same notation for $F'_{\PPP(E)}$ and $F''_{\PPP(E)}$. By definition $K_{\zeta}(X) := \bar{\Gamma}(F_{\PPP(E)})$ and we set $K'_{\zeta}(X) := \bar{\Gamma}(F'_{\PPP(E)})$. The maps \eqref{HomEquivBdles} induce the corresponding ones between sections, that are isomorphisms up to homotopy, since the maps \eqref{HomEquivH} are homotopy equivalences. We get the following canonical isomorphism:
\begin{equation}\label{IsoTheta}
	\theta_{\zeta} := \varphi_{*} \circ \psi_{*}^{-1} \colon K_{\zeta}(X) \overset{\!\simeq}\longrightarrow K'_{\zeta}(X).
\end{equation}
The group $K'_{\zeta}(X)$ turns out to be more suitable than $K_{\zeta}(X)$ in order to define the corresponding differential extension.

\paragraph{\textbf{Natural connection on $\UU(\HH)$.}} Since $\PU(\HH) := \UU(\HH)/\UU(1)$, the projection $\pi \colon \UU(\HH) \to \PU(\HH)$ naturally induces a principal $\UU(1)$-bundle structure, which can be endowed with the universal connection for line bundles $\theta \colon T\UU(\HH) \to \R$. We recall some basic properties of $\theta$, that we are going to use in the following sections. First of all, given two functions $f, g \colon Z \to \UU(\HH)$, for any smooth manifold $Z$, we have that
\begin{equation}\label{ThetaFG}
	(fg)^{*}\theta = f^{*}\theta + g^{*}\theta,
\end{equation}
where $fg$ denotes the point-wise product. Moreover, if $\zeta \colon Z \to \UU(1)$, then, thinking of $\UU(1) \subset \UU(\HH)$ as the centre, we have:
\begin{equation}\label{ThetaCenter}
	\zeta^{*}\theta = \textstyle \frac{1}{2\pi i} \, \zeta^{-1}d\zeta.
\end{equation}
Such a formula easily follows from the fact that $\theta$, restricted to the centre, coincides with the Maurier-Cartan 1-form of $\UU(1)$. Moreover, for any 1-form $\Lambda \colon TZ \to \R$, there exists $h \colon Z \to \UU(\HH)$ such that
\begin{equation}\label{ThetaAnyLambda}
	h^{*}\theta = \Lambda.
\end{equation}
In fact, since $\theta$ is the universal connection for line bundles and since $\Lambda$ represents any connection $\nabla_{\Lambda}$ on $Z \times \C$, we can find a homotopically-trivial function $\bar{h} \colon Z \to \PU(\HH)$, covered by $h' \colon \bar{h}^{*}\UU(\HH) \to \UU(\HH)$, such that the connection $\nabla_{\Lambda}$ corresponds to $(h')^{*}\theta$. Choosing a trivialization $s \colon Z \to \bar{h}^{*}\UU(\HH)$, inducing $\Lambda$ as the potential representing $\nabla_{\Lambda}$, we set $h := h' \circ s$ and we get that $h^{*}\theta = \Lambda$.

\paragraph{\textbf{Realizing a Deligne cocycle.}} Given a $\zeta$-twisted Hilbert bundle $E = (\{E_{i}\}, \{\varphi_{ij}\})$, let us fix a system of local sections $\{s_{i}^{n} \colon U_{i} \to E_{i}\}_{n \in \N}$, that locally trivializes $E$ (i.e.\ that trivializes each $E_{i}$). Such sections determine a Deligne cocycle $(\zeta, \Lambda) \in \check{Z}^{2}(S^{1}_{X})$ as follows. The first component is determined by $E$ itself, that is $\zeta$-twisted by construction. Moreover, the fixed sections determine the corresponding transition functions $g_{ij} \colon U_{ij} \to \UU(\HH)$, therefore, from the universal connection $\theta$ on $\UU(\HH)$, we set $\Lambda_{ij} := g_{ij}^{*}\theta$. It follows from \eqref{ThetaFG} and \eqref{ThetaCenter} that $(\zeta, \Lambda)$ is a cocycle.

Let us show that any cocycle $(\zeta, \Lambda)$ can be reached in this way. In fact, we already know that for any $\zeta$ there exists a $\zeta$-twisted bundle $E$. Inducing any $(\zeta, \Lambda_{0})$ through sections $\{s_{i}^{n}\}$, for any fixed $\Lambda$ there exists a cochain $(1, \lambda)$ such that $(\zeta, \Lambda) = (\zeta, \Lambda_{0}) \cdot \check{D}^{1}(1, \lambda) = (\{\zeta_{ijk}\}, \{\Lambda_{0\, ij} - \lambda_{i} + \lambda_{j}\})$. We can replace the sections $\{s_{i}^{n}\}$, inducing transition functions $g_{ij}$, through any change of basis $h_{i} \colon U_{i} \to \UU(\HH)$, so that we get the transition functions $g'_{ij} := h_{i}g_{ij}h_{j}^{-1}$. We choose $h_{i}$ such that $h_{i}^{*}\theta = -\lambda_{i}$, that is always possible, as we have shown in the remarks after formula \eqref{ThetaAnyLambda}. It follows from \eqref{ThetaFG} that $(g'_{ij})^{*}\theta = \Lambda_{0\, ij} - \lambda_{i} + \lambda_{j} = \Lambda_{ij}$, as desired.

\paragraph{\textbf{Differential twisted K-theory.}} The spaces $\Gr^{p}$ and $\Fred(\HH)$ are Banach manifolds, smoothly homotopically equivalent among each other. In particular, they are all homotopic to the Hilbert manifold $\Gr^{2}$, therefore their de-Rham cohomology is well-defined using the complex of smooth differential forms, as in the finite-dimensional setting, and there is a canonical isomorphism from the de-Rham cohomology with complex coefficients to singular cohomology with complex coefficients. On $\Gr^{p}$, using Quillen's superconnections, we can fix smooth even-degree differential forms $\Phi_{2n} \in \Omega^{2n}(\Gr^{p})$ such that $\Phi_{\ev} := \sum_{n=0}^{\infty} \Phi_{2n}$ represents the Chern character of the canonical K-theory class, the latter being the class represented by the identity \cite[Section 2.2]{CMW}.

Let us fix a $\zeta$-twisted Hilbert bundle $E = (\{E_{i}\}, \{\varphi_{ij}\})$ and a section $\psi \in \Gamma(F'_{\PPP(E)})$. Given a Deligne cocycle $(\zeta, \Lambda, B)$, we fix a system of local sections $\{s_{i}^{n} \colon U_{i} \to E_{i}\}_{n \in \N}$, inducing the cocycle $(\zeta, \Lambda)$ as we have seen above. Such sections identify $\psi$ with a family of local functions $\psi_{i} \colon U_{i} \to \Gr^{p}$, therefore we get the local pull-backs $\psi_{i}^{*}\Phi_{\ev} \in \Omega^{\ev}(U_{i})$. This implies that the local sections determine at the same time the local forms $\psi_{i}^{*}\Phi_{\ev}$ and the local potentials $B_{i}$ up to a global form $\tilde{B}$, therefore it is not surprising that these two data glue to the global form $e^{B_{i}} \wedge \psi_{i}^{*}\Phi_{\ev}$, that is $d_{H}$-closed. Let us show more in detail why. If a function $h_{i} \colon U_{i} \to \UU(\HH)$ acts on $\psi_{i}$ by conjugation (i.e.\ if we apply the representation $\rho'$ defined above), then the behaviour of $\psi_{i}^{*}\Phi_{\ev}$ is the following one:
\begin{equation}\label{BehaviourPhiEvPB}
	\bigl(\rho'(h_{i})(\psi_{i})\bigr)^{*}\Phi_{\ev} = \psi_{i}^{*}\Phi_{\ev} \wedge e^{-d(h_{i}^{*}\theta)}.
\end{equation}
It follows that, conjugating $\psi_{i}$ by the transition function $g_{ij}$ (i.e.\ acting with $\rho'(g_{ij})$), we have that $\psi_{i}^{*}\Phi_{\ev} \mapsto \psi_{i}^{*}\Phi_{\ev} \wedge e^{-d\Lambda_{ij}}$. Moreover, $e^{B_{i}} \mapsto e^{B_{i}} \wedge e^{d\Lambda_{ij}}$ because of the Deligne cocycle condition $B_{j} - B_{i} = d\Lambda_{ij}$, hence $e^{B_{i}} \wedge \psi_{i}^{*}\Phi_{\ev}$ does not change. This justifies the following definition.

\begin{Def}\label{DefTwistedKT} Given a Deligne 2-cocycle $(\zeta, \Lambda, B)$ on $X$, with curvature $H$, we fix a $\zeta$-twisted bundle $E = (\{E_{i}\}, \{\varphi_{ij}\})$ and a set of local trivializations $\{s_{i}^{n}\}$ inducing $\Lambda$ as above. We define the group $\hat{K}_{(\zeta, \Lambda, B)}(X)$ as follows.  An element of this group is a homotopy class of pairs $(\psi, \eta)$, where:
\begin{itemize}
	\item $\psi$ is a smooth section of $F'_{\PPP(E)}$;
	\item $\eta \in \Omega^{\odd}(X)/\IIm(d_{H})$;
	\item a homotopy between $(\psi, \eta)$ and $(\psi', \eta')$ is a homotopy of sections $\Psi \colon \psi \sim \psi'$ such that
	\begin{equation}\label{DefHomotopyTwistedKT}
		\eta' - \eta \sim_{d_{H}} \int_{I} e^{B_{i}} \wedge \Psi_{i}^{*}\Phi_{\ev},
	\end{equation}
	where `$\sim_{d_{H}}$' denotes that they are equal up to a $d_{H}$-exact form.\footnote{Of course we are identifying $B_{i}$ and $H$ with $\pi^{*}B_{i}$ and $\pi^{*}H$, where $\pi \colon X \times I \to X$ is the natural projection. Moreover, $\Psi_{i}$ comes from the obvious local trivialization of $\pi^{*}E$, induced by the one of $E$.}
\end{itemize}
\end{Def}
We get the following functors:
\begin{itemize}
	\item $I \colon \hat{K}_{(\zeta, \Lambda, B)}(X) \to K_{\zeta}(X)$, $[\psi, \eta] \mapsto \theta_{\zeta}^{-1}[\psi]$, where $[\psi] \in K'_{\zeta}(X)$ and $\theta_{\zeta}$ is the isomorphism \eqref{IsoTheta};
	\item $R \colon \hat{K}_{(\zeta, \Lambda, B)}(X) \to \Omega^{\ev}(X)$, $[\psi, \eta] \mapsto e^{B_{i}} \wedge \psi_{i}^{*}\Phi_{\ev} - d_{H}\eta$;
	\item $a \colon \Omega^{\odd}(X)/\IIm(d_{H}) \to \hat{K}_{(\zeta, \Lambda, B)}(X)$, $\eta \mapsto [\psi, 0] - [\psi, \eta]$ for any section $\psi$.
\end{itemize}
The extension to any negative degree is realized through definition \ref{KMinusN} in this context as well, and, because of the Bott periodicity, we set $\hat{K}^{n}_{(\zeta, \Lambda, B)}(X) := \hat{K}^{-n}_{(\zeta, \Lambda, B)}(X)$ for any $n > 0$. The reader can verify that axioms R1--R3 are satisfied. Product and $S^{1}$-integration are defined similarly to the finite-order framework.

\SkipTocEntry \subsubsection{Dependence on the cocycle}\label{DepCocycleGTwC}

Let us analyse the dependence of $\hat{K}_{(\zeta, \Lambda, B)}(X)$ on $(\zeta, \Lambda, B)$ in steps. We will get the same picture of section \ref{DepCocycleFL}.

\paragraph{\textbf{Dependence on the cocycle -- Part I.}} Definition \ref{DefTwistedKT} seems to depend on the triple $(E, s, B)$, but we can show that it canonically depends only on $(\zeta, \Lambda, B)$ as follows. We call $\hat{K}_{(E, s, B)}(X)$ the group defined in \ref{DefTwistedKT}, assuming that it depends on $E$ and $s$. Moreover, we call $\Gamma(F'_{\PPP(E)})$ the set of smooth sections of $F'_{\PPP(E)}$ (we recall that $F'_{\PPP(E)} := P_{\PPP(E)} \times_{\rho'} \Gr^{p}$) and we set $\Gamma'(F'_{\PPP(E)}) := \Gamma(F'_{\PPP(E)}) \times (\Omega^{\odd}(X)/\IIm \, d_{H})$. Using the notation of definition \ref{DefTwistedKT}, it follows that $(\psi, \eta) \in \Gamma'(F'_{\PPP(E)})$, hence $\hat{K}_{(E, s, B)}(X)$ is the quotient of $\Gamma'(F'_{\PPP(E)})$ up to homotopy. We observe that, fixing $(\zeta, \Lambda, B)$, the space $\Gamma'(F'_{\PPP(E)})$ only depends on $E$, while the notion of homotopy is determined by $s$, since $s$ induces the local forms $\Psi_{i}$ in equation \eqref{DefHomotopyTwistedKT}. Therefore, we write:
\begin{equation}\label{KEsBQuotient}
	\hat{K}_{(E, s, B)}(X) = \Gamma'(F'_{\PPP(E)}) /\! \sim_{s},
\end{equation}
where `$\sim_{s}$' is the equivalence relation induced by the existence of a homotopy. Now we have to analyse the dependence on $E$ and on $s$.

\texttt{Dependence on $E$.} Let us consider two triples $(E, s, B)$ and $(E', s', B)$, both inducing $(\zeta, \Lambda, B)$. If (and only if) the transition functions induced by $s$ and $s'$ coincide, then there exists an isomorphism $f \colon E \to E'$, inducing $\bar{f} \colon \PPP(E) \to \PPP(E')$ and $\bar{\bar{f}} \colon F'_{\PPP(E)} \to F'_{\PPP(E')}$, such that $s' = f_{*}s$. In this case such isomorphism is unique, since the condition $s' = f_{*}s$ completely determines $f$. We get the isomorphism:
\begin{equation}\label{IsoInducedF}
\begin{split}
	f_{\#} \colon & \hat{K}_{(E, s, B)}(X) \to \hat{K}_{(E', s', B)}(X) \\
	& [(\psi, \eta)] \mapsto [(\bar{\bar{f}} \circ \psi, \eta)].
\end{split}
\end{equation}
The reader can verify that it is well-defined, since, if $(\psi_{1}, \eta_{1}) \sim_{s} (\psi_{2}, \eta_{2})$ through the homotopy $\Psi$, then $(\bar{\bar{f}} \circ \psi_{1}, \eta_{1}) \sim_{s'} (\bar{\bar{f}} \circ \psi_{2}, \eta_{2})$ through the homotopy $\bar{\bar{f}} \circ \Psi$.

The isomorphism $f_{\#}$, induced by the unique isomorphism $f$, would be enough for our purposes if the condition $g_{ij}^{*}\theta = \Lambda_{ij}$ completely determined the transition functions $g_{ij}$, but this is not the case in general. Nevertheless, at least it shows that, fixing $(\zeta, \Lambda, B)$, we are free to choose any $\zeta$-twisted bundle $E$. In fact, given any triple $(E', s', B)$ and fixing $E$, we can choose an isomorphism $f \colon E \to E'$ and set $s := f^{*}s'$, so that $\hat{K}_{(E', s', B)}(X)$ is canonically isomorphic to $\hat{K}_{(E, s, B)}(X)$ through $f_{\#}$.

\texttt{Dependence on $s$.} Given two triples of the form $(E, s, B)$ and $(E, s', B)$, both inducing $(\zeta, \Lambda, B)$, we fix a base change $\{h_{i} \colon U_{i} \to \UU(\HH)\}$ from $s$ to $s'$ and we set $\alpha_{i} := h_{i}^{*}\theta$. Calling $g$ and $g'$ the transition functions induced by $s$ and $s'$ respectively, since $g_{ij}^{*}\theta = {g'}_{ij}^{*}\theta = \Lambda_{ij}$ and $g'_{ij} = h_{i}g_{ij}h_{j}^{-1}$, it follows from \eqref{ThetaFG} that $h_{i}^{*}\theta = h_{j}^{*}\theta$, hence the local forms $\alpha_{i}$ glue to a global $1$-form $\alpha$ on $X$. We get the following isomorphism:
\begin{equation}\label{IsoInducedH}
\begin{split}
	\Xi_{s, s'} \colon & \hat{K}_{(E, s, B)}(X) \to \hat{K}_{(E, s', B)}(X) \\
	& [(\psi, \eta)] \mapsto [(\psi, \eta \wedge e^{-d\alpha})].
\end{split}
\end{equation}
The reader can verify that, if $(\psi_{1}, \eta_{1}) \sim_{s} (\psi_{2}, \eta_{2})$ through the homotopy $\Psi$, then $(\psi_{1}, \eta_{1} \wedge e^{d\alpha}) \sim_{s'} (\psi_{2}, \eta_{2} \wedge e^{-d\alpha})$ through the same homotopy $\Psi$, hence $\Xi_{s, s'}$ is well-defined.\footnote{The same isomorphism would hold fixing $s$ and replacing $B$ by $B' := B - d\alpha$.}

\texttt{Canonicity.} We did not show the canonical dependence on $(\zeta, \Lambda, B)$ yet, since the isomorphisms \eqref{IsoInducedF} and \eqref{IsoInducedH} do not agree when they are both defined. In fact, given $(E, s, B)$ and $(E, s', B)$, such that $g = g'$, the isomorphism $f_{\#}$, induced by the unique automorphism $f$ of $E$ such that $f_{*}s = s'$, is different from $\Xi_{s, s'}$ in general. For this reason, we argue as follows. We have seen in the previous paragraph that, from $s$ and $s'$, we get the global 1-form $\alpha$, that here we denote by $\alpha_{s, s'}$. If $d\alpha_{s, s'} = 0$, then it follows from formula \eqref{BehaviourPhiEvPB} that the equivalence relations `$\sim_{s}$' and `$\sim_{s'}$' in formula \eqref{KEsBQuotient} coincide (since the induced forms $\Psi_{i}$, appearing in the definition of homotopy, are the same), thus $\hat{K}_{(E, s, B)}(X) = \hat{K}_{(E, s', B)}(X)$. Coherently, it easily follows from definition \eqref{IsoInducedH} that $\Xi_{s, s'}$ is the identity in this case. For this reason, we introduce the following equivalence relation in the set of local trivializations of $E$ inducing $\Lambda$: we set $s \sim s'$ if and only $\alpha_{s, s'} = 0$ (up to now it would be enough to require $d\alpha_{s, s'} = 0$, but we will need $\alpha$ itself vanishing). It follows that $\hat{K}_{(E, [s], B)}(X)$ is well-defined.

Given an automorphism $f \colon E \to E$, we set $\alpha_{f} := \alpha_{s, f_{*}s}$ for any local trivialization $s$. We introduce the following equivalence relation in $\Aut(E)$: we set $f \sim f'$ if and only if $\alpha_{f} = \alpha_{f'}$ (equivalently, $f \sim \id$ if and only if $\alpha_{f} = 0$). Coherently, given two $\zeta$-twisted bundles $E$ and $E'$ and two isomorphisms $f, f' \colon E \to E'$, we set $f \sim f'$ if and only if $(f')^{-1} \circ f \sim \id$. Let us show that the isomorphism $f_{\#}$, defined in \eqref{IsoInducedF}, only depends on the equivalence class $[f]$. Equivalently, we have to show that, if $f \colon E \to E$ is equivalent to the identity, then $f_{\#}$ is the identity. In fact, $f_{\#}[(\psi, \eta)] = [(\bar{\bar{f}} \circ \psi, \eta)]$ and, since $\bar{f}$ lifts to $f$, there exists a homotopy $\bar{F} \colon \bar{f} \sim \id_{\PPP(E)}$, inducing $\bar{\bar{F}} \colon \bar{\bar{f}} \sim \id_{F'_{\PPP(E)}}$. We have that
\begin{equation}\label{IsoFHomot1}
	\textstyle f_{\#}[(\psi, \eta)] = [(\bar{\bar{f}} \circ \psi, \eta)] = \bigl[\bigl(\psi, \eta + \int_{I} \exp(B_{i}) \wedge (\bar{\bar{F}} \circ \psi)_{i}^{*}\Phi_{\ev} \bigr) \bigr],
\end{equation}
the last equality being due to formula \eqref{DefHomotopyTwistedKT}. The functions $\psi_{i}$ are defined from $F'_{\PPP(E)} := \PPP(E) \times_{\rho} \Gr^{p}$ as $\psi = [s_{i}, \psi_{i}]$, hence
	\[\bar{\bar{F}} \circ \psi = [\bar{F}_{*}s_{i}, \psi_{i}] \overset{(\star)}= [H_{i}(s_{i}), \psi_{i}] = [s_{i}, \rho'(H_{i})(\psi_{i})],
\]
where, in the equality $(\star)$, we called $H_{i} \colon U_{i} \to \UU(\HH)$ the change of basis from $s_{i}$ to $\bar{F}_{*}s_{i}$. It follows that $(\bar{\bar{F}} \circ \psi)_{i} = \rho'(H_{i})(\psi_{i})$, thus
	\[(\bar{\bar{F}} \circ \psi)_{i}^{*}\Phi_{\ev} = \bigl(\rho'(H_{i})(\psi_{i})\bigr)^{*}\Phi_{\ev} \overset{(\star\star)}= \psi_{i}^{*}\Phi_{\ev} \wedge e^{-dA},
\]
where, in the equality $(\star\star)$, we applied formula \eqref{BehaviourPhiEvPB}, the local forms $H_{i}^{*}\theta$ glueing to the global 1-form $A$. Hence, from formula \eqref{IsoFHomot1} we get:
\begin{equation}\label{IsoFHomot12}
	\textstyle f_{\#}[(\psi, \eta)] = \bigl[\bigl(\psi, \eta + \int_{I} \exp(B_{i}) \wedge \psi_{i}^{*}\Phi_{\ev} \wedge e^{-dA} \bigr)\bigr].
\end{equation}
We have that
\begin{align*}
	\textstyle e^{-dA} & \textstyle = 1 + \sum_{n=1}^{+\infty} \frac{(-dA)^{\wedge n}}{n!} = 1 + \sum_{n=1}^{+\infty} \frac{(-dA) \wedge (-dA)^{n-1}}{n!} \\
	& \textstyle = 1 - d \Bigl( A \wedge \sum_{n=1}^{+\infty} \frac{(-dA)^{\wedge(n-1)}}{n!} \Bigr) = 1 - d(A \wedge A'),
\end{align*}
where we set $A' := \sum_{n=1}^{+\infty} \frac{(-dA)^{\wedge(n-1)}}{n!}$. Hence, the integral in formula \eqref{IsoFHomot12} becomes:
\begin{align*}
	\int_{I} \exp(B_{i}) \wedge \psi_{i}^{*}\Phi_{\ev} \wedge e^{-dA} & = \exp(B_{i}) \wedge \psi_{i}^{*}\Phi_{\ev} \wedge \int_{I} \bigl( 1 - d(A \wedge A') \bigr) \\
	& = -\exp(B_{i}) \wedge \psi_{i}^{*}\Phi_{\ev} \wedge \int_{I} d(A \wedge A').
\end{align*}
From Stoke's formula we have:
	\[\int_{I} d(A \wedge A') = d \int_{I} A \wedge A' + \int_{\partial I} A \wedge A'.
\]
Moreover, from the relation $d_{H}(\omega \wedge \xi) = (d_{H}\omega) \wedge \xi + (-1)^{\abs{\omega}}\omega \wedge d\xi$ we get
	\[\exp(B_{i}) \wedge \psi_{i}^{*}\Phi_{\ev} \wedge d \int_{I} A \wedge A' = d_{H} \Bigl( \exp(B_{i}) \wedge \psi_{i}^{*}\Phi_{\ev} \wedge \int_{I} A \wedge A' \Bigr)
\]
and a $d_{H}$-exact term can be cut by definition \ref{DefTwistedKT} (since $\eta \in \Omega^{\odd}(X)/\IIm(d_{H})$), hence only the integral on $\partial I$ is meaningful. Since $A$ is the 1-form induced by the change of basis from $s$ to $F_{*}s$, where $F$ restricts to the identity on $X \times \{1\}$ and to $f$ on $X \times \{0\}$, it follows that $A$ restricts to $0$ on $X \times \{1\}$ and to the global $1$-form $\alpha := \alpha_{s, f_{*}s}$ on $X \times \{0\}$, hence $\int_{\partial I} A \wedge A' = -\alpha \wedge \alpha'$. Therefore, formula \eqref{IsoFHomot12} becomes:
\begin{equation}\label{IsoFHomot13}
	\textstyle f_{\#}[(\psi, \eta)] = \bigl[\bigl(\psi, \eta - \exp(B_{i}) \wedge \psi_{i}^{*}\Phi_{\ev} \wedge \alpha \wedge \alpha' \bigr)\bigr].
\end{equation}
Formula \eqref{IsoFHomot13} holds for \emph{any} automorphism $f \colon E \to E$, but now we can apply the hypothesis $f \sim \id$, i.e.\ $\alpha = 0$. In this case we immediately get $f_{\#}[(\psi, \eta)] = [(\psi, \eta)]$, i.e.\ $f_{\#}$ is the identity, as required.

It follows that, given an equivalence class of isomorphisms $[f] \colon E \to E'$, such that $[f]_{*}[s] = [s']$, the isomorphism $f_{\#} \colon \hat{K}_{(E, s, B)}(X) \to \hat{K}_{(E', s', B)}(X)$ is well-defined. Now only one step is missing: we must prove that, fixing $(E, [s], B)$ and $(E', [s'], B)$, there exits a unique class $[f] \colon E \to E'$ such that $[f]_{*}[s] = [s']$. Uniqueness is straightforward: if $f, g \colon E \to E'$ are such that $f_{*}s \sim g_{*}s$, then $g = (gf^{-1})f$, where $(gf^{-1})_{*}s' \sim s'$, i.e.\ $gf^{-1} \sim 1$, thus $f \sim g$. Existence is a consequence of the following property:
\begin{quote}
	($*$) For any 1-form $\alpha$ on $X$ and any $\zeta$-twisted bundle $E$, there exits an automorphism $f \colon E \to E$ such that $\alpha_{f} = \alpha$. 
\end{quote}
In fact, given $(E, s, B)$ and $(E', s', B)$, we choose any isomorphism $f' \colon E \to E'$ and, applying $(*)$, we choose an automorphism $g' \colon E' \to E'$ such that $\alpha_{g'} = -\alpha_{f'_{*}s, s'}$. The isomorphism $f := g' \circ f'$ satisfies $f_{*}s \sim s'$.

In order to prove $(*)$, we start from the following property of the universal connection $\theta$. Let us fix a $\zeta$-twisted bundle $E$, a $\zeta'$-twisted bundle $E'$, an automorphism $f \colon E \to E$ and an automorphism $f' \colon E' \to E'$. The automorphism $f \otimes f' \colon E \otimes E' \to E \otimes E'$ satisfies $\alpha_{f \otimes f'} = \alpha_{f} + \alpha_{f'}$. Let us call $\HH$ the trivial (non-twisted) bundle. For any $\zeta$-twisted bundle $E$, we have that $E \simeq E \otimes \HH$. An automorphism of $\HH$ is a global function $f \colon X \to \UU(\HH)$ and, since $\theta$ is a universal connection, for any 1-form $\alpha$ on $X$ there exists $f$ such that $f^{*}\theta = \alpha$. It follows that $\alpha_{\id \otimes f} = \alpha$ as required.

This completes the proof that $\hat{K}_{(\zeta, \Lambda, B)}(X)$ is canonically defined. Summarizing, given $(E, s, B)$ and $(E', s', B)$, we have a canonical isomorphism $\hat{K}_{(E, s, B)}(X) \simeq \hat{K}_{(E', s', B)}(X)$ defined as follows: we fix any isomorphism $f \colon E \to E'$, such that $f_{*}s \sim s'$ (equivalently, $[f]_{*}[s] = [s']$) and we apply $f_{\#}$, the latter being independent of the representative $f$.

\paragraph{\textbf{Dependence on the cocycle -- Part II.}} Now we can show that $\hat{K}_{(\zeta, \Lambda, B)}(X)$ only depends on the cohomology class $[(\zeta, \Lambda, B)]$ up to non-canonical isomorphism. We fix a cochain $(\xi, \lambda)$ and we set $(\zeta', \Lambda', B') := (\zeta, \Lambda, B) + \check{D}^{1}(\xi, \lambda)$. We get the isomorphism
\begin{equation}\label{IsoXiLambdaInfinite}
	\Phi_{(\xi, \lambda)} \colon \hat{K}_{(\zeta, \Lambda, B)}(X) \to \hat{K}_{(\zeta', \Lambda', B')}(X)
\end{equation}
defined as follows. We consider separately cochains of the form $(\xi, 0)$ and $(1, \lambda)$.

\texttt{Action of $(\xi, 0)$.} We fix a representative $\hat{K}_{(E, s, B)}(X)$ of $\hat{K}_{(\zeta, \Lambda, B)}(X)$. We set $E' := \Phi_{\xi}(E)$, i.e., if $E = (\{E_{i}\}, \{\varphi_{ij}\})$, then $E' := (\{E_{i}\}, \{\varphi_{ij}\xi_{ij}\})$. It follows that $\PPP(E) = \PPP(E')$, hence $\hat{K}_{(E, s, B)}(X) = \hat{K}_{(E', s, B)}(X)$, therefore we define $\Phi_{(\xi, 0)}$ as the identity between these two representatives. We remark that the transition functions determined by $s$ in the two cases are not the same. In particular, we have that $g'_{ij} = g_{ij}\xi_{ij}$, hence $(g'_{ij})^{*}\theta = g_{ij}^{*}\theta + \xi_{ij}^{*}\theta = \Lambda_{ij} + \tilde{d}\xi_{ij} = \Lambda'_{ij}$, as desired. 

\texttt{Action of $(1, \lambda)$.} We fix a representative $\hat{K}_{(E, [s], B)}(X)$ of $\hat{K}_{(\zeta, \Lambda, B)}(X)$. We fix any change of basis $h_{i} \colon U_{i} \to \UU(\HH)$ such that $h_{i}^{*}\theta = \lambda_{i}$ and we set $s'_{i} := h_{i}(s_{i})$. We have that $\hat{K}_{(E, s, B)}(X) = \hat{K}_{(E, s', B')}(X)$, therefore we define $\Phi_{(1, \lambda)}$ as the identity between these two representatives. In fact, considering formula \eqref{KEsBQuotient}, we have that $\Gamma'(F'_{\PPP(E)})$ is the same in the two cases, since it only depends on $E$ and $H$. It remains to show that $\sim_{s}$ and $\sim_{s'}$ coincide. In fact, given a section $\psi = [\bar{s}_{i}, \psi_{i}]$ with respect to $(E, s, B)$, we have that $\psi = [\bar{h}_{i}(\bar{s}_{i}), \psi'_{i}] = [\bar{s}_{i}, \rho(\bar{h}_{i})(\psi'_{i})]$ with respect to $(E, s', B')$, thus $\psi_{i} = \rho(\bar{h}_{i})(\psi'_{i})$. It follows that the expression $\psi_{i}^{*}\Phi_{\ev} \wedge e^{B_{i}}$ becomes
\begin{align*}
	{\psi'}_{i}^{*}\Phi_{\ev} \wedge e^{B'_{i}} &\,= \bigl(\rho(\bar{h}^{-1}_{i})(\psi_{i})\bigr)^{*}\Phi_{\ev} \wedge e^{B_{i} + d\lambda_{i}} \\
	&\overset{\eqref{BehaviourPhiEvPB}}= \psi_{i}^{*}\Phi_{\ev} \wedge e^{-d\lambda_{i}} \wedge e^{B_{i}} \wedge e^{d\lambda_{i}} = {\psi}_{i}^{*}\Phi_{\ev} \wedge e^{B_{i}},
\end{align*}
hence $\sim_{s}$ and $\sim_{s'}$ coincide. This also shows that \eqref{IsoXiLambdaInfinite} does not change the curvature.

\texttt{Coboundaries and canonicity.} Let us show that, if $(\xi, \lambda) = \check{D}^{0}(\upsilon)$, then \eqref{IsoXiLambdaInfinite} is the identity. In fact, we have that $(\xi, \lambda) = (\check{\delta}^{0}\upsilon, \tilde{d}\upsilon)$. Starting from $\hat{K}_{(E, s, B)}(X)$ and applying $\Phi_{(\check{\delta}^{0}\upsilon, 0)}$ we get the identity from $\hat{K}_{(E, s, B)}(X)$ to $\hat{K}_{(\Phi_{\check{\delta}^{0}\upsilon}(E), s, B)}(X)$. Then, we choose $h_{i} = \upsilon_{i}$, so that $h_{i}^{*}\theta = \tilde{d}\upsilon$, hence, applying $\Phi_{(1, \tilde{d}\upsilon)}$, we get the identity to $\hat{K}_{(\Phi_{\check{\delta}^{0}\upsilon}(E), \upsilon \cdot s, B)}(X)$. Let us show that $\id \colon \hat{K}_{(E, s, B)}(X) \to \hat{K}_{(\Phi_{\check{\delta}^{0}\upsilon}(E), \upsilon \cdot s, B)}(X)$ is exactly the canonical isomorphism considered in the definition of $\hat{K}_{(\zeta, \Lambda, B)}(X)$. In fact, we fix $f \colon E \to E' := \Phi_{\check{\delta}^{0}\upsilon}(E)$, defined by $f_{i} \colon E_{i} \to E_{i}$, $v \mapsto \upsilon_{i}v$. By definition, $[(\psi, \eta)] \mapsto [(\bar{\bar{f}}_{*}\psi, \eta)] = [(\psi, \eta)]$, since $\bar{f} = \id_{\PPP(E)}$. Moreover, in the domain we quotient up to $s$, while in the codomain we quotient up to $f_{*}s = \upsilon s$, coherently with $\Phi_{(\check{\delta}^{0}\upsilon, \tilde{d}\upsilon)}$. It follows that the set of isomorphisms of the form \eqref{IsoXiLambdaInfinite} is a torsor over $\check{H}^{1}(X; \R/\Z)$. In particular, if $H_{1}(X; \Z) = 0$, then $\hat{K}_{[(\zeta, \Lambda, B)]}(X)$ is canonically defined.

\paragraph{\textbf{Dependence on the cocycle -- Part III.}} If we replace $B$ by $B + \tilde{B}$, where $\tilde{B}$ is a global form (even changing the cohomology class), then we have the isomorphism $[(\psi, \eta)] \mapsto [(\psi, \eta \wedge e^{\tilde{B}})]$. This shows that we can define canonically $\hat{K}_{(\zeta, \Lambda)}(X)$ as in the finite-order setting. When $\tilde{B}$ is an integral form and $(1, 0, \tilde{B}) = \check{D}^{1}(\xi, \lambda)$, this isomorphism does not coincide with \eqref{IsoXiLambdaInfinite}, therefore it is not relevant for our purposes.

\paragraph{\textbf{Compact support.}} All of the constructions shown in this section can be easily extended to the compactly-supported framework. In particular, we define a compactly-supported representative as a pair $(\psi, \eta)$ such that there exists a compact subset $K \subset X$ such that $\psi_{i} = \epsilon$ for every $U_{i} \subset X \setminus K$ and any local sections $s_{i}$. Since $\epsilon$ is invariant by conjugation, this condition does not depend on $s_{i}$. A homotopy between two such pairs must be compactly-supported too. In this way we define $\hat{K}_{(\zeta, \Lambda, B), \cpt}(X)$, that only depends on $[(\zeta, \Lambda, B)]$ up to isomorphism, the latter being non-canonical unless $H_{1}(X; \Z) = 0$.

\SkipTocEntry \subsubsection{Equivalence for torsion twisting}\label{EqTorsTwSec}

We sketch how to show that, when the twisting cocycle represents a finite-order class, the models through Fredholm operators or Schatten Grasmannians agree with the model through twisted vector bundles. We defer the details to a future paper.

\paragraph{\textbf{Topological framework.}} We assume that $X$ is compact and $\U$ is finite. Moreover, we assume that each element of $\U$ has compact closure. We fix a cocycle $\zeta \in \check{Z}^{2}(\U, \underline{\UU}(1))$, that represents a finite-order cohomology class, and a $\zeta$-twisted rank-$N$ vector bundle $\bar{E} := (\{U_{i} \times \C^{N}\}, \{g_{ij}\})$, for any suitable $N \in \N$.\footnote{We can identify $\varphi_{ij}$ with the transition function $g_{ij}$, since we have chosen each bundle $\bar{E}_{i} = U_{i} \times \C^{N}$ as the product bundle.} We consider the $\zeta$-twisted Hilbert bundle $E := \bar{E} \otimes \HH$, where $\HH$ is the trivial bundle. This means that $E := (\{U_{i} \times (\C^{N} \otimes \HH)\}, \{g_{ij} \otimes 1\})$. Since $\C^{N} \otimes \HH \simeq \HH$, the bundle $E$ respects definition \ref{DefTwistedHilbBdl}.

In this section we set $K_{E}(X) := \bar{\Gamma}(F_{\PPP(E)})$, following definition \ref{DefKuX}, while we denote by $K_{\zeta}(X)$ the model through finite-dimensional twisted bundles, as defined in section \ref{TopFiniteOrder}. Let us consider a section $\psi \in \Gamma(F_{\PPP(E)})$. It follows that $[\psi] \in \bar{\Gamma}(F_{\PPP(E)}) = K_{E}(X)$. Since the local bundles $U_{i} \times (\C^{N} \otimes \HH)$ are already trivialized, the section $\psi$ corresponds to a family of sections $\psi_{i} \colon U_{i} \to \Fred(\C^{N} \otimes \HH)$ such that $\psi_{i} = g_{ij} \cdot \psi_{j} \cdot g_{ij}^{-1}$. We have that $\C^{N} \otimes \HH \simeq \HH^{\oplus N}$ and we call $\pi_{1}, \ldots, \pi_{N} \colon \HH^{\oplus N} \to \HH$ the canonical projections. For each $x \in \bar{U}_{i}$, we consider the following space $V_{x, i} \subset \HH$:
	\[V_{x, i} := \bigl(\pi_{1}(\Ker \, \psi_{i}(x))\bigr)^{\bot} \cap \ldots \cap \bigl(\pi_{N}(\Ker \, \psi_{i}(x))\bigr)^{\bot}.
\]
Such a space is closed and finite-codimensional. In fact, $\Ker \, \psi_{i}(x)$ is finite-dimensional, since $\psi_{i}(x)$ is Fredholm, hence each projection $\pi_{k}(\Ker \, \psi_{i}(x))$ is finite-dimensional too. It follows that the orthogonal complement is closed and finite-codimensional, hence the same holds about the finite intersection $V_{x, i}$. Thus, $V_{x, i}^{\oplus N}$ is closed and finite-codimensional in $\HH^{\oplus N}$. Moreover, we have that $(V_{x, i}^{\oplus N}) \cap \Ker\,\psi_{i}(x) = \{0\}$. This follows from the fact that $V_{x, i}^{\oplus N} \subset (\Ker\,\psi_{i}(x))^{\bot}$. In fact, if $v := (v_{1}, \ldots, v_{N}) \in V_{x, i}^{\oplus N}$ and $w := (w_{1}, \ldots, w_{N}) \in \Ker\,\psi_{i}(x)$, then $w_{k} \in \pi_{k}(\Ker\,\psi_{i}(x))$ and $v_{k} \in \bigl(\pi_{k}(\Ker \, \psi_{i}(x))\bigr)^{\bot}$, hence $\langle v_{k}, w_{k} \rangle = 0$, therefore $\langle v, w \rangle = 0$.

Following the proof of \cite[Prop.\ A5]{Atiyah}, for each $x \in \bar{U}_{i}$ there exists a neighbourhood $U'_{x, i} \subset U_{i}$ such that $V_{x, i}^{\oplus N} \cap \Ker\,\psi_{i}(y) = \{0\}$ for every $y \in U'_{x, i}$. We extract a finite sub-cover of $\{U'_{x, i}\}_{x \in \bar{U}_{i}}$, that we denote by $\{U'_{x_{1}, i}, \ldots, U'_{x_{n_{i}}, i}\}$, and we set $V := \bigcap_{i \in I} V_{x_{1}, i} \cap \ldots \cap V_{x_{n_{i}}, i}$. It follows $V^{\oplus N}$ is closed and finite-codimensional and $V^{\oplus N} \cap \Ker(\psi_{i}(x)) = \{0\}$ for every $x \in U_{i}$ and for every $i$. Moreover, $g_{ij}(V^{\oplus N}) = V^{\oplus N}$ for every $i$ and $j$, since the transition functions act as $N \times N$ complex matrices on $\HH^{\oplus N}$. Projecting to the quotient, we get the pointwise isomorphism $\bar{g}_{ij} \colon \HH^{\oplus N}/V \to \HH^{\oplus N}/V$, therefore we get the following well-defined $\zeta$-twisted finite-dimensional vector bundle on $X$:
\begin{equation}\label{TwistedES}
	F_{\psi} := (\{U_{i} \times (\HH^{\oplus N}/V)\}, \{\bar{g}_{ij}\}).
\end{equation}
We set $\HH^{\oplus N}/\psi_{i}(V) := \bigsqcup_{x \in U_{i}} \HH^{\oplus N}/(\psi_{i})_{x}(V)$, as a quotient space of $U_{i} \times \HH^{\oplus N} \simeq \bigsqcup_{x \in U_{i}} \HH^{\oplus N}$. By \cite[Prop.\ A3]{Atiyah} the space $\HH^{\oplus N}/\psi_{i}(V)$ is a vector bundle on $U_{i}$, hence, since $U_{i}$ is contractible, it is a trivial vector bundle. Moreover, we get a well-defined isomorphism $\bar{\bar{g}}_{ij} \colon \HH^{\oplus N}/\psi_{i}(V) \to \HH^{\oplus N}/\psi_{j}(V)$, since $(g_{ij})_{x}((\psi_{i})_{x}(V)) = (\psi_{j})_{x}((g_{ij})_{x}V) = (\psi_{j})_{x}(V)$, therefore we get the following $\zeta$-twisted finite-dimensional vector bundle on $X$:
\begin{equation}\label{TwistedFS}
	G_{\psi} := (\{\HH^{\oplus N}/\psi_{i}(V)\}, \{\bar{\bar{g}}_{ij}\}).
\end{equation}
With these data, we get the following isomorphism:
\begin{equation}\label{IsoThetaEZeta}
\begin{split}
	\Theta \colon & K_{E}(X) \overset{\!\simeq}\longrightarrow K_{\zeta}(X) \\
	& [\psi] \mapsto F_{\psi} - G_{\psi}.
\end{split}
\end{equation}
The proof that \eqref{IsoThetaEZeta} is actually an isomorphism follows the same line of the appendix of \cite{Atiyah}, adapted to the twisted framework.

\paragraph{\textbf{Differential extension.}} Now we extend the isomorphism \eqref{IsoThetaEZeta} to the differential framework. For this aim, we choose a cocyle of the form $(\zeta, 0, B)$, with $\zeta$ constant. Since $e^{B}$ is a global form, the local form $\psi_{i}^{*}\Phi_{\ev}$ in definition \ref{DefTwistedKT} glue to a global form $\psi^{*}\Phi_{\ev}$. Let us show that, in this case, we can give a definition similar to \ref{DefTwistedKT}, considering the space $\Fred(\HH)$ (as in the topological setting) instead of $\Gr^{p}$.

We fix two homotopy equivalences $\xi \colon \Fred(\HH^{\oplus N}) \to \Gr^{p}(\HH^{\oplus N})$ and $\tilde{\xi} \colon \Gr^{p}(\HH^{\oplus N}) \to \Fred(\HH^{\oplus N})$ inverse to each other, so that we have two homotopies $\Theta \colon \xi \circ \tilde{\xi} \simeq \id_{\Gr^{p}}$ and $\tilde{\Theta} \colon \tilde{\xi} \circ \xi \simeq \id_{\Fred(\HH^{\oplus N})}$. We construct these maps in such a way that they are $\UU(N)$-equivariant. Moreover, we set $\tilde{\Phi}_{\ev} := \xi^{*}\Phi_{\ev}$ on $\Fred(\HH^{\oplus N})$ and $\phi_{\odd} := \int_{I} \Theta^{*}\Phi_{\ev}$ on $\Gr^{p}$. We define $\hat{\hat{K}}_{(E, s, B)}(X)$ as in definition \ref{DefTwistedKT}, but replacing $F'_{\PPP(E)}$ by $F_{\PPP(E)}$, which means replacing the fibre $\Gr^{p}$ by $\Fred(\HH^{\oplus N})$, and $e^{B_{i}} \wedge \Psi_{i}^{*}\Phi_{\ev}$ by $e^{B} \wedge \Psi^{*}\tilde{\Phi}_{\ev}$. The following functions are isomorphisms inverse to each other:
	\[\begin{array}{rl}
	F \colon & \hat{\hat{K}}_{(E, s, B)}(X) \overset{\!\simeq}\longrightarrow \hat{K}_{(E, s, B)}(X) \\
	& [(\psi, \eta)] \mapsto [(\xi \circ \psi, \eta)]
\end{array} \qquad\; \begin{array}{rl}
	F^{-1} \colon & \hat{K}_{(E, s, B)}(X) \overset{\!\simeq}\longrightarrow \hat{\hat{K}}_{(E, s, B)}(X) \\
	& [(\psi, \eta)] \mapsto [(\tilde{\xi} \circ \psi, \eta - \psi^{*}\phi_{\odd} \wedge e^{-B})],
\end{array}\]
the compositions $\xi \circ \psi$ and $\tilde{\xi} \circ \psi$ and the pull-back $\psi^{*}\phi_{\odd}$ being defined through the local functions $\psi_{i}$ induced by $s$. The twisted bundles \eqref{TwistedES} and \eqref{TwistedFS} can be endowed with a $\zeta$-twisted connection, induced by orthogonal projection by any fixed $\zeta$-twisted connection on $\bar{E}$, extended to $\bar{E} \otimes \HH$ by tensor product with the trivial connection. Therefore, we extend \eqref{IsoThetaEZeta} to the following isomorphism:
\begin{equation}\label{IsoThetaDiff}
\begin{split}
	\hat{\Theta} \colon & \hat{\hat{K}}_{(E, s, B)}(X) \overset{\!\simeq}\longrightarrow \hat{K}_{(\zeta, \Lambda, B)}(X) \\
	& [(\psi, \eta)] \mapsto [(F_{\psi}, \nabla^{F_{\psi}}, \eta)] - [(G_{\psi}, \nabla^{G_{\psi}}, 0)].
\end{split}
\end{equation}
Similar isomorphisms hold for compactly-supported classes.

\subsection{Topological K-theory twisted by a Deligne class}\label{TopTwistedDeligne}

We have shown the definition of $\zeta$-twisted topological K-theory and of $(\zeta, \Lambda, B)$-twisted differential K-theory. Actually, we can define \emph{topological} $(\zeta, \Lambda, B)$-twisted K-theory, that we denote by $K^{\bullet}_{(\zeta, \Lambda, B)}(X)$, both in the model through twisted bundles and through Schatten Grassmannians. Considering the transformation $a \colon \Omega^{\odd}(X)/\IIm(d_{H}) \to \hat{K}_{(\zeta, \Lambda, B)}(X)$, introduced after definition \ref{DefTwistedKT}, we set:
\begin{equation}\label{DefTopKTDeligne}
	K^{\bullet}_{(\zeta, \Lambda, B)}(X) := \hat{K}^{\bullet}_{(\zeta, \Lambda, B)}(X) \,/\, \IIm(a).
\end{equation}
We get the natural isomorphism $K^{\bullet}_{(\zeta, \Lambda, B)}(X) \overset{\!\simeq}\longrightarrow K^{\bullet}_{\zeta}(X)$, that, choosing a model $(E, s, B)$ for $\hat{K}_{(\zeta, \Lambda, B)}(X)$, is realized by $K_{(E, s, B)}(X) \overset{\!\simeq}\longrightarrow K_{E}(X)$, $[(\psi, \zeta)] \mapsto \theta_{\zeta}^{-1}[\psi]$, where $\theta_{\zeta}$ is the isomorphism \eqref{IsoTheta}. Analogously, in the twisted-bundle model, the isomorphism is given by $[(E, \nabla, \omega)] \mapsto [E]$.

The advantage of definition \eqref{DefTopKTDeligne} is that, when $H_{1}(X; \Z) = 0$, it depends canonically on the Deligne class $[(\zeta, \Lambda. B)]$, since $\hat{K}^{\bullet}_{(\zeta, \Lambda, B)}(X)$ does. Therefore, in this case, the transformation $I \colon \hat{K}^{\bullet}_{(\zeta, \Lambda, B)}(X) \to K^{\bullet}_{\zeta}(X)$ can be written in the canonical form $I \colon \hat{K}^{\bullet}_{[(\zeta, \Lambda, B)]}(X) \to K^{\bullet}_{[(\zeta, \Lambda, B)]}(X)$, completing the canonical definition of $\hat{K}_{[(\zeta, \Lambda, B)]}$ as a twisted cohomology theory. In fact, we get:
\begin{itemize}
	\item $I \colon \hat{K}^{\bullet}_{[(\zeta, \Lambda, B)]}(X) \to K^{\bullet}_{[(\zeta, \Lambda, B)]}(X)$;
	\item $R \colon \hat{K}^{\bullet}_{[(\zeta, \Lambda, B)]}(X) \to \Omega^{^{\bullet}}_{H\textnormal{-}\cl}(X)$;
	\item $a \colon \Omega^{\bullet-1}(X)/\IIm(d_{H}) \to \hat{K}^{\bullet}_{[(\zeta, \Lambda, B)]}(X)$,
\end{itemize}
satisfying the axioms of differential cohomology. In particular, if we choose the class $[(\zeta, 0, 0)]$, we get the groups $K^{\bullet}_{[\zeta]}(X)$, with $\zeta$ constant, already considered in the last paragraph before section \ref{RelativeVersion}.

%%%%%%%%%%%%%%%%%%%%%%%%%%%%%%%%%%%%%%%%%%%%%%%%%%%%%%%%%%%%%%%%%%%%%%%%%%%%%%%%%%%%%%%%%%%%%%%

\section{Twisted K-characters and K-homology}\label{TwKChar}

We now introduce the notion of \emph{twisted K-character}, that is the suitable tool in order to define rigorously the world-volume of a D-brane and the Wess-Zumino action. We start recalling the notions of orientation and integration in the framework of twisted differential K-theory, following \cite[sec.\ 4.8-4.10]{Bunke} and \cite[chap.\ 6]{FRRB}.

\subsection{Orientation and integration}\label{SecOrientInt}

We show how to orient a vector bundle, a smooth map and a smooth manifold, then we discuss the integration map. In this section every manifold is simply-connected by hypothesis.\footnote{Actually, requiring that the first-homology group vanishes would be enough, but it would be less elegant and it will not be enough in the next section.}

\paragraph{\textbf{I. Orientation of vector bundles.}} 

We fix a real vector bundle $\pi \colon E \to X$ of rank $n$ on a compact manifold $X$. We have seen in section \ref{SecThomIso} that a topological $K^{\bullet}$-orientation of $E$ is a $w_{2}(E)$-twisted Thom class $u$ of $E$. We get the Thom isomorphism
\begin{equation}\label{ThomIsoTwistedTop}
\begin{split}
	T \colon & K_{[(\zeta, \Lambda, B)]}^{\bullet}(X) \overset{\!\simeq}\longrightarrow K^{\bullet+n}_{[(\zeta, \Lambda, B)] + w_{2}(E),\cpt}(E) \\
	& \alpha \mapsto u \cdot \pi^{*}\alpha,
\end{split}
\end{equation}	
that is the canonical version of \eqref{ThomIso} through definition \eqref{DefTopKTDeligne}. We call \emph{integration map} its inverse $\int_{E/X} \colon K^{\bullet+n}_{[(\zeta, \Lambda, B)] + w_{2}(E), \cpt}(E) \to K^{\bullet-n}_{[(\zeta, \Lambda, B)]}(X)$, $u \cdot \pi^{*}\alpha \mapsto \alpha$. The $n$-degree component of the Chern character $\ch \,u$ (see formulas \eqref{ChTwistedK} and \eqref{ChTwistedK1} with $H = 0$) defines an orientation of $E$ in the usual sense, hence it is possible to integrate a compactly-supported form fibre-wise. We define the \emph{Todd class} $\Td(u) := \int_{E/X} \ch \, u \in H^{\ev}_{\dR}(X)$. The following formula holds:
\begin{equation}\label{ChIntegral}
	\int_{E/X} \ch \, \alpha = \Td(u) \cdot \biggl( \ch \int_{E/X} \alpha \biggr).
\end{equation}
In order to orient a vector bundle with respect to \emph{differential} K-theory, one just has to refine a Thom class $u$ to a \emph{differential Thom class}.
\begin{Def} A \emph{differential Thom class} of $E$ is a class $\hat{u} \in \hat{K}^{n}_{w_{2}(E),\cpt}(E)$ such that $I(\hat{u})$ is a topological Thom class.
\end{Def}
We define the differential Thom morphism, which is not surjective any more, as
\begin{equation}\label{ThomIsoTwisted}
\begin{split}
	T \colon & \hat{K}_{[(\zeta, \Lambda, B)]}^{\bullet}(X) \hookrightarrow \hat{K}^{\bullet+n}_{[(\zeta, \Lambda, B)] + w_{2}(E), \cpt}(E) \\
	& \hat{\alpha} \mapsto \hat{u} \cdot \pi^{*}\hat{\alpha}.
\end{split}
\end{equation}	
We define the \emph{Todd class} $\Td(\hat{u}) := \int_{E/X} R(\hat{u}) \in \Omega^{\ev}_{\cl}(X)$. It follows that $\dR(\Td(\hat{u})) = \Td(I(\hat{u}))$.

\begin{Def}\label{HomotopyThom} Let $\pi_{X} \colon I \times X \to X$ be the natural projection and $i_{0}, i_{1} \colon X \to I \times X$ the natural embeddings. Two differential Thom classes $\hat{u}, \hat{u}' \in \hat{K}^{n}_{w_{2}(E), \cpt}(E)$ are \emph{homotopic} if there exists a Thom class $\hat{U} \in \hat{K}^{n}_{w_{2}(E), \cpt}(\pi_{X}^{*}E)$ such that $i_{0}^{*}\hat{U} = \hat{u}$, $i_{1}^{*}\hat{U} = \hat{u}'$ and $\Td(\hat{U}) = \pi_{X}^{*}\Td(\hat{u})$.
\end{Def}

\begin{Lemma}[2x3 principle]\label{Rule23Diff} Given two real vector bundles $E, F \to X$, with projections $p_{E} \colon E \oplus F \to E$ and $p_{F} \colon E \oplus F \to F$, we consider a triple $(\hat{u}, \hat{v}, \hat{w})$ of differential Thom classes on $E$, $F$ and $E \oplus F$ respectively, such that $\hat{w}$ is homotopic to $p_{E}^{*}\hat{u} \cdot p_{F}^{*}\hat{v}$. Two elements of such a triple uniquely determine the third one up to homotopy.
\end{Lemma}

For the proofs see \cite[prob.\ 4.187]{Bunke} and \cite[cor.\ 3.19]{FR}, adapted to the twisted framework.

\paragraph{\textbf{II. Orientation of smooth maps.}} As in the non-twisted framework, we start defining a \emph{representative} of an orientation as follows.

\begin{Def}\label{TopOrientedMap} A \emph{representative of a $\hat{K}^{\bullet}$-orientation} of a smooth neat map $f \colon Y \to X$ between compact manifolds is the datum of:
\begin{itemize}
	\item a neat embedding $\iota \colon Y \hookrightarrow X \times \R^{N}$, for any $N \in \N$, such that $\pi_{X} \circ \iota = f$;
	\item a differential Thom class $\hat{u}$ of the normal bundle $\NN := N_{\iota(Y)}(X \times \R^{N})$;
	\item a diffeomorphism $\varphi \colon N_{\iota(Y)}(X \times \R^{N}) \to U$, for $U$ a neat tubular neighbourhood of $\iota(Y)$ in $X \times \R^{N}$.
\end{itemize}
\end{Def}
We have the following natural map on differential forms, called \emph{curvature map}:
\begin{equation}\label{CurvatureMap}
\begin{split}
	R_{(\iota, \hat{u}, \varphi)} \colon & \Omega^{\bullet}(Y) \to \Omega^{\bullet-n}(X) \\
	& \omega \mapsto \int_{X \times \R^{N}/X} i_{*}\varphi_{*}(R(\hat{u}) \wedge \pi^{*}\omega).
\end{split}
\end{equation}
When $f$ is a submersion, we can choose a \emph{proper} representative \cite{FRRB}, i.e.\ we can choose the diffeomorphism $\varphi$ in such a way that the image of the fibre of $\NN_{y}$ is contained in $\{f(y)\} \times \R^{N}$. In this case:
\begin{equation}\label{IntProp}
	R_{(\iota, \hat{u}, \varphi)}(\omega) = \int_{Y/X} \Td(\hat{u}) \wedge \omega.
\end{equation}
We can introduce a suitable equivalence relation on representatives, induced by \emph{homotopy} and \emph{stabilization} \cite{FRRB}. In particular, the curvature map \eqref{CurvatureMap} must be constant along a homotopy. By definition, a \emph{$\hat{K}^{\bullet}$-orientation} on $f \colon Y \to X$ is an equivalence class $[\iota, \hat{u}, \varphi]$. Because of the uniqueness up to homotopy and stabilization of the tubular neighbourhood and of the diffeomorphism with the normal bundle, the class $[\iota, \hat{u}, \varphi]$ does not depend on $\varphi$, hence we denote it by $[\iota, \hat{u}]$. Moreover, any two embeddings $\iota$ and $\iota'$ are equivalent, therefore the meaningful datum is $\hat{u}$.

\begin{Rmk}\label{OrInducedBdFunction} \emph{If $X$ and $Y$ are manifolds with boundary, such that $Y$ and any component of $\partial Y$ have vanishing homology group of degree $1$, an orientation on $f \colon Y \to X$ canonically induces an orientation on $\partial f := f\vert_{\partial Y} \colon \partial Y \to \partial X$. In fact, fixing a representative $(\iota, \hat{u}, \varphi)$ for $f$, by neatness $\iota$ restricts to $\iota' \colon \partial Y \hookrightarrow \partial X \times \R^{N}$. The normal bundle and the tubular neighbourhood, being neat, restrict to the boundary too, hence we get a representative $(\iota', \hat{u}', \varphi')$ for $\partial f$. Any homotopy of representatives determines a homotopy on the boundary, therefore the resulting orientation of $\partial f$ is well-defined.}
\end{Rmk}

Given two $\hat{K}^{\bullet}$-oriented neat maps $f \colon Y \to X$ and $g \colon X \to W$, with orientation respectively $[\iota, \hat{u}]$ and $[\kappa, \hat{v}]$, there is a naturally induced $\hat{K}^{\bullet}$-orientation on $g \circ f \colon Y \to W$, that we denote by $[\kappa, \hat{v}][\iota, \hat{u}]$ \cite{FRRB}. The following lemma is a consequence of lemma \ref{Rule23Diff} and of the uniqueness up to homotopy and stabilization of the embedding $\iota$.
\begin{Lemma}[2x3 principle]\label{Rule23Subm} Let $f \colon Y \to X$ and $g \colon X \to W$ be $\hat{K}^{\bullet}$-oriented neat submersions, with orientations $[\iota, \hat{u}]$ and $[\kappa, \hat{v}]$, and let $[\xi, \hat{w}]$ be the orientation induced on $g \circ f$. Two elements of the triple $([\iota, \hat{u}], [\kappa, \hat{v}], [\xi, \hat{w}])$ uniquely determine the third one.
\end{Lemma}
For the proof see \cite[theorem 5.24 p.\ 233]{Karoubi}. There is also a natural notion of \emph{homotopy of $\hat{K}^{\bullet}$-oriented} maps. The reader can find the details in \cite[Definition 6.9]{FRRB}.

\paragraph{\textbf{III. Orientation of smooth manifolds.}}  In this subsection we discuss separately the cases of manifolds without boundary and with boundary.
\begin{Def}\label{OrientedManifold} A $\hat{K}^{\bullet}$-orientation of a manifold without boundary $X$ is a $\hat{K}^{\bullet}$-orientation of the map $p_{X} \colon X \to \{pt\}$.
\end{Def}
By definition, an orientation of $p_{X}$ is essentially a Thom class $\hat{u}$ on the (stable) normal bundle of $X$; when $\hat{u}$ has been fixed, we set $\Td(X) := \Td(\hat{u})$.

Given a manifold with boundary $X$, we recall that a \emph{defining function for the boundary} is a smooth neat map $\Phi \colon X \to I$ such that $\partial X = \Phi^{-1}\{0\}$ (by neatness, it follows that $\Phi^{-1}\{1\} = \emptyset$). It is easy to verify that any two defining functions are neatly homotopic. Nevertheless, a defining function is not a submersion in general, hence we have to modify slightly the definition of orientation. Following definition \eqref{CurvatureMap}, the curvature map should be $\omega \mapsto \int_{I \times \R^{N}/I} i_{*}\varphi_{*}(R(\hat{u}) \wedge \pi^{*}\omega)$, but we also integrate on $I$ the result:
\begin{equation}\label{CurvMapBd}
\begin{split}
	R^{\partial}_{(\iota, \hat{u}, \varphi)} \colon & \Omega^{\bullet}(X; \R) \to \Omega^{\bullet-n}(pt; \R) \\
	& \omega \mapsto \int_{0}^{1} \int_{I \times \R^{N}/I} i_{*}\varphi_{*}(R(\hat{u}) \wedge \pi^{*}\omega).
\end{split}
\end{equation}

\begin{Def}\label{OrientedManifoldBoundary} A \emph{$\hat{K}^{\bullet}$-orientation} on a smooth manifold with boundary $X$ is a homotopy class of $\hat{K}^{\bullet}$-oriented defining functions for the boundary, considering the curvature map \eqref{CurvMapBd} in the definition of homotopy.
\end{Def}
Formula \eqref{IntProp} keeps on holding, in the following form:
\begin{equation}\label{RBdTodd}
	R^{\partial}_{(\iota, \hat{u}, \varphi)}(\omega) = \int_{X} \Td(X) \wedge \omega.
\end{equation}
With this definition of the curvature map, an orientation of a manifold without boundary can be thought of as a particular case of an orientation of a manifold with boundary. 

\begin{Rmk}\label{OrInducedBd} \emph{It follows from remark \ref{OrInducedBdFunction} that an orientation on a manifold with boundary canonically induces an orientation on the boundary. In particular, let us fix a defining function $\Phi \colon X \to I$ and an orientation $[\iota, \hat{u}]$, with $\iota \colon X \hookrightarrow I \times \R^{N}$. We call $i_{\partial X} \colon \partial X \hookrightarrow X$ the natural embedding and we set $\iota' := \iota \circ i_{\partial X} \colon \partial X \hookrightarrow \{0\} \times \R^{N}$ and $\hat{u}' := \hat{u}\vert_{\partial X}$. We get the orientation $[\iota', \hat{u}']$ of $\partial X$.}
\end{Rmk}

\paragraph{\textbf{IV. Integration.}} Let $f \colon Y \to X$ be a $\hat{K}^{\bullet}$-oriented neat submersion. We set $w_{2}(f) := w_{2}(Y) + f^{*}w_{2}(X)$ and $W_{3}(f) := W_{3}(Y) + f^{*}W_{3}(X)$. Moreover, we fix twisting classes $[(\xi, \Theta, C)]$ in $Y$ and $[(\zeta, \Lambda, B)]$ in $X$ and we denote by $w'_{2}(f)$ the image of $w_{2}(f)$ trough the morphism in cohomology induced by $\Z_{2} \hookrightarrow \UU(1)$. The integration map (or Gysin map) induced by $f$ is defined under the following hypothesis:
\begin{equation}\label{HypGysinMapDiff}
	[(\xi, \Theta, C)] + w'_{2}(f) = f^{*}[(\zeta, \Lambda, B)].
\end{equation}
Computing the first Chern class of each term, we get the topological condition
\begin{equation}\label{HypGysinMapTop}
	[\xi] + W_{3}(f) = f^{*}[\zeta].
\end{equation}
We remark that, for a fixed $[(\zeta, \Lambda, B)]$, there exists a unique class $[(\xi, \Theta, C)]$ such that \eqref{HypGysinMapDiff} holds. In this case, we define:
\begin{equation}\label{GysinMapDiff}
\begin{split}
	f_{!} \colon & \hat{K}^{\bullet}_{[(\xi, \Theta, C)]}(Y) \to \hat{K}^{\bullet-n}_{[(\zeta, \Lambda, B)]}(X) \\
	& \hat{\alpha} \mapsto \int_{\R^{N}}i_{*}\varphi_{*}(\hat{u} \cdot \pi^{*}\hat{\alpha}),
\end{split}
\end{equation}
where $n = \dim(Y) - \dim(X)$. The integration map with respect to $\R^{N}$ is defined as follows. The open embedding $j \colon \R^{N} \hookrightarrow (S^{1})^{N}$, defined through the embedding $\R \hookrightarrow \R^{+} \simeq S^{1}$ in each coordinate, induces the push-forward $(\id \times j)_{*}$ in compactly-supported K-theory, thus we set $\int_{\R^{N}}\hat{\beta} := \int_{S^{1}} \cdots \int_{S^{1}} (\id \times j)_{*}\hat{\beta}$.

Condition \eqref{HypGysinMapDiff} is due to the fact that the class $w_{2}(\NN)$, where $\NN$ is the normal bundle considered in definition \ref{TopOrientedMap}, coincides with $w_{2}(f)$, therefore the class $\hat{u} \cdot \pi^{*}\hat{\alpha}$ is twisted by $[(\xi, \Theta, C)] + w'_{2}(f)$, that must coincide with the twisting of the codomain, i.e.\ $[(\zeta, \Lambda, B)]$.

The map \eqref{GysinMapDiff} only depends on the $\hat{K}^{\bullet}$-orientation $[\iota, \hat{u}]$, not on the specific representative (\cite[theorem 5.24 p.\ 233]{Karoubi}, \cite[sec.\ 4.9, 4.10]{Bunke}). Moreover, if $X$ and $Y$ are manifolds with boundary, considering remark \ref{OrInducedBdFunction}, one has, for every $\hat{\alpha} \in \hat{K}^{\bullet}_{[(\xi, \Theta, C)]}(Y)$:
\begin{equation}\label{RestrictionBoundaryGysin}
	(\partial f)_{!}(\hat{\alpha}\vert_{\partial Y}) = (f_{!}\hat{\alpha})\vert_{\partial X}.
\end{equation}
Such a formula is due to the fact that all the structures involved in the definition of $(\partial f)_{!}$ are the restrictions to the boundary of the corresponding structures for $f_{!}$.

It is easy to prove from the axioms that:
	\[R(f_{!}\hat{\alpha}) = R_{[\iota, \hat{u}]}(R(\hat{\alpha})) \qquad f_{!}a(\omega) = a(R_{[\iota, \hat{u}]}(\omega)),
\]
thus the following diagram commutes:
\begin{equation}\label{CurvAMaps}
	\xymatrix{
	\Omega^{\bullet-1}(Y)/\IIm(d_{H}) \ar[r]^(.55){a} \ar[d]^{R_{[\iota, \hat{u}]}} & \hat{K}^{\bullet}_{[(\xi, \Theta, C)]}(Y) \ar[r]^{I} \ar[d]^{f_{!}} \ar@/^2pc/[rr]^{R} & K^{\bullet}_{[(\xi, \Theta, C)]}(Y) \ar[d]^{f_{!}} & \Omega_{H\textnormal{-}\cl}^{\bullet}(Y) \ar[d]^{R_{[\iota, \hat{u}]}} \\
		\Omega^{\bullet-n-1}(X)/\IIm(d_{H}) \ar[r]^(.55){a} & \hat{K}^{\bullet-n}_{[(\zeta, \Lambda, B)]}(X) \ar[r]^{I} \ar@/_2pc/[rr]_{R} & K^{\bullet-n}_{[(\zeta, \Lambda, B)]}(X) & \Omega_{H\textnormal{-}\cl}^{\bullet-n}(X).
	}
\end{equation}

\begin{Theorem}\label{fpDifProperties} Let $f \colon Y \to X$ be a neat $\hat{K}^{\bullet}$-oriented submersion between compact manifolds.
\begin{itemize}
	\item The Gysin map $f_{!}$ only depends on the homotopy class of $f$ as an $\hat{K}^{\bullet}$-oriented map.
	\item For any $\hat{\alpha} \in \hat{K}^{\bullet}_{[(\xi, \Theta, C)]}(Y)$ and $\hat{\beta} \in \hat{K}^{\bullet}_{[(\nu, \Xi, D)]}(X)$, where $[(\nu, \Xi, D)]$ is any twisting class:
	\[f_{!}(\hat{\alpha} \cdot f^{*}\hat{\beta}) = f_{!}\hat{\alpha} \cdot \hat{\beta}.
\]
	\item Given another neat $\hat{K}^{\bullet}$-oriented map $g \colon Z \to Y$ and endowing $f \circ g$ of the naturally induced orientation, we have $(f \circ g)_{!} = f_{!} \circ g_{!}$.
	\item We have that:
	\begin{equation}\label{RAIntSubmersion}
	R(f_{!}\hat{\alpha}) = \int_{Y/X} \Td(\hat{u}) \wedge R(\hat{\alpha}) \qquad f_{!}(a(\omega)) = a \biggl( \int_{Y/X} \Td(\hat{u}) \wedge \omega \biggr).
	\end{equation}
\end{itemize}
\end{Theorem}
For the proof see \cite[Lemmas 3.24 and 3.27]{FR} and \cite[Problems 4.219 and 4.233]{Bunke}.

Up to now we supposed that $f$ is oriented. If $f \colon Y \to X$ is a neat submersion between $\hat{K}^{\bullet}$-oriented manifolds without boundary, then, since $p_{Y} = p_{X} \circ f$, it follows from lemma \ref{Rule23Subm} that $f$ inherits a unique orientation from the ones of $X$ and $Y$. Hence, the integration map $f_{!}$ is well-defined. If $X$ and $Y$ have boundary, the same result follows from the analogous 2x3 principle.

\paragraph{\textbf{V. Topological Integration.}} In the differential framework it is essential to consider proper submersions, since only in this setting some natural tools of algebraic topology (in particular, the 2x3 rule and homotopy invariance) keep on holding. The main reason is that the curvature is relevant as a single differential form, not only as a representative of a cohomology class. Of course this does not happen in the topological framework, therefore all of the previous definitions and constructions can be applied to any $K^{\bullet}$-oriented smooth map (and to any bundle or manifold as well). In particular, if condition \eqref{HypGysinMapDiff} holds, we get the Gysin map
\begin{equation}\label{GysinMapTop}
	f_{!} \colon K^{\bullet}_{[(\xi, \Theta, C)]}(Y) \to K^{\bullet-n}_{[(\zeta, \Lambda, B)]}(X)
\end{equation}
for any $K^{\bullet}$-oriented $f \colon Y \to X$, applying definition \eqref{DefTopKTDeligne}, i.e.\ quotienting out by the image of the natural transformation $a$. If $X$ and $Y$ are $K^{\bullet}$-oriented, then $f$ inherits canonically an orientation, so that \eqref{GysinMapDiff} is well-defined. The integration map, stated with the present language, is coherent with the the topological one constructed in \cite{CW} and with the the differential one constructed (for embeddings) in \cite{CMW}. We defer the details of the comparison to a future paper.

\subsection{Twisted K-Characters}

In \cite{Jakob} the author provides a geometric model of the homology theory dual to a given cohomology theory. One item of that definition uses the ``vector bundle modification'', that has been replaced in \cite{FR} by the (more natural) integration map. We briefly recall the definition of this model, in the specific case of (non-twisted) K-homology. We only consider smooth manifolds, since we are interested in the differential extension, but the construction holds for any space with the homotopy type of a finite CW-complex.
\begin{Def}\label{DefKHom} On a compact manifold $X$ we define:
\begin{itemize}
	\item the group of \emph{$n$-pre-cycles} as the free abelian group generated by quadruples of the form $(M, u, \alpha, f)$, with:
\begin{itemize}
	\item $M$ a smooth compact spin$^{c}$ manifold (without boundary) with Thom class $u$, whose connected components $\{M_{i}\}$ have dimension $n+q_{i}$, with $q_{i}$ arbitrary;
	\item $\alpha \in K^{\bullet}(M)$ such that $\alpha\vert_{M_{i}} \in K^{q_{i}}(M_{i})$;
	\item $f \colon M \to X$ a smooth map;
\end{itemize}
	\item the group of \emph{$n$-cycles}, denoted by $Z_{n}(X)$, as the quotient of the group of $n$-pre-cycles by the free subgroup generated by elements of the form:
\begin{itemize}
	\item $(M, u, \alpha + \beta, f) - (M, u, \alpha, f) - (M, u, \beta, f)$;
	\item $(M, u, \alpha, f) - (M_{1}, u\vert_{M_{1}}, \alpha\vert_{M_{1}}, f\vert_{M_{1}}) - (M_{2}, u\vert_{M_{2}}, \alpha\vert_{M_{2}}, f\vert_{M_{2}})$, for $M = M_{1} \sqcup M_{2}$;
	\item $(M, u, \varphi_{!}\alpha, f) - (N, v, \alpha, f \circ \varphi)$ for $\varphi \colon N \to M$ a smooth map;
\end{itemize}
	\item the group of \emph{$n$-boundaries}, denoted by $B_{n}(X)$, as the subgroup of $Z_{n}(X)$ generated by the cycles which are representable by a pre-cycle $(M, u, \alpha, f)$, such that there exits a quadruple $(W, U, A, F)$, where $W$ is a manifold and $M = \partial W$, $U$ is a Thom class of $W$ and $U\vert_{M} = u$, $A \in K^{\bullet}(W)$ and $A\vert_{M} = \alpha$, $F \colon W \to X$ is a smooth map satisfying $F\vert_{M} = f$.
\end{itemize}
We set $K_{n}(X) := Z_{n}(X) / B_{n}(X)$.
\end{Def}
We cannot generalize \ref{DefKHom} to the twisted framework directly, since we have seen that the integration map is canonically defined only in the category of manifolds with vanishing first-degree homology. A stronger but more natural condition consists of simply-connectedness, and it turns out to be the correct choice. Let us show how. We define the groups $Z_{n}^{\simpc}(X)$ and $B_{n}^{\simpc}(X)$ as in definition \ref{DefKHom}, but imposing that $M$ is simply-connected in every pre-cycle (hence in every cycle). In the definition of boundaries, $W$ must be simply-connected too. We set $K^{\simpc}_{n}(X) := Z^{\simpc}_{n}(X) / B^{\simpc}_{n}(X)$. The inclusion of simply-connected pre-cycles in the set of all pre-cycles induces a natural morphism $\upsilon \colon K^{\simpc}_{n}(X) \to K_{n}(X)$. The following lemma justifies the definition of $K^{\simpc}_{n}(X)$.

\begin{Lemma}\label{LemmaSimpC} If $X$ is simply-connected, then $\upsilon \colon K^{\simpc}_{n}(X) \overset{\!\simeq}\longrightarrow K_{n}(X)$ is an isomorphism.
\end{Lemma}
\textsc{Sketch of the proof:} We begin with surjectivity, i.e.\ we show that any class $[(M, u, \alpha, f)]$ $\in K_{n}(X)$ can be represented by $(M', u', \alpha', f')$ such that $M'$ is simply-connected. We prove it in steps.

\texttt{Step I.} \emph{We can suppose that $M$ is connected, $\dim(M) \geq 4$ and $\alpha \in K^{0}(M)$.} In fact, if $M$ is not connected, we just analyse separately each connected component. About the dimension, if it is not high enough, we replace $M$ by $\bar{M} := M \times \mathbb{T}^{n}$, for any $n$, and $\alpha$ by a class $\bar{\alpha} \in K^{\bullet}(\bar{M})$ such that $\int_{\mathbb{T}^{n}} \bar{\alpha} = \alpha$. Moreover, we choose the canonically induced Thom class $\bar{u}$ and $\bar{f} := f \circ \pi$, where $\pi \colon \bar{M} \to M$ is the natural projection. It follows that $[(\bar{M}, \bar{u}, \bar{\alpha}, \bar{f})] = [(M, u, \pi_{!}\bar{\alpha}, \bar{f})] = [(M, u, \alpha, f)]$, hence $M$ has been replaced by $\bar{M}$, of arbitrarily high dimension. Moreover, if $\bar{\alpha} \in K^{1}(\bar{M})$, then we just replace $\mathbb{T}^{n}$ by $\mathbb{T}^{n+1}$, so that $\bar{\alpha} \in K^{0}(\bar{M})$.

\texttt{Step II.} \emph{$M$ is cobordant to a simply-connected manifold $M'$.} The following argument\footnote{See the link https://mathoverflow.net/questions/14903/every-manifold-cobordant-to-a-simply-connected-manifold.} holds since we can suppose $\dim(M) \geq 4$ by step I. Let us fix a loop $\gamma$ in $M$ that represents a generator of $\pi_{1}(M)$. We fix a tubular neighbourhood of $\gamma$, that we identify with $S^{1} \times D^{n-1}$. We consider the trivial cobordism $M \times I$ and, in $M \times \{1\}$, we glue $D^{2} \times D^{n-1}$ to $S^{1} \times D^{n-1}$. Applying this construction to each generator of $\pi_{1}(M)$, we get a manifold $W$ that realizes a cobordism between $M \times \{0\}$ and $M'$, the latter being the manifold obtained from $M \times \{1\}$ by the surgery that replaces each $S^{1} \times D^{n-1}$ by $D^{2} \times S^{n-2}$. Since $M'$ is simply connected, we get the result.

\texttt{Step III.} \emph{We can complete $W$ to a quadruple $(W, U, A, F)$ restricting to $(M, u, \alpha, f)$.} In fact, we fix any Thom class $U$ of $W$ and we set $v := U\vert_{M}$. If $v \neq u$, we consider the identity $\id \colon (M, u) \to (M, v)$ and we have that $[(M, u, \alpha, f)] = [(M, v, \id_{!}\alpha, f)]$, hence, up to equivalence, we can suppose that $u = U\vert_{M}$. Let us show that we can extend $\alpha$ to a class $A$ on $W$. The manifold $W$ has been obtained from $M \times I$ gluing spaces of the form $D^{2} \times D^{n-1}$. Let us suppose for simplicity to have glued only one piece of that form (in general, we apply the following argument inductively). We apply the Mayer-Vietoris sequence to the cover $\{U, V\}$ of $W$, where $U$ is an open neighbourhood of $M \times I$, retracting by deformation on the latter, and $V$ is a subspace of the form $B^{2} \times D^{n-1} \subset D^{2} \times D^{n-1}$, where $B^{2}$ is an open ball, in such a way that $U \cap V$ has the homotopy type of $S^{1} \times D^{n-1}$. From the sequence
	\[\cdots \longrightarrow \tilde{K}^{\bullet}(W) \longrightarrow \tilde{K}^{\bullet}(U) \oplus \tilde{K}^{\bullet}(V) \longrightarrow \tilde{K}^{\bullet}(U \cap V) \longrightarrow \cdots,
\]
since $U \simeq M \times I \simeq M$, $V$ is contractible and $U \cap V \simeq S^{1}$, we get
	\[\cdots \longrightarrow \tilde{K}^{0}(W) \longrightarrow \tilde{K}^{0}(M) \longrightarrow 0.
\]
This shows that the restriction map $\tilde{K}^{0}(W) \to \tilde{K}^{0}(M)$ is surjective. By step I, we can assume $\alpha \in K^{0}(M)$, hence $\alpha = \tilde{\alpha} + k$, with $\tilde{\alpha} \in \tilde{K}^{0}(M)$ and $k \in \Z$. Because of the surjectivity of the restriction, the class $\tilde{\alpha}$ lift to a class $\tilde{A} \in \tilde{K}^{0}(W)$, hence $A := \tilde{A} + k$ lifts $\alpha$ as desired. It only remains to extend $f$ to $F$. We trivially extend $f$ to the domain $M \times I$. Since the codomain $X$ is simply-connected by hypothesis, the restriction of $f$ to $S^{1} \times D^{n-1}$ can be extended continuously to $D^{2} \times D^{n-1}$, hence, smoothing such an extension if necessary, we get $F$.

\texttt{Step IV.} \emph{$\nu$ is surjective.} From the quadruple $(W, U, A, F)$, defined in step III, we set $(M', u', \alpha', f') := (W, U, A, F)\vert_{M'}$. It follows fhat $[(M, u, \alpha, f)] = [(M', u', \alpha', f')]$, with $M'$ simply-connected by construction.

\texttt{Injectivity.} It remains to prove that $\nu$ is injective, i.e.\ that, if $(M, u, \alpha, f)$ and $(M', u', \alpha', f')$, with $M$ and $M'$ simply-connected, are equivalent through a generic quadruple $(W, U, A, F)$, then they are also equivalent through a quadruple $(W', U', A', F')$ with $W'$ simply-connected. We apply an argument similar to the previous one, constructing a cobordism (with corners) $Z$ from $W$ to a simply-connected manifold $W'$, that is constant on $M$ and $M'$ (this is possible since $M$ and $M'$ are already simply-connected, hence we do not need to apply any surgery on them). We extend $(W, U, A, F)$ to $(Z, V, B, G)$ as above and we set $(W', U', A', F') := (Z, V, B, G)\vert_{W'}$. In this way we get the desired equivalence. $\hfill\diamondsuit$

\begin{Rmk} \emph{In the previous lemma the hypothesis that $X$ is simply-connected is essential to extend the function $f$, while assuming $H_{1}(X; \Z) = 0$ is not sufficient.}
\end{Rmk}

In \cite{FR} and \cite{FRRB} we introduced a suitable differential refinement of cycles and boundaries, inducing the same homology groups, but allowing for the definition of \emph{holonomy} of a differential cohomology class on a cycle. This construction naturally leads to the definition of \emph{Cheeger-Simons character} for any cohomology theory. In particular, in definition \ref{DefKHom}, one just has to replace each orientation $u$ by a differential orientation $\hat{u}$ and each K-theory class $\alpha$ by a differential K-theory class $\hat{\alpha}$. Of course the topological Gysin map is replaced by its differential version. For this reason, in the third item of the definition of cycle, the map $\varphi \colon N \to M$ must be a submersion, as we have seen above. We call $\hat{Z}_{n}(X)$ and $\hat{B}_{n}(X)$ the corresponding groups of differential cycles and boundaries and we set $K'_{n}(X) := \hat{Z}_{n}(X) / \hat{B}_{n}(X)$. The natural morphism $K'_{n}(X) \to K_{n}(X)$, $[(M, \hat{u}, \hat{\alpha}, f)] \mapsto [(M, I(\hat{u}), I(\hat{\alpha}), f)]$, turns out to be an isomorphism, hence we can identify $K'_{n}(X)$ with $K_{n}(X)$. Moreover, we define $\hat{Z}^{\simpc}_{n}(X)$ and $\hat{B}^{\simpc}_{n}(X)$ in the same way, but only considering simply-connected manifolds, and lemma \ref{LemmaSimpC} keeps on holding.

Now we adapt this construction to the case of twisted K-theory. Considering definition \eqref{DefTopKTDeligne}, it is natural to start from the differential framework. Since twisted K-theory canonically depends on the cohomology class only on simply-connected manifolds, lemma \ref{LemmaSimpC} naturally suggests the following definition.
\begin{Def}\label{TwistedKHomDiff} On a compact simply-connected manifold $X$ with a fixed Deligne class $[(\zeta, \Lambda, B)] \in \hat{H}^{2}(X)$, we define:
\begin{itemize}
	\item the group of \emph{$n$-pre-cycles} as the free abelian group generated by quadruples of the form $(M, \hat{u}, \hat{\alpha}, f)$, with:
\begin{itemize}
	\item $M$ a smooth compact simply-connected manifold (without boundary) with ($w_{2}(M)$-twisted) Thom class $\hat{u}$, whose connected components $\{M_{i}\}$ have dimension $n+q_{i}$, with $q_{i}$ arbitrary;
	\item $f \colon M \to X$ a smooth map;
	\item $\hat{\alpha} \in \hat{K}^{\bullet}_{[(\xi, \Theta, C)]}(M)$, where
	\begin{equation}\label{FWCondition}
		[(\xi, \Theta, C)] := w'_{2}(M) - f^{*}[(\zeta, \Lambda, B)],
	\end{equation}
	such that $\hat{\alpha}\vert_{M_{i}} \in K^{q_{i}}_{[(\xi, \Theta, C)]}(M_{i})$;
\end{itemize}
	\item the group of \emph{$n$-cycles}, denoted by $\hat{Z}_{[(\zeta, \Lambda, B)], n}(X)$, as the quotient of the group of $n$-pre-cycles by the free subgroup generated by elements of the form:
\begin{itemize}
	\item $(M, \hat{u}, \hat{\alpha} + \hat{\beta}, f) - (M, \hat{u}, \hat{\alpha}, f) - (M, \hat{u}, \hat{\beta}, f)$;
	\item $(M, \hat{u}, \hat{\alpha}, f) - (M_{1}, \hat{u}\vert_{M_{1}}, \hat{\alpha}\vert_{M_{1}}, f\vert_{M_{1}}) - (M_{2}, \hat{u}\vert_{M_{2}}, \hat{\alpha}\vert_{M_{2}}, f\vert_{M_{2}})$, for $M = M_{1} \sqcup M_{2}$;
	\item $(M, \hat{u}, \varphi_{!}\hat{\alpha}, f) - (N, \hat{v}, \hat{\alpha}, f \circ \varphi)$ for $\varphi \colon N \to M$ a submersion;
\end{itemize}
	\item the group of \emph{$n$-boundaries}, denoted by $\hat{B}_{[(\zeta, \Lambda, B)], n}(X)$, as the subgroup of $\hat{Z}_{[(\zeta, \Lambda, B)], n}(X)$ generated by the cycles which are representable by a pre-cycle $(M, \hat{u}, \hat{\alpha}, f)$, such that there exits a quadruple $(W, \hat{U}, \hat{A}, F)$, where $W$ is a simply-connected manifold and $M = \partial W$, $\hat{U}$ is a Thom class of $W$ and $\hat{U}\vert_{M} = \hat{u}$, $F \colon W \to X$ is a smooth map satisfying $F\vert_{M} = f$ and $\hat{A} \in \hat{K}^{\bullet}_{[(\Xi, \Tau, \CC)]}(W)$, where $[(\Xi, \Tau, \CC)] := w'_{2}(W) - F^{*}[(\zeta, \Lambda, B)]$, such that $\hat{A}\vert_{M} = \hat{\alpha}$.
\end{itemize}
We set $K'_{[(\zeta, \Lambda, B)], n}(X) := \hat{Z}_{[\zeta, \Lambda, B)], n}(X) / \hat{B}_{[(\zeta, \Lambda, B)], n}(X)$.
\end{Def}

Condition \eqref{FWCondition} will be clarified soon. Now we are ready to define Cheeger-Simons characters in the framework of twisted differential K-theory.

\begin{Def}\label{GeneralizedCS} A \emph{Cheeger-Simons differential $\hat{K}^{\bullet}_{[(\zeta, \Lambda, B)]}$-character} of degree $n$ on $X$ is a couple $(\chi_{n}, \omega_{n})$, where:
\begin{equation}\label{GeneralizedCSDef}
	\chi_{n} \colon \hat{Z}_{[(\zeta, \Lambda, B)], n-1}(X) \to \R/\Z; \quad\qquad \omega_{n} \in \Omega^{n \,\textnormal{mod}\, 2}(X)
\end{equation}
such that, if $(M, \hat{u}, \hat{\beta}, f) = \partial(W, \hat{U}, \hat{B}, F)$, then:
\begin{equation}\label{CSFormula}
	\chi_{n}[(M, \hat{u}, \hat{\beta}, f)] = \int_{W} \Td(W) \wedge R(\hat{B}) \wedge F^{*}\omega_{n} \mod \Z.
\end{equation}
We denote by $\check{K}^{n}_{[(\zeta, \Lambda, B)]}(X)$ the group of characters of degree $n$.
\end{Def}

The following theorem can be considered as the definition of holonomy of a differential K-theory class on a cycle.

\begin{Theorem}\label{CShnThm} There is a natural graded-group isomorphism:
\begin{equation}\label{CShn}
\begin{split}
	\textnormal{CS} \colon & \hat{K}^{\bullet}_{[(\zeta, \Lambda, B)]}(X) \overset{\!\simeq}\longrightarrow \check{K}^{\bullet}_{[(\zeta, \Lambda, B)]}(X) \\
	& \hat{\alpha} \mapsto (\chi_{\hat{\alpha}}, R(\hat{\alpha})),
\end{split}
\end{equation}
where $\chi_{\hat{\alpha}}$ is defined, with respect to $[(M, \hat{u}, \hat{\beta}, f)] \in \hat{Z}_{[(\zeta, \Lambda, B)], \bullet-1}(X)$, by:
\begin{equation}\label{DefChi}
	\chi_{\hat{\alpha}}[(M, \hat{u}, \hat{\beta}, f)] := (p_{M})_{!}(\hat{\beta} \cdot f^{*}\hat{\alpha}).
\end{equation}
\end{Theorem}
The map $(p_{M})_{!}$ in formula \eqref{DefChi}, where $p_{M} \colon M \to \{pt\}$, is well-defined because of condition \eqref{FWCondition}. In fact, the class $\hat{\beta} \cdot f^{*}\hat{\alpha}$ is twisted by $[(\xi, \Theta, C)] + f^{*}[(\zeta, \Lambda, B)] = w'_{2}(M)$, hence, applying the Thom isomorphism, we get a class twisted by $w'_{2}(M) + w'_{2}(M) = 0$. It follows that, applying $(p_{M})_{!}$, we get a non-twisted class on the point, that belongs to $\hat{K}^{1}(\{pt\}) \simeq \R/\Z$. The proof of the fact that \eqref{CShn} is an isomorphism is analogous to \cite[Theorems 5.5 and 6.6]{FR}.

\subsection{Twisted K-homology}

We can provide a definition analogous to \ref{TwistedKHomDiff}, but replacing every differential class by the underlying topological one, generalizing \ref{DefKHom} to twisted K-homology through definition \eqref{DefTopKTDeligne}. We get the group $K_{[(\zeta, \Lambda, B)], n}(X)$, that is canonically isomorphic to the group $K'_{[(\zeta, \Lambda, B)], n}(X)$ defined in \ref{TwistedKHomDiff}. Such a group will be the natural one to classify topological D-brane charges. Fixing an orientation $\bar{u}$ of $X$, we get the twisted Poincar\'e duality, defined as follows (we set $x := \dim X$):
\begin{equation*}
\begin{split}
	\PD \colon & K^{\bullet}_{[(\zeta, \Lambda, B)]}(X) \overset{\!\simeq}\longrightarrow K_{w'_{2}(X) - [(\zeta, \Lambda, B)], x-\bullet}(X) \\
	& \alpha \mapsto [(X, \bar{u}, \alpha, \id_{X})],
\end{split}
\end{equation*}
the inverse map being the following one:
\begin{equation}\label{PDTwisted}
\begin{split}
	\PD^{-1} \colon & K_{w'_{2}(X) - [(\zeta, \Lambda, B)], \bullet}(X) \overset{\!\simeq}\longrightarrow K^{x-\bullet}_{[(\zeta, \Lambda, B)]}(X) \\
	& [(M, u, \alpha, f)] \mapsto f_{!}(\alpha).
\end{split}
\end{equation}
More generally, we define the cap-product as follows:
	\[\begin{split}
	\cap \colon & K_{[(\zeta, \Lambda, B)], k}(X) \times K^{h}_{[(\zeta', \Lambda', B')]}(X) \to K_{[(\zeta, \Lambda, B)] - [(\zeta', \Lambda', B')], k-h}(X) \\
	& [(M, u, \alpha, f)] \cap \beta := [(M, u, \alpha \cdot f^{*}\beta, f)],
\end{split}\]
the Poincar\'e duality being the cap-product with the fundamental class $[(X, \bar{u}, 1, \id_{X})] \in K_{w'_{2}(X), x}(X)$, where $1 := X \times \C$ is the trivial line bundle (i.e.\ the unit of ordinary K-theory). We will compare this definition of K-homology with other ones in literature (see \cite{BCW, Liu, Wang} and many others) in a future paper.

\subsection{Twisted-Integral Forms}

We call $\Omega_{[(\zeta, \Lambda, B)], \Kint}^{\bullet}(X)$ the graded group of closed (poly)forms with integral twisted $K$-periods, in the sense that we now specify. Given a form $\omega \in \Omega^{n \,\textnormal{mod}\, 2}_{H\textnormal{-}\cl}(X)$ and a twisted $K$-homology class $[(M, u, \alpha, f)] \in K_{[(\zeta, \Lambda, B)], n}(X)$, we consider the following pairing:
\begin{equation}\label{Pairing}
	\langle \omega, [(M, u, \alpha, f)] \rangle := \int_{M} f^{*}\omega \wedge \ch(\alpha) \wedge \Td(M).
\end{equation}
The form $f^{*}\omega \wedge \ch(\alpha) \wedge \Td(M)$ is closed (since $\omega$ is $H$-closed), a representative of $\ch(\alpha)$ is $-H$-closed (since $\alpha$ is twisted by $w'_{2}(M) - f^{*}[(\zeta, \Lambda, B)]$ because of condition \eqref{FWCondition}), and a representative of $\Td(M)$ is closed (since it is twisted by $w'_{2}(M)$), therefore the integral on $M$ is well-defined. We say that $\omega$ is \emph{twisted $K$-integral} or that it \emph{has integral twisted $K$-periods} if the pairing \eqref{Pairing} takes an integral value for any $K$-homology class. As in the non-twisted framework, a form is $K$-integral if and only if its cohomology class belongs to the image of the Chern character. The curvature of a twisted $K$-theory class is $K$-integral, since:
\begin{align*}
	\langle R(\hat{\beta}), [(M, u, \alpha, f)] \rangle &= \int_{M} R(f^{*}\hat{\beta}) \wedge \ch(\alpha) \wedge \Td(M) \\
	&= \int_{M} \ch\bigl(f^{*}I(\hat{\beta}) \cdot \alpha\bigr) \wedge \Td(M).
\end{align*}
By condition \eqref{FWCondition} the class $f^{*}\beta \cdot \alpha$ is $w'_{2}(M)$-twisted, hence we can use the Freed-Lott model, representing it as a difference of two classes of the form $[(E, \nabla, \omega)]$. We have that
	\[\int_{M} \ch(E) \wedge \Td(M) \in \Z
\]
because of the (twisted) index theorem.

%%%%%%%%%%%%%%%%%%%%%%%%%%%%%%%%%%%%%%%%%%%%%%%%%%%%%%%%%%%%%%%%%%%%%%%%%%%%%%%%%%%%%%%%%%%%%%%

\section{D-Branes charges and Wess-Zumino action}\label{DBraneCWZ}

Now we have all the tools to correctly define the world-volume, the Wess-Zumino action and the topological charge of a D-brane with the language of twisted K-theory. We start with a brief review of the cohomological classification, in order to compare it with the K-theoretical framework. We assume for simplicity that the space-time $X$ is compact, but this hypothesis can be removed quite easily.

\subsection{Cohomological classification}

Using the language of ordinary (i.e.\ singular) differential cohomology, we suppose that the $B$-field vanishes on the whole space-time, since the construction of the twisted version is more problematic than the K-theoretical counter-part. In this setting, the local Ramond-Ramond potentials $C_{p}$ are (a part of) a connection on an abelian $(p-1)$-gerbe, the latter being described by a Deligne cohomology class $\hat{\alpha} \in \hat{H}^{p+1}(X)$. Concretely, $\hat{\alpha}$ is described by the local data \eqref{PGerbesCocycle}, the corresponding curvature being the field strength $G_{p+1}$. Moreover, a D$(p-1)$-brane world-volume is thought of as a $p$-dimensional submanifold $W$ of the space-time $X$, that, via a suitable triangulation, defines a singular $p$-cycle. When the numerical charge is $q \in \mathbb{Z}$, we think of a stack of $q$ D-branes (anti-branes if $q < 0$), whose underlying cycle is $qW$. The topological charge of the D-brane is the Poincar\'e dual of the underlying homology class $[qW] \in H_{p}(X; \mathbb{Z})$, therefore it belongs to $H^{n-p}(X; \Z)$, where $n := \dim \, X$. The Wess-Zumino action, usually written as $\int_{W} C_{p}$, is the holonomy of $\hat{\alpha}$ on the cycle $W$. In the particular case $I(\hat{\alpha}\vert_{W}) = 0$, we have that $\hat{\alpha}\vert_{W} = a(C_{p})$, where $C_{p}$ is a global form on $W$, and the holonomy of $\hat{\alpha}$ coincides with $\int_{W} C_{p} \mod \Z$. Nevertheless, for a generic class $\hat{\alpha}$, such an integral is meaningless. Moreover, assuming for simplicity that $W$ is the only world-volume in $X$, the violated Bianchi identity is:
	\[dG_{n-p-1} = q \cdot \delta(W) \qquad\qquad dG_{p+1} = 0.
\]
This implies that $G_{n-p-1}$ is a closed form in the complement of $W$ and, if $L$ is a linking manifold of $W$, with linking number $l$, we get $\frac{1}{l}\int_{L} G_{n-p-1} = q \in \Z$. That's why the Ramond-Ramond field strength is quantized, so that it is correct to think of it as the curvature of a connection.

\subsection{$K$-Theoretical classification}

In the $K$-theoretical framework, we impose no constraints on the $B$-field, but we start supposing that the space-time and each D-brane world-volume are simply-connected, since, in this case, all of the twisted groups are canonically defined, in the sense that they only depend on the cohomology classes involved.

\paragraph{\textbf{Ramond-Ramond fields.}} In order to describe the Ramond-Ramond fields, the two meaningful groups are $\hat{K}^{0}_{[(\zeta, \Lambda, B)]}(X)$ and $\hat{K}^{1}_{[(\zeta, \Lambda, B)]}(X)$. More precisely, in type IIB theory the Ramond-Ramond fields have even-degree field-strength, thus they are jointly classified by a class $\hat{\alpha} \in \hat{K}_{[(\zeta, \Lambda, B)]}(X)$; similarly, in type IIA theory they have odd-degree field-strength, hence they are classified by $\hat{\beta} \in \hat{K}_{[(\zeta, \Lambda, B)]}^{1}(X)$. The twisting class $[(\zeta, \Lambda, B)] \in \hat{H}^{2}(X)$ is the $B$-field in the space-time.  We discuss the features of $\hat{\alpha}$, the discussion about $\hat{\beta}$ being analogous. The curvature $R(\hat{\alpha})$ is a twisted-K-integral form $G_{\ev} \in \Omega^{\ev}_{H\textnormal{-}\cl}(X)$. The component of degree $2p$ is the field-strength $G_{2p}$. If we consider a local chart $U$ of $X$, then $\hat{\alpha}\vert_{U}$ is topologically trivial, hence it can be represented in the form $a(C_{\odd})$, with $C_{\odd} \in \Omega^{\odd}(U)$, unique up to the addition of a twisted $K$-integral form. The component of degree $2p-1$ is the local potential $C_{2p-1}$. This means that the potentials are a local expression of a global twisted differential $K$-theory class, which is the complete datum encoded in the space-time.

\paragraph{\textbf{World-volume and Wess-Zumino action.}} The following definition will allow for the mathematical description of D-brane world-volumes, starting from statement $(\star)$ in remark \ref{RmkSCBTilde}.

\begin{Def}\label{DefBundlePreC} Let $X$ be a simply-connected manifold with a fixed class $[(\zeta, \Lambda, B)] \in \hat{H}^{2}(X)$. A \emph{bundle-$n$-pre-cycle} on $X$ is a quadruple $(M, \hat{u}, [(E, \nabla), \tilde{B}], f)$, where:
\begin{itemize}
	\item $M$ is a smooth compact simply-connected $n$-manifold (without boundary) and $\hat{u}$ is a ($w'_{2}(M)$-twisted) Thom class of $M$;
	\item $f \colon M \to X$ is a smooth map;
	\item $(E, \nabla)$ is a $\xi$-twisted non-integral vector bundle with connection and $\tilde{B}$ a global 2-form on $M$ such that
	\begin{equation}\label{FWConditionTorsion}
		[(\xi, 0, \tilde{B})] = w'_{2}(M) - f^{*}[(\zeta, \Lambda, B)];
	\end{equation}
	\item two pairs $((E, \nabla), \tilde{B})$ and $((E', \nabla'), \tilde{B}')$ are equivalent if there exits a non-integral line bundle $(L, \nabla_{L})$ on $M$, with curvature $F$, such that $(E', \nabla') = (E, \nabla) \otimes (L, \nabla_{L})$ up to isomorphism and $\tilde{B}' = \tilde{B} + F$.
\end{itemize}
\end{Def}
Condition \eqref{FWConditionTorsion} coincides with \eqref{FWConditionSC} and it imposes that $f^{*}[\zeta]$ is a torsion class, since $\xi$ must be constant. Equivalently, it imposes that $d\tilde{B} = f^{*}H$, hence $f^{*}H$ must be exact.
\begin{Rmk}\label{RmkTorsionPreCCanonical} \emph{If $f^{*}H = 0$, then we have the canonical choice $\tilde{B} = 0$, hence we can represent a bundle-pre-cycle in the form $(M, \hat{u}, (E, \nabla), f)$, with a fixed non-integral vector bundle. This is coherent with remark \ref{RmkSCBTilde}.}
\end{Rmk}

\begin{Rmk}\label{BdlCyclesToCycles} \emph{We have a natural function from bundle-$n$-pre-cycles to differential $n$-pre-cycles, that sends $(M, \hat{u}, [(E, \nabla), \tilde{B}], f)$ to $(M, \hat{u}, \hat{\alpha}, f)$, where $\hat{\alpha} \in \hat{K}_{[(\xi, 0, \tilde{B})]}(M)$ is represented in the Freed-Lott model by $[(E, \nabla, 0)]$. In particular, the corresponding cycle is well-defined, hence we can compute the holonomy of a differential class on it.}
\end{Rmk}

We observe that, if $M$ is a Riemannian manifold, then, from a (topological)  twisted spin structure of $M$, we get canonically a differential Thom class $\hat{u}$ through the spin connection in the Freed-Lott model. Hence, in this case, it is enough to fix a topological orientation $u$. If $X$ is Riemannian and $f$ is an embedding, then $M$ can be endowed with the pull-back metric, hence we give the following definition.

\begin{Def}\label{DefRiemBundlePreC} Let $X$ be a simply-connected Riemannian manifold with a fixed class $[(\zeta, \Lambda,$ $B)] \in \hat{H}^{2}(X)$. A \emph{Riemannian bundle-$n$-pre-cycle} on $X$ is a quadruple $(M, u, [(E, \nabla), \tilde{B}], \rho)$, defined as in \ref{DefBundlePreC}, but with a topological Thom class $u$ and requiring that $\rho$ is an embedding.
\end{Def}

The previous definition perfectly fits to formalize the notion of D-brane world-volume. In fact, $M$ is the underlying submanifold, that we usually denote by $W$ when referring to a D-brane, $u$ is a spin structure, that must be fixed as a part of the background data, $\rho$ is the embedding in the space time (the latter being endowed with the space-time metric), $(E, \nabla)$ is the Chan-Patton bundle with the corresponding gauge theory and, when $\rho^{*}H$ does not vanish, the form $\tilde{B}$ is necessary to fix the bundle up to large gauge transformations, as we have seen in remark \ref{RmkSCBTilde}. Actually, we consider one bundle pre-cycle made by all the even-dimensional world-volumes or one made by all the odd-dimensional ones, depending whether we are considering the type IIA or type IIB theory. Moreover, we have seen in remark \ref{BdlCyclesToCycles} that a bundle pre-cycle canonically induces a differential cycle, on which we can compute the holonomy of a differential K-theory class through formula \eqref{DefChi}. When the differential K-theory class represents the Ramond-Ramond fields of the corresponding degree, such a holonomy is by definition the Wess-Zumino action. Summarizing:

\begin{Def}\label{DefWWWZ} A \emph{D-brane world-volume} in the space time $X$ is a $[(\zeta, \Lambda, B)]$-twisted Riemannian bundle-$(n-1)$-pre-cycle $(W, u, [(E, \nabla), \tilde{B}], \rho)$, where $n = 0$ for IIB theory and $n = 1$ for IIA theory. We denote by $\hat{z} \in \hat{Z}_{[(\zeta, \Lambda, B)], n-1}(X)$ the corresponding cycle and by $\hat{\alpha} \in \hat{K}_{[(\zeta, \Lambda, B)]}^{n}(X)$ the class representing the Ramond-Ramond fields. The \emph{Wess-Zumino action} of $\hat{\alpha}$ on $\hat{z}$ is the corresponding holonomy through formula \eqref{DefChi}, i.e.\ $(p_{W})_{!}(\hat{\beta} \cdot \rho^{*}\hat{\alpha})$, where $\hat{\beta} := [(E, \nabla, 0)] \in \hat{K}_{[(\xi, 0, \tilde{B})]}^{0}(W)$ using the Freed-Lott model, $\xi$ being any representative that realizes condition \eqref{FWConditionTorsion}.
\end{Def}

Usually the Wess-Zumino action is written supposing that $\hat{\alpha}$ is topologically trivial, so that it can be described by a global form $C$ up to twisted K-integral ones. It has the following form \cite{MM}, using the notation of definition \ref{DefWWWZ}:
\begin{equation}\label{PairingC}
	\int_{W} \rho^{*}C \wedge \ch(\nabla) \wedge \Td(W) \wedge \rho^{*}\Td(X)^{-\frac{1}{2}}.
\end{equation}
This easily follows from the right formula of \eqref{RAIntSubmersion}, normalizing $C$ with $\Td(X)^{-\frac{1}{2}}$. In fact:
\begin{align*}
	(p_{W})_{!}\bigl(\hat{\beta} \cdot \rho^{*}a(C \wedge \Td(X)^{-\frac{1}{2}})\bigr) &= (p_{W})_{!}a \bigl(R(\hat{\beta}) \wedge \rho^{*}(C \wedge \Td(X)^{-\frac{1}{2}})\bigr) \\
	&= (p_{W})_{!}a \bigl(\ch(\nabla) \wedge \rho^{*}(C \wedge \Td(X)^{-\frac{1}{2}})\bigr) \\
	&\!\!\overset{\eqref{RAIntSubmersion}}= \int_{W} \Td(W) \wedge \ch(\nabla) \wedge \rho^{*}(C \wedge \Td(X)^{-\frac{1}{2}}).
\end{align*}
We stated definition \ref{DefWWWZ} under the hypothesis that the space-time $X$ and the world-volume submanifold $W$ are simply-connected. Actually, definitions \ref{TwistedKHomDiff} and \ref{GeneralizedCS} and theorem \ref{CShnThm} hold with no hypotheses on $X$ and only requiring that $H_{1}(W; \Z) = 0$, the disadvantage being that they do not correctly generalize the analogous notions in the non-twisted setting under these weaker assumptions. Nevertheless, this does not affect definition \ref{DefWWWZ} in any way, therefore the mathematical formalization of the notions of world-volume and Wess-Zumino action, that we provided in this section, hold for any space-time and for world-volumes with vanishing first-degree homology. When $H_{1}(W; \Z)$ does not vanish, definition \ref{DefWWWZ} is canonical up to a flat line bundle on $W$, this ambiguity being physically motivated, since it perfectly corresponds to the ambiguity arising from the classification of the possible gauge theories on $W$, realized in section \ref{ClassGauge}. Of course, when the $B$-field vanishes in the whole $X$ and $W$ is spin, then we have no ambiguity in any case, since ordinary $K$-homology and $K$-characters are always well-defined.

\begin{Rmk} \emph{This picture completes (and partially corrects) some statements in \cite{FR4}. In fact, first of all it is not necessary to choose the differential refinement of the orientation $u$ on $W$ as a part of the background datum, since, as we have seen above, it is induced by the space-time metric. Moreover, in \cite{FR4} we stated that, if the $B$-field vanishes, then we can use ordinary K-theory and K-homology. This is ``canonically true'' only if the world-volume is spin, while the Freed-Witten anomaly in this case only imposes that it is spin$^{c}$. In \cite{FR4} we got ordinary K-theory fixing a spin$^{c}$ structure on $W$, instead of a $w_{2}(W)$-twisted spin structure. As we have seen in the paragraph before section \ref{FLModelTwisted}, this corresponds to the choice of an isomorphism between $w_{2}(W)$-twisted K-theory and ordinary K-theory. Nevertheless, the half-integral line bundle, involved in the choice of a spin$^{c}$-structure, is better included in the gauge theory, hence it is more natural to fix a twisted spin structure. It follows that, only if $w_{2}(W) = 0$, we can canonically avoid any twisting.}
\end{Rmk}

\begin{Rmk} \emph{As we anticipated in the introduction, definition \ref{DefWWWZ} shows that the Freed-Witten anomaly cancellation, that has been introduced with respect to the world-sheet action, turns out to be the condition for the Wess-Zumino action to be well-defined too. In fact, condition \eqref{FWCondition} is essentially equivalent to condition \eqref{FWConditionSA}. More precisely, the latter concerns the cocycles in a partially non-abelian relative Deligne class $[(\zeta, \Lambda, B, \bar{g}, -\bar{A})]$, hence it only involves the cochains $(\zeta, \Lambda)$ and $(\xi, \Theta)$, while the former involves the complete classes $[(\zeta, \Lambda, B)]$ and $[(\xi, \Theta, C)]$. Nevertheless, there is no contradiction, since it follows from definition \ref{DefTwistedDiffVB} that the bundle $[(g, A)]$ is only $(\xi, \Theta)$-twisted, while the corresponding $K$-theory class (through the Fredd-Lott model) is $(\xi, \Theta, C)$-twisted, independently of $\Tr\,F$. Hence, condition \eqref{FWCondition} can always be imposed if \eqref{FWConditionSA} is satisfied and vice-versa.}
\end{Rmk}

\paragraph{\textbf{Topological charge.}} Given a D-brane world-volume as in definition \ref{DefWWWZ}, the Poincar\'e dual of the underlying $K$-homology class is the topological charge, exactly as in the ordinary cohomology framework. More precisely, let us consider a world-volume $(W, u, [(E, \nabla), \tilde{B}], \rho)$. As above, we set $\hat{\beta} := [(E, \nabla, 0)] \in \hat{K}_{[(\xi, 0, \tilde{B})]}^{0}(W)$. The underlying topological cycle is $z = [(W, u, \beta, \rho)]$, inducing the corresponding K-homology class $[z] \in K_{[(\zeta, \Lambda, B)], n-1}(X)$. Applying the Poincar\'e duality \eqref{PDTwisted}, with $w_{2}(X) = 0$ since the space-time is spin, we get the charge $\rho_{!}(\beta) \in K^{x-n+1}_{-[(\zeta, \Lambda, B)]}(X)$. It follows that the latter group is the natural one to classify D-branes topologically. If $H_{1}(X; \Z) = 0$, such a group is canonically defined, otherwise we have an ambiguity up to a torsion line bundle on $X$. Again this ambiguity is physically motivated, since it is the topological version of the residual gauge freedom arising from the classification in \ref{ClassGauge}, explained in detail in \cite{FR2, FR4}.

\paragraph{\textbf{Quantization.}} Using classical cohomology, the integral of the field-strength along a linking manifold is the numerical charge of the D-brane. A linking manifold $L$ of $W$ is the boundary of a manifold $S$ that intersects $W$ transversely in a finite number of points of the interior. The number of such points is the linking number. In the twisted $K$-theoretical framework, we consider a \emph{linking $K$-cycle} $(L, u, F, \iota)$. Here $L$ is a ``generalized'' linking manifold, i.e., $L$ is the boundary of a manifold $S$ such that $S$ and $W$ intersect transversely in a submanifold (without boundary) contained in the interior of $S$. If $S \cap W$ is $0$-dimensional, we get a linking manifold in the usual sense. Moreover, the twisted bundle $F$ represents a $K$-theory class and $\iota \colon L \hookrightarrow X$ is the embedding. We consider the even-dimensional field-strengths $G_{ev}$, the discussion about $G_{odd}$ being analogous. The violated Bianchi identity is \cite{MM}:
\begin{equation}\label{BianchiK}
	d_{H}G_{ev} = \delta(W) \wedge \ch\nabla \wedge \Td(W) \wedge \Td(X)^{-\frac{1}{2}}.
\end{equation}
Equation \eqref{BianchiK} implies that $G_{ev} \wedge \Td(X)^{-\frac{1}{2}}$ is $K$-quantized and the pairing with a linking $K$-cycle gives the corresponding charge. In fact, choosing any connection $\nabla'$ on $F$, since $G_{ev}$ is $H$-closed and $\ch(\nabla')$ is $-H$-closed, we have:
	\[\begin{split}
	\langle G_{ev} \wedge \Td(X)^{-\frac{1}{2}}, &(L, u, F, \iota) \rangle = \int_{L} G_{ev} \wedge \Td(X)^{-\frac{1}{2}} \wedge \ch(F) \wedge \Td(L) \\
	& = \int_{S} d\bigl(G_{ev} \wedge \Td(X)^{-\frac{1}{2}} \wedge \ch(\nabla') \wedge \Td(S)\bigr) \\
	& = \int_{S} d_{H}G_{ev} \wedge \Td(X)^{-\frac{1}{2}} \wedge \ch(\nabla') \wedge \Td(S) \\
	& = \int_{S} \delta(W) \wedge \ch(\nabla \otimes \nabla') \wedge \frac{\Td(W) \wedge \Td(S)}{\Td(X)} \\
	& = \int_{S \cap W} \ch(E \otimes F) \wedge \Td(S \cap W) \in \mathbb{Z}.
\end{split}\]

\subsection{Comparing the two frameworks}

Now we have all the elements in order to draw a complete parallel between the two classification schemes of D-branes. Table \ref{fig:ComparisonPhys} shows such a parallel.

\begin{table*}[h!]
	\centering
		\begin{tabular}{|l|l|l|}
			\hline & & \\ & \textbf{Singular cohomology} & \textbf{Twisted $K$-theory} \\ & & \\ \hline
			& & \\ \textbf{World-vol.} & Singular cycle $qW$ & Riem.\ bundle-pre-cycle \\ & & $(W, u, [(E, \nabla), \tilde{B}], \rho)$ \\ & & \\ \hline
			& & \\ \textbf{Top.\ charge} & Sing.\ coh.\ class $\PD_{X}[qW]$ & Twisted $K$-th.\ class \\ & & $\PD_{X}[(W, u, \hat{\beta}, \rho)] = \rho_{!}\hat{\beta}$ \\ & & \\ \hline
			& & \\ \textbf{RR fields} & Ordinary diff.\ cohom.\ class & Diff.\ Twisted $K$-th.\ class \\  & & \\ & Integral field strength & $K$-Integral field strength \\ & & \\ \hline
			& & \\ \textbf{WZ action} & Holonomy of the RR fields & Twisted $K$-Holonomy of \\ & & the RR fields \\ & & \\ \hline
			& & \\ \textbf{Num.\ charge} & $\int$ f.s.\ over a linking manifold & $\int$ f.s.\ over a linking $K$-cycle \\ & & \\ \hline
		\end{tabular}
\caption{Comparison.}\label{fig:ComparisonPhys}
\end{table*}

%%%%%%%%%%%%%%%%%%%%%%%%%%%%%%%%%%%%%%%%%%%%%%%%%%%%%%%%%%%%%%%%%%%%%%%%%%%%%%%%%%%%%%%%%%%%%%%

\section*{Acknowledgements}

The first author was partially supported by FAPESP (Funda\c{c}\~ao de Amparo \`a Pesquisa do Estado de S\~ao Paulo), processo 2014/03721-3. The second author was financed by the Coordena\c c\~ao de Aperfei\c coamento de Pessoal de N\'ivel Superior - Brasil (CAPES) - Finance Code 001.

\end{document}